\documentclass[a4paper,11pt]{article}
\pdfoutput=1
\usepackage[utf8]{inputenc}
\usepackage[T1]{fontenc}
\usepackage[english]{babel}
\usepackage[colorlinks=true
,urlcolor=blue
,anchorcolor=blue
,citecolor=blue
,filecolor=blue
,linkcolor=blue
,menucolor=blue
,linktocpage=true
,pdfproducer=medialab
,pdfa=true
]{hyperref}
\usepackage{amsmath,amsthm,amsbsy,amssymb,amsfonts,graphicx,mathrsfs,bm,cite,color,accents}
\usepackage[boxsize=1em]{ytableau}
\usepackage[customcolors]{hf-tikz}
\hfsetfillcolor{blue!5}
\hfsetbordercolor{white}
\hyphenchar\font=\string"7F
\usepackage{tikz}
\usetikzlibrary{calc}
\usetikzlibrary{patterns}
\DeclareMathAlphabet{\mathdsl}{U}{bbm}{m}{sl}
\DeclareMathAlphabet{\numbb}{U}{BOONDOX-ds}{m}{n}

\usepackage[utf8]{inputenc}
\renewcommand{\thefootnote}{\arabic{footnote}}

\makeatletter
\catcode`\@=11
\@addtoreset{equation}{section}

\makeatother

\newcommand{\dd}{\rmd}

\newcommand{\Ah}{\hat{A}}
\newcommand{\Bh}{\hat{B}}
\newcommand{\Ch}{\hat{C}}
\newcommand{\Dh}{\hat{D}}
\newcommand{\Eh}{\hat{E}}
\newcommand{\Fh}{\hat{F}}
\newcommand{\Gh}{\hat{G}}

\newcommand{\Ih}{\hat{I}}
\newcommand{\Jh}{\hat{J}}
\newcommand{\Kh}{\hat{K}}
\newcommand{\Lh}{\hat{L}}

\newcommand{\rmd}{\mathrm{d}}

\newcommand{\cA}{\mathcal A}\newcommand{\cB}{\mathcal B}
\newcommand{\cD}{\mathcal D}
\newcommand{\cE}{\mathcal E}\newcommand{\cF}{\mathcal F}
\newcommand{\cG}{\mathcal G}\newcommand{\cH}{\mathcal H}
\newcommand{\cI}{\mathcal I}\newcommand{\cJ}{\mathcal J}
\newcommand{\cK}{\mathcal K}\newcommand{\cL}{\mathcal L}
\newcommand{\cM}{\mathcal M}\newcommand{\cN}{\mathcal N}
\newcommand{\cO}{\mathcal O}\newcommand{\cP}{\mathcal P}
\newcommand{\cQ}{\mathcal Q}\newcommand{\cR}{\mathcal R}

\newcommand{\cU}{\mathcal U}\newcommand{\cV}{\mathcal V}

\newcommand{\nn}{\nonumber}
\newcommand{\OO}{\text{O}}
\newcommand{\SL}{\text{SL}}
\newcommand{\GL}{\text{GL}}



\makeatletter
\newcommand{\Exp}[1]{\mathrm{e}^{#1}}
\makeatother

\newcommand{\gLie}{\cL}
\newcommand{\gLieDFT}{\mathbb{L}}
\newcommand{\tr}{\text{tr}}

\newcommand{\Aht}{\mathbb{A}}
\newcommand{\Bht}{\mathbb{B}}
\newcommand{\Cht}{\mathbb{C}}
\newcommand{\Dht}{\mathbb{D}}
\newcommand{\Eht}{\mathbb{E}}
\newcommand{\Fht}{\mathbb{F}}

\newcommand{\Iht}{\mathbb{I}}
\newcommand{\Jht}{\mathbb{J}}
\newcommand{\Kht}{\mathbb{K}}

\newcommand{\gA}{\mathcal{A}}
\newcommand{\gB}{\mathcal{B}}
\newcommand{\gC}{\mathcal{C}}
\newcommand{\gD}{\mathcal{D}}

\newcommand{\gI}{\mathcal{I}}
\newcommand{\gJ}{\mathcal{J}}
\newcommand{\gK}{\mathcal{K}}

\newcommand{\bat}{\bm{a}}
\newcommand{\bbt}{\bm{b}}
\newcommand{\bct}{\bm{c}}

\newcommand{\bba}{\alpha}
\newcommand{\bbb}{\beta}
\newcommand{\bbc}{\gamma}
\newcommand{\bbd}{\delta}

\renewcommand{\equiv}{:=}

\newcommand{\Gduality}{\mathbf{\mathrm{G}}_{\mathrm{D}}}
\newcommand{\Gdualityt}{\tilde{\mathbf{\mathrm{G}}}_{\mathrm{D}}}

\newcommand{\Vtheta}{\bm{\vartheta}}

\allowdisplaybreaks[3]

\begin{document}

\hypersetup{pageanchor=false}
\begin{titlepage}
	\renewcommand{\thefootnote}{\fnsymbol{footnote}}

	\vspace*{1.0cm}

	\centerline{\LARGE\textbf{Generalized Dualities for Heterotic and Type I Strings}}

	\vspace{1.0cm}

	\centerline{
		{Falk Hassler}%
		\footnote{E-mail address: falk.hassler@uwr.edu.pl},
		{Yuho Sakatani}%
		\footnote{E-mail address: yuho@koto.kpu-m.ac.jp}
		and
			{Luca Scala}%
		\footnote{E-mail address: luca.scala@uwr.edu.pl}
	}

	\begin{center}
		${}^\ast {}^\ddagger${\it University of Wroc\l{}aw, Faculty of Physics and Astronomy,}\\
		{\it Maksa Borna 9, 50-204 Wrocław, Poland}

		${}^\dagger${\it Department of Physics, Kyoto Prefectural University of Medicine,}\\
		{\it 1-5 Shimogamohangi-cho, Sakyo-ku, Kyoto 606-0823, Japan\\}
	\end{center}

	\begin{abstract}
	We define generalized dualities for heterotic and type I strings based on consistent truncations to half-maximal gauged supergravities in more than three dimensions. The latter are constructed from a generalized Scherk-Schwarz ansatz in heterotic double field theory that satisfies the strong constraint. Necessary and sufficient conditions on the resulting embedding tensor are discussed, showing that only certain gaugings, called geometric, can arise from this procedure. For all of them, we explicitly construct the internal geometry and gauge potentials. In general, this construction is not unique and permits different uplifts which are used to define generalized T-duality. Two examples are worked out underlying the utility of our approach to explore new dualities and uplifts of half-maximal gauged supergravities.
	\end{abstract}

	\thispagestyle{empty}
\end{titlepage}
\hypersetup{pageanchor=true}

\setcounter{footnote}{0}

\newpage
\hrule
\tableofcontents\vspace{2em}
\hrule

\section{Introduction}
Generalized dualities originate from the desire to extend abelian T-duality on the string's worldsheet to a larger class of target spaces. In this way, less and less restrictive families of dualities were revealed over time. These include, for example, non-abelian T-duality\cite{delaOssa:1992vci}, Poisson-Lie T-duality\cite{Klimcik:1995ux} and dressing cosets\cite{Klimcik:1996np}. All of them are summarized under the term \textit{generalized T-dualities}. As the name T(arget space)-duality suggests, they relate the classical dynamics of closed strings in different, dual target spaces by canonical transformations of the two-dimensional worldsheet theory. In contrast to abelian T-duality, which is known to be a genuine symmetry of string theory, their fate under quantum corrections is still under active investigation. Still, they received a new wave of interest over the last years since their underlying structures seem to be vital to integrablity in string theory and to gauged supergravities (gSUGRA) in various dimensions. The latter allow to go beyond strings, and hint, through solution preserving transformations for M-theory, that generalized U-dualities might also apply to membranes \cite{Sakatani:2020iad}.

To approach generalized dualities from the low-energy effective description of string and M-theory, consistent truncations have proven to be a valuable tool \cite{Hassler:2017yza,Demulder:2018lmj,Sakatani:2019zrs,Malek:2019xrf,Butter:2022iza}. They split the effective theory's spacetime into an external and internal part. The latter is fixed in terms of an ansatz, whose parameters become fields on the external spacetime. Effectively, this removes degrees of freedom from the theory and explains the name truncation. This process, however, is only consistent if the field equations of the truncated theory obtained by
\begin{enumerate}
	\item applying the truncation ansatz to the action before computing the field equations by variation of the truncated action,
	\item plugging the truncation ansatz into the field equations of the initial theory,
\end{enumerate}
are equivalent. Remarkably, this consistency condition results in severe constraints on admissible ans\"atze for truncations, making them in general very hard to find. In this framework, generalized dualities arise as equivalence relations identifying different consistent truncation ans\"atze which result in the same truncated theory. Therefore, in order to explore generalized dualities, one might start from the lower-dimensional, truncated theory and ask what are its higher-dimensional origins, called uplifts.

Implementing this programme for maximal gSUGRAs in ten or less dimensions gives rise to all the currently known generalized U-dualities \cite{Inverso:2017lrz,Bugden:2021wxg,Hassler:2022egz}. Since the understanding of generalized dualities for membrane models is still very limited, their connection to consistent truncation is currently the best way to access them. This is in stark contrast to the bosonic string case, where at first T-dualities were discovered on the worldsheet and later related to the low-energy effective theory. 

When it comes to generalized dualities, there is still a family of string theories that has not been studied extensively, namely, heterotic and type I. Recently, the first explorations in this direction have shown that $\cE$-models can be defined for them \cite{Hatsuda:2022zpi} and used to construct new integrable deformations \cite{Osten:2023cza}. Motivated by these hints towards interesting new physics, we take the complementary approach and use insights from half-maximal gSUGRAs as a natural way to define generalized dualities for heterotic and type I strings. Like their maximal relatives, they admit a full classification in terms of the embedding tensor \cite{Nicolai:2001ac,deWit:2003ja,Schon:2006kz,Dibitetto:2012rk,Dibitetto:2015bia,Dibitetto:2019odu}, and their uplifts are the concern of this article.

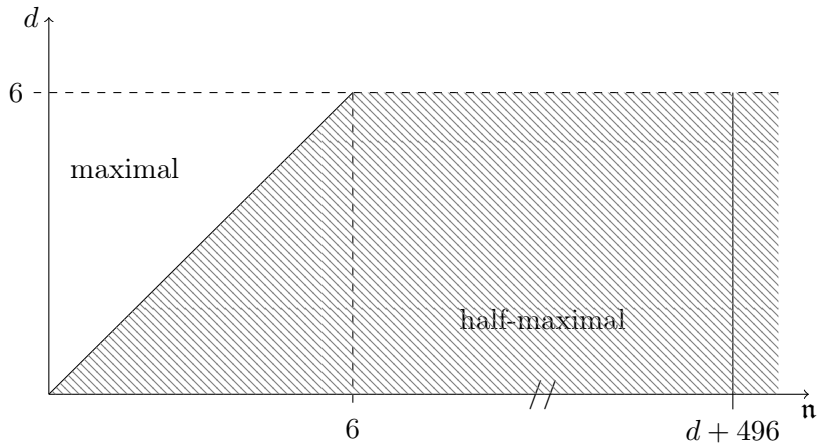
\begin{figure}[h]\label{fig:thSpace}
	\begin{tikzpicture}
		\node at (1,3) {maximal};
		\node at (6.5,1) {half-maximal};
		\draw[->] (0,0) -- (0,5) node[left]{$d$};
		\draw[->] (0,0) -- (10,0) node[below]{$\mathfrak{n}$};
		\draw[dashed]  (9.6,4) -- (-0.20,4) node[above,left]{6};
		\draw (0,0) -- (4,4);
		\draw[dashed] (4,4) -- (4,-0.20) node[below]{6};
		\draw (9,4) -- (9,-0.20) node[below]{$d+496$};
		\draw (6.5,0) node{$//$};
		\fill[pattern=north west lines, pattern color=gray] (0,0) -- (9.6,0) -- (9.6,4) -- (4,4);
	\end{tikzpicture}
	\centering
	\caption{Theory space of half-maximal gSUGRAs.}
\end{figure}

Half-maximal gSUGRAs in $10-d$ dimensions, for $1<d\le 6$, are classified by the embedding of their gauge group into the Lie group
\begin{equation}\label{eqn:halfmaxGduality}
	\Gduality = \Gdualityt(d) \times \mathrm{O}(d,\mathfrak{n})
	\qquad \text{with} \qquad
	\Gdualityt(d) = \begin{cases}
		\mathbb{R}^+ & 0 < d \le 5 \\
		\SL(2)        & d = 6
	\end{cases},
\end{equation}
where $\mathfrak{n}\ge 0$ counts the number of vector multiplets they have in addition to the gravity one. In three dimensions (where $d=7$), the product form of $\Gduality$ in \eqref{eqn:halfmaxGduality} is replaced by $\Gduality=$O(8,$\mathfrak{n}$). Conceptually this case is not different, but it is algebraically much more demanding and we thus restrict our discussion to $d\le 6$. Concerning possible uplifts, one may consider all the five perturbative superstring theories and M-theory. Depending on the amount of supersymmetry in the low-energy limit, they can be assigned to the two categories:
\begin{center}
	\begin{tabular}{p{0.4\textwidth}|p{0.4\textwidth}}
		Maximal (32 real supercharges) & Half-maximal (16 real supercharges) \\
		\begin{itemize}
			\item M-theory / type IIA
			\item Type IIB
		\end{itemize}      &
		\begin{itemize}
			\item Heterotic E$_8 \times$ E$_8$
			\item Heterotic SO(32)
			\item Type I.
		\end{itemize}
	\end{tabular}
\end{center}
Both cases are interesting but they require different strategies. To uplift to any theory in the first column, the truncation ansatz has to break half of the supersymmetry. This scenario has been analyzed in Exceptional Field Theory (ExFT) \cite{Berman:2011jh,Berman:2012vc,Hohm:2013vpa,Hohm:2013uia}, where the consistency of the truncation imposes strong restrictions on the number $\mathfrak{n} \le d$ of allowed vector multiplets \cite{Malek:2017njj}. This bound arises because the largest subgroup of E$_{d+1(d+1)}$ (the duality group of ExFT on a $d$-dimensional internal space) capable of hosting the embedding tensor of a half-maximal gSUGRA is O($d$, $d$). On the other hand, reproducing the low-energy limit of any of the theories in the second column requires generalized Scherk-Schwarz-type \cite{Dabholkar:2002sy,Aldazabal:2011nj,Geissbuhler:2011mx} uplifts on generalized parallelizable spaces \cite{Grana:2008yw,Lee:2014mla}. Here, $\mathfrak{n}\ge d$ is required, showing that these two approaches are not redundant but complementary. A priori, there is no upper bound on the number of vector multiplets; however, cancellation of the gauge anomaly in ten dimensions by the Green-Schwarz mechanism requires $\mathfrak{n}=d+496$ \cite{Green:1984sg}. Therefore, we focus on uplifts of consistent truncations of the shaded region in the theory space of half-maximal gSUGRAs, Fig.\ref{fig:thSpace}. They are central in defining generalized dualities for heterotic and type I strings.

Our presentation in this article follows the steps outlined above. We first introduce generalized Scherk-Schwarz reductions of half-maximal SUGRAs in Section~\ref{sec:genSSred}. Next, we ask which gSUGRAs can be realized in terms of these compactifications. More specifically, we define in Section~\ref{sec:halfmaxgSUGRAs} their embedding tensor in terms of generalized Lie derivatives for the different duality groups \eqref{eqn:halfmaxGduality} and then check how they relate to the generalized Scherk-Schwarz reductions presented before. It turns out that the central object in all these considerations is a generalized frame field. Initially, this latter object is used implicitly, but in Section~\ref{sec:frames} we eventually turn to its explicit construction, which extends the results of \cite{Sakatani:2021eqo}. For a generic upliftable gSUGRA with gauge group $G$, the frame is not unique but depends on the choice of certain admissible subgroups $H\subset G$. This is an hallmark of generalized dualities, as we further explain in Subsection~\ref{sec:gendualities}. To better clarify our construction, in Section~\ref{sec:examples} we present two examples. The first one explicitly demonstrates heterotic generalized T-duality at work, while the second one underlines that our construction is completely systematic and general, and that it can be applied also to complicated settings with various non-trivial fluxes.
A final Section~\ref{sec:conclusion} deals, in the end, with the conclusions of our analysis. 

\section{Generalized Scherk-Schwarz reduction}\label{sec:genSSred}
In the following, we present how half-maximal gSUGRAs can be derived from a generalized Scherk-Schwarz reduction of the ten-dimensional theory. Although conceptually not very different from the maximal setup, this kind of reduction has received much less attention in the literature to now. The ten-dimensional starting point is double field theory (DFT) with duality group $\OO(10, 10+n)$, where $n$ denotes the dimension of the gauge group $\cG$ of heterotic/type I supergravity. This theory is also known as heterotic DFT \cite{Hohm:2011ex} and can be either formulated in terms of a generalized dynamical metric or in terms of a generalized frame, containing all the dynamical fields except for the dilaton. For our purposes, the latter is more suited and we will review, now, this construction before specifying the reduction ansatz.

\subsection{Heterotic double field theory}\label{sec:hetDFT}
In the frame formulation of heterotic DFT, all the bosonic fields are contained in the T-duality-invariant dilaton $d$ and in the generalized vielbein $\mathbb{E}_{\Aht}{}^{\Iht}\in \OO(10,10+n)$. The latter relates flat indices, like $\Aht$, to their curved counterparts, here $\Iht$\footnote{Since in the paper we will make use of a large number of different kind of indices, we summarized them for the reader's convenience in Appendix \ref{app:index}.}. Both these kind of indices take values in the fundamental representation of the duality group. Instead of using the fundamental fields directly, the action and the associated field equations can be written in terms of the following generalized fluxes
\begin{align}
	\mathbb{F}_{\Aht\Bht\Cht} \equiv -3\,\mathbb{D}_{[\Aht}\mathbb{E}_{\Bht}{}^{\Iht}\,\mathbb{E}_{|\Iht|\Cht]}\,,\qquad
	\mathbb{F}_{\Aht} \equiv 2\,\mathbb{D}_{\Aht} d - \partial_{\Jht}\mathbb{E}_{\Aht}{}^{\Jht}\,,
\end{align}
with $\mathbb{D}_{\Aht}\equiv \mathbb{E}_{\Aht}{}^{\Iht}\,\partial_{\Iht}$.
Here, the indices $\Aht$ and $\Iht$ are raised or lowered with the $\OO(10,10+n)$-invariant metrics $\eta_{\Aht\Bht}$ and $\eta_{\Iht\Jht}$, respectively. In all the equations we have to impose the section condition $\eta^{\Iht\Jht}\,\partial_{\Iht} \,\cdot\, \partial_{\Jht} \, \cdot \, = 0$, solved by the canonical choice
\begin{equation}\label{eqn:solSC}
	\partial_{\Iht} = \begin{pmatrix} \partial_m & \partial_{\cI} & \partial^m\end{pmatrix}=\begin{pmatrix} \partial_m & 0 & 0\end{pmatrix}\,.
\end{equation}
This dictates how we decompose $\OO(10,10+n)\rightarrow\GL(10)\times\cG$ by putting the theory on section. The $m$ indices label the fundamental of the first factor in the decomposition, while $\cI$ correspond to the adjoint of the second factor. Flat indices follow the same convention, namely $\Fht_{\Aht} = \begin{pmatrix} \Fht_{\hat{a}} & \Fht_{\cA} & \Fht^{\hat{a}} \end{pmatrix}$\,.

The Lagrangian density of heterotic DFT is given by $\cL_{\text{DFT}} = \Exp{-2d}\,\mathbb{R} $ with
\begin{align}
	\mathbb{R} = -\tfrac{1}{12}\,\mathbb{H}^{\Aht\Dht}\,\mathbb{H}^{\Bht\Eht}\,\mathbb{H}_{\Cht\Fht}\,\hat{\mathbb{F}}_{\Aht\Bht}{}^{\Cht}\,\hat{\mathbb{F}}_{\Dht\Eht}{}^{\Fht} - \tfrac{1}{4}\,\mathbb{H}^{\Aht\Bht}\, \hat{\mathbb{F}}_{\Aht\Dht}{}^{\Cht}\,\hat{\mathbb{F}}_{\Bht\Cht}{}^{\Dht} + 2\,\mathbb{H}^{\Aht\Bht}\,\mathbb{D}_{\Aht} \mathbb{F}_{\Bht} - \mathbb{H}^{\Aht\Bht}\, \mathbb{F}_{\Aht} \, \mathbb{F}_{\Bht} \,.
\end{align}
Here, $\mathbb{H}_{\Aht\Bht}\in \OO(10,10+n)$ is a constant matrix (chosen for convenience to be diagonal),
\begin{equation}\label{eqn:10dhatF}
	\hat{\mathbb{F}}_{\Aht\Bht}{}^{\Cht}\equiv \mathbb{F}_{\Aht\Bht}{}^{\Cht} + \mathbb{E}_{\Aht}{}^{\Iht}\,\mathbb{E}_{\Bht}{}^{\Jht}\,\mathbb{E}_{\Kht}{}^{\Cht}\,\Sigma_{\Iht\Jht}{}^{\Kht}\,,
\end{equation}
and $\Sigma_{\Iht\Jht}{}^{\Kht}$ is a constant torsion implementing the gauge group $\cG$ in ten dimensions. This torsion is fixed by
\begin{equation}\label{eqn:torsiondef}
	\Sigma_{\Iht\Jht}{}^{\Kht} = \begin{cases}
		f_{\cI\cJ}{}^{\cK}      \\
		0 & \text{otherwise\,,}
	\end{cases}
\end{equation}
where $f_{\cI\cJ}{}^{\cK}$ denote the structure constants of $\cG$'s Lie algebra. The reason to introduce this contribution in a separate term is that it cannot be generated by a frame $\Eht_{\Aht}{}^{\Iht}$ which is compatible with the section \eqref{eqn:solSC}. To keep track of this special contribution we decorate quantities that contain it with a hat. Additionally, one needs to adapt the generalized Ricci scalar $\mathbb{R}$ by adding the generalized scalar
\begin{align}
	\hat{\mathbb{Z}} \equiv \eta^{\Aht\Bht}\,\bigl(- \tfrac{1}{6}\,\hat{\mathbb{F}}_{\Aht\Cht}{}^{\Dht}\,\hat{\mathbb{F}}_{\Bht\Dht}{}^{\Cht} + 2\,\mathbb{D}_{\Aht}\mathbb{F}_{\Bht}- \mathbb{F}_{\Aht}\,\mathbb{F}_{\Bht} \bigr)\,,
\end{align}
which reduces to $\frac{1}{3!}\,\Sigma_{\Aht\Bht\Cht}\,\Sigma^{\Aht\Bht\Cht}$ under the section condition. Subtracting this term, we define the modified generalized Ricci scalar as
\begin{align}\label{eqn:defRhat}
	\hat{\mathbb{R}} \equiv -\tfrac{1}{12}\,\mathbb{H}^{\Aht\Dht}\,\mathbb{H}^{\Bht\Eht}\,\mathbb{H}_{\Cht\Fht}\,\hat{\mathbb{F}}_{\Aht\Bht}{}^{\Cht}\,\hat{\mathbb{F}}_{\Dht\Eht}{}^{\Fht} - \tfrac{1}{4}\,\mathbb{H}^{\Aht\Bht}\, \hat{\mathbb{F}}_{\Aht\Cht}{}^{\Dht}\,\hat{\mathbb{F}}_{\Bht\Dht}{}^{\Cht} + 2\,\mathbb{H}^{\Aht\Bht}\,\mathbb{D}_{\Aht} \mathbb{F}_{\Bht} - \mathbb{H}^{\Aht\Bht}\, \mathbb{F}_{\Aht} \, \mathbb{F}_{\Bht} - \hat{\mathbb{Z}} \,.
\end{align}

To see how this quantity relates to the action of ten-dimensional half-maximal supergravity
\begin{equation}\label{eqn:hmaxSUGRA10d}
	\begin{aligned}
		S        & =\int \dd^{10} x\, \sqrt{-g}\,\cL_{10} \qquad \text{with} \qquad g = \det(g_{mn})\,,
		\\
		\cL_{10} & \equiv \Exp{-2\Phi}\bigl(R + 4\,g^{mn}\,\partial_m\Phi\,\partial_n\Phi
		-\tfrac{1}{12}\, \hat{H}_{mnp}\,\hat{H}^{mnp}
		-\tfrac{1}{4}\, \kappa_{\gI\gJ}\,F_{mn}{}^{\gI}\,F^{mn\gJ}\bigr)\,,
	\end{aligned}
\end{equation}
with the field strengths 
\begin{align}
        \hat{H}_3 & \equiv \frac{1}{3!}\,\hat{H}_{mnp}\,\rmd x^m\wedge\rmd x^n\wedge \rmd x^p\,,\nn\\
        F_2^{\gI} & \equiv \frac{1}{2!}\,F_{mn}{}^{\gI}\,\rmd x^m\wedge\rmd x^n\,,\\ 
        \intertext{defined by}
    	\hat{H}_3 & \equiv \rmd B_2 - \tfrac{1}{2}\,\kappa_{\gI\gJ}\,A^{\gI}\wedge \rmd A^{\gJ} - \tfrac{1}{3!}\,f_{\gI\gJ\gK}\,A^{\gI}\wedge A^{\gJ}\wedge A^{\gK}\,,\nn
    	\\
    	F_2^{\gI} & \equiv \rmd A^{\gI} + \tfrac{1}{2}\,f_{\gJ\gK}{}^{\gI}\,A^{\gJ}\wedge A^{\gK}\,,
\end{align}
we assume the parametrization
\begin{equation}
    \begin{aligned}
    	\mathbb{E}_{\Aht}{}^{\Iht} & \equiv \begin{pmatrix} e_{\hat{a}}^n & 0 & 0 \\ 0 & \nu_{\cA}^{\cN} & 0 \\ 0 & 0 & e^{\hat{a}}_n
    	                                    \end{pmatrix}\begin{pmatrix} \delta_n^p & - A_n{}^{\cP} & -\frac{1}{2}\,A_n{}^{\cQ}\,A_{p\cQ} \\ 0 & \delta_{\cN}^{\cP} & A_{p\cN} \\ 0 & 0 & \delta^n_p
    	                                                 \end{pmatrix}
    	\begin{pmatrix} \delta_p^m & 0 & - B_{pm} \\ 0 & \delta_{\cP}^{\cM} & 0 \\ 0 & 0 & \delta^p_m
    	\end{pmatrix},
    	\\
    	\Exp{-2d}                  & \equiv \Exp{-2\Phi}\sqrt{-g}\,,
    \end{aligned}
\end{equation}
of the generalized frame and dilaton. Plugging it into \eqref{eqn:defRhat}, we find
\begin{equation}
	S = \int \dd x^{10}\, \Bigr[\Exp{-2d} \hat{\mathbb{R}} - \partial_m\bigl(4\,\Exp{-2\Phi}\sqrt{-g}\,g^{mn}\,\partial_n\Phi\bigr)\Bigr]\,,
\end{equation}
and thereby match the SUGRA action up to a boundary term. Finally, we can slightly modify $\hat{\mathbb{R}}$ into
\begin{equation}
	\hat{\mathbb{R}}'= -\tfrac{1}{12}\,\mathbb{H}^{\Aht\Dht}\,\mathbb{H}^{\Bht\Eht}\,\mathbb{H}_{\Cht\Fht}\,\hat{\mathbb{F}}_{\Aht\Bht}{}^{\Cht}\,\hat{\mathbb{F}}_{\Dht\Eht}{}^{\Fht} - \tfrac{1}{4}\,\mathbb{H}^{\Aht\Bht}\, \hat{\mathbb{F}}_{\Aht\Dht}{}^{\Cht}\,\hat{\mathbb{F}}_{\Bht\Cht}{}^{\Dht}
	+ \mathbb{H}^{\Aht\Bht}\, \mathbb{F}_{\Aht} \, \mathbb{F}_{\Bht} - \hat{\mathbb{Z}}
\end{equation}
to have all derivatives contained in the generalized fluxes and $\hat{\mathbb{Z}}$. As a result we can write the action of the ten-dimensional half-maximal SUGRA as
\begin{equation}
	S = \int \dd^{10} x \, \Bigr[\Exp{-2d} \hat{\mathbb{R}}' - \partial_m\bigl(2\,\Exp{-2\Phi}\sqrt{-g}\,K_{n}{}^{nm}\bigr)\Bigl]\,,
\end{equation}
where we have defined
\begin{equation}\label{eqn:Kdefinition}
	K_{mn}{}^p \equiv \tfrac{1}{2}\,\bigl(F_m{}^p{}_n + F_n{}^p{}_m - F_{mn}{}^p\bigr)\,,\qquad
	F_{mn}{}^p \equiv -2\,\partial_{[m} e_{n]}^{\hat{a}}\,e_{\hat{a}}^p\,.
\end{equation}
In this paper we will consider $S = \int \dd^{10} x \, \Exp{-2d} \hat{\mathbb{R}}'$ without the boundary term as the ten-dimensional action.

\subsection{Hohm-Samtleben-like split formulation}\label{sec:splitHS}
In order to prepare the ground for the dimensional reduction of this action on an internal, $d$-dimensional, space we follow \cite{Hohm:2013nja} and choose a different splitting of the $\OO(10,10+n)$ indices,
\begin{align}\label{eqn:HSsplit}
	x^{\Iht} = \begin{pmatrix} x^\mu & x^I & x_\mu \end{pmatrix},\qquad \text{with} \qquad x^I \equiv \begin{pmatrix} x^i & x^{\gI} & x_i\end{pmatrix}\,.
\end{align}
There are several things to note here: First, we decompose
\begin{align}
	\OO(10,10+n) & \rightarrow \GL(D)\times \OO(d,d+n)\,,
	\intertext{where $D:=10-d$ and then further}
	\OO(d,d+n)   & \rightarrow \GL(d) \times \cG\,,
\end{align}
with $\mu=1,\dotsc,D$ and $i=D+1,\dotsc,10$\,. We also change the parametrization of the generalized frame and dilaton accordingly to
\begin{equation}
    \begin{aligned}
    	\mathbb{E}_{\Aht}{}^{\Iht} & = \begin{pmatrix}
    		                               e_{\bat}^\mu & - e_{\bat}^\nu\,A_\nu{}^I & -e_{\bat}^\nu\,(B_{\nu\mu}+\tfrac{1}{2}\,A_\nu{}^{K}\,A_{\mu K}) \\[1mm]
    		                               0            & \cV_A{}^{I}               & \cV_A{}^{K}\,A_{\mu K}                                           \\[1mm]
    		                               0            & 0                         & e_\mu^{\bat}
    	                               \end{pmatrix}, \\
    	\Exp{-2d}                  & =\Exp{-2\phi}\sqrt{-\det(g_{\mu\nu})}\,.
    \end{aligned}
\end{equation}
With this choice, one eventually obtains the fluxes
\begin{align}
    \begin{alignedat}{2}
    	\hat{\mathbb{F}}_{\bat\bbt\bct}    & = e_{\bat}^\mu\,e_{\bbt}^\nu\,e_{\bct}^\rho\,\cH_{\mu\nu\rho}\,,                      &  
    	\hat{\mathbb{F}}_{\bat\bbt}{}^C & = e_{\bat}^\mu\,e_{\bbt}^\nu\,\cF_{\mu\nu}{}^I\,\cV_I{}^C\,,
    	\\
    	\hat{\mathbb{F}}_{\bat\bbt}{}^{\bct} & = 2\,e_{[\bat}^\mu\,e_{\bbt]}^\nu\,D_\mu e_\nu^{\bct}\,,                          &   \hat{\mathbb{F}}_{\bat B}{}^C & = e_{\bat}^\mu\,\cV_B{}^I\,\bigl(D_\mu \cV_I{}^C - A_\mu{}^J\,\Sigma_{JI}{}^K \cV_K{}^C\bigr)\,,
    	\\
    	\hat{\mathbb{F}}^{\bat}{}_{\bbt C} & = e_\mu^{\bat}\,\cD_C e_{\bbt}^\mu \,,                                           &   \hat{\mathbb{F}}_{ABC} & = \hat{\cF}_{ABC} \equiv \cF_{ABC} + \cV_A{}^I\,\cV_B{}^J\,\cV_C{}^K\,\Sigma_{IJK}\,,
    	\\
    	\mathbb{F}_{\bat}            & = 2\,e_{\bat}^\mu\,D_\mu \phi -2\,e_{[\bat}^\mu\,e_{\bbt]}^\nu\,D_\mu e_\nu^{\bbt} \,, \hspace{0.7cm} &  
    	\mathbb{F}_{A} & = \cF_A - e^{-1}\,\cD_A e \,,
	\end{alignedat}
\end{align}
where $\cD_A\equiv \cV_A{}^I\,\partial_I$\,, $e\equiv \det (e_\mu^{\bat})$, and $\cF_{AB}{}^C$, $\cF_A$ are the generalized fluxes associated with $\cV_A{}^I$. Furthermore, we have defined
\begin{equation}
    \begin{aligned}
    	\cH_{\mu\nu\rho} & \equiv 3\,D_{[\mu}B_{\nu\rho]} - 3\, A_{[\mu}{}^I\,\partial_{\nu}A_{\rho]I} + \eta_{IJ}\,A_{[\mu}{}^I\,[A_\nu,\,A_{\rho]}]^J_{\text{D}} -\Sigma_{IJK}\,A_\mu{}^I\,A_\nu{}^J\,A_\rho{}^K\,,
    	\\
    	\cF_{\mu\nu}{}^I & \equiv 2\,\partial_{[\mu} A_{\nu]}{}^I - [A_{[\mu},\,A_{\nu]}]_{\text{D}}^I + \Sigma_{JK}{}^I\,A_\mu{}^J\,A_\nu{}^K + \partial^I B_{\mu\nu}\,,
    \end{aligned}
\end{equation}
where $[\cdot,\cdot]_\text{D}$ is the DFT D-bracket and we made use of the covariant derivative $D_\mu \equiv \partial_\mu - \gLieDFT_{A_\mu}$. The latter is defined in terms of the generalized Lie derivative for the internal space, the one for the duality group $\OO(d,d+n)$, which acts on a generalized vector $W^J$ of weight zero by another vector $V^I$ as
\begin{equation}\label{eqn:genLieDef}
	\gLieDFT_{V} W^I = V^J \partial_J W^I - \left( \partial_J V^I - \partial^I V_J \right) W^J\,.
\end{equation}
From this, we find that the action of the covariant derivative on the fundamental tensors is given by
\begin{equation}
    \begin{aligned}
    	D_\mu e_\nu^{\bat}     & = \partial_\mu e_\nu^{\bat} - A_\mu{}^I \partial_I e_\nu^{\bat}\,,
    	\\
    	D_\rho B_{\mu\nu} & = \partial_\rho B_{\mu\nu} - A_\rho{}^I \partial_I B_{\mu\nu}\,,
    	\\
    	D_\mu \cV_{I}{}^A & = \partial_\mu \cV_{I}{}^A - \bigl(A_\mu{}^J \partial_J \cV_{I}{}^A + \cV_{J}{}^A\,\partial_I A_\mu{}^J - \cV_{J}{}^A\,\partial^J A_{\mu I} \bigr)\,,
    	\\
    	D_\mu \phi        & = \partial_\mu \phi - A_\mu{}^I\,\partial_I\phi + \tfrac{1}{2}\,\partial_I A_\mu{}^I\,,
    \end{aligned}
\end{equation}
and can be easily extended to all the other relevant fields or combinations of them.

Taking into account the splitting $x^m = \begin{pmatrix} x^\mu & x^i \end{pmatrix}$, which follows from \eqref{eqn:HSsplit}, the section fixed in \eqref{eqn:solSC} implies $\partial^\mu=0$\,. With that, the ten-dimensional SUGRA action \eqref{eqn:hmaxSUGRA10d} can be rewritten as
\begin{equation}\label{eq:DFT-action-d}
	\begin{aligned}
		S     & = \int\rmd^{10} x\,\bigl[\cL_D - \partial_\mu\bigl(2\,e\,\Exp{-2\phi}\,\cK_{\nu}{}^{\nu\mu}\bigr)\bigr]\,,
		\\
		\cL_D & \equiv e\,\Exp{-2\,\phi}\bigl(\widehat{R}+4\,D_\mu\phi\,D^\mu\phi -\tfrac{1}{12}\,\cH_{\mu\nu\rho}\,\cH^{\mu\nu\rho}
		+ \tfrac{1}{8}\,\hat{D}_\mu \mathbb{H}^{IJ} \hat{D}^\mu \mathbb{H}_{IJ} - \tfrac{1}{4}\,\mathbb{H}_{IJ}\,\cF^{\mu\nu I} \cF_{\mu\nu}{}^{J} -V\bigr) \,,
	\end{aligned}
\end{equation}
where $e\equiv\sqrt{-\det(g_{\mu\nu})}$\,, $\cK_{\mu}{}^{\nu\rho}$ is defined in terms of $e_{\bat}^\mu$ similarly to $K_{m}{}^{np}$ (equation \eqref{eqn:Kdefinition}), and
\begin{equation}
    \begin{aligned}
    	\mathbb{H}_{IJ}             & \equiv \cV^A{}_I \mathbb{H}_{AB} \cV^B{}_J\,,                                   \\
    	\hat{D}_\mu \mathbb{H}_{IJ} & \equiv D_\mu \mathbb{H}_{IJ} - 2\,A_\mu{}^K\,\Sigma_{K(I}{}^L\,\mathbb{H}_{J)L}\,.
    \end{aligned}
\end{equation}
Furthermore, we have defined the following quantities
\begin{equation}\label{eqn:ingr-DFT-action-d}
	\begin{aligned}
		\widehat{R} & \equiv R - e^{\bat\mu}\,\cF_{\mu\nu}{}^I\, \partial_{I} e^\nu_{\bat}\,,
		\\
		V           & \equiv -\bigl(\hat{\cR} + \tfrac{1}{4}\,\mathbb{H}^{IJ}\,\partial_{I}g^{\mu\nu}\,\partial_{J}g_{\mu\nu} + e^{-2}\,\mathbb{H}^{AB}\,\cD_A e\,\cD_B e -2\,e^{-1}\,\mathbb{H}^{AB}\,\cF_A\,\cD_B e\bigr)\,,
		\\
		\hat{\cR}   & \equiv -\tfrac{1}{12}\,\mathbb{H}^{AD}\,\mathbb{H}^{BE}\,\mathbb{H}_{CF}\,\hat{\cF}_{AB}{}^{C}\,\hat{\cF}_{DE}{}^{F} - \tfrac{1}{4}\,\mathbb{H}^{AB}\, \hat{\cF}_{AD}{}^{C}\,\hat{\cF}_{BC}{}^{D}
		+ \mathbb{H}^{AB}\, \cF_{A} \, \cF_{B} - \hat{\mathbb{Z}} \,.
	\end{aligned}
\end{equation}
Note that the last two terms in the potential $V$ have been missed in \cite{Hohm:2013nja}, but they are important to reproduce the scalar potential for non-unimodular gaugings (we will explain the meaning of this in Section \ref{sec:reltogSS}, when we will compute the scalar potential of reduced theories).

\subsection{Scherk-Schwarz ansatz}\label{sec:gSSansatz}
Each half-maximal gauged supergravity can be characterized by its embedding tensor $X_{\Ah\Bh}{}^{\Ch}$ \cite{Nicolai:2001ac,deWit:2003ja,Schon:2006kz,Dibitetto:2012rk,Dibitetto:2015bia,Dibitetto:2019odu}, where hatted indices are in the fundamental representation of the duality group $\Gduality$ defined in \eqref{eqn:halfmaxGduality}. There are various consistent choices for this tensor, but some of them cannot be uplifted to ten-dimensional SUGRA. Here, we will not consider those and instead focus on theories which result from a reduction of the action \eqref{eqn:hmaxSUGRA10d}, whose corresponding classes of embedding tensors (or gaugings) are called geometric. 
The embedding tensor $X_{\Ah\Bh}{}^{\Ch}$ always contain the $\OO(d,d+n)$ tensor $X_{AB}{}^{C}$, which can be decomposed into irreducible representations, $F_{AB}{}^C$ and $\xi_A$ (see~\eqref{eqn:Xdecomp} for the explicit decomposition). Comparing this $\OO(d,d+n)$ group with \eqref{eqn:halfmaxGduality}, we find $\mathfrak{n}=d+n$, and since $n\ge 0$, we immediately see how the restriction to the shaded region in the theory space (Fig.\ref{fig:thSpace}) arises. 

To reproduce all gauged supergravities compatible with the requirements elaborated above we need to make an appropriate Scherk-Schwarz reduction ansatz.
For this purpose, we consider a generalized frame field $E_A{}^I\in \mathbb{R}^+\times \OO(d,\mathfrak{n})$ that satisfies
\begin{align}\label{eqn:framealg}
	\gLieDFT_{E_A} E_B{}^I = - X_{AB}{}^{C}\,E_{C}{}^I\,,
\end{align}
where the generalized Lie derivative is given in \eqref{eqn:genLieDef}. How to explicitly construct such a frame will be the objective of Section \ref{sec:frames}; here, we just assume that it exists. 
To present a general expression independent of dimensions, here we focus on the case where the resulting gauged supergravity admits an action, but we will relax this assumption in the following sections. 
According to \eqref{eqn:halfmaxGduality}, the matrix $E_A{}^I$ is an element of $\mathbb{R}^+\times \OO(d,\mathfrak{n})$ and thus we need to rescale it to obtain an $\OO(d,\mathfrak{n})$ matrix:
\begin{align}\label{eqn:cEfromE}
	\cE_A{}^I \equiv \Exp{\Delta}\,E_A{}^I \in \OO(d,\mathfrak{n})\,.
\end{align}
From our construction of $E_A{}^I$, we find that $\cE_A{}^I$ and $\Delta$ satisfy
\begin{align}\label{eqn:defcF1}
	F_A & \equiv - \Exp{2\phi}\,\partial_I\bigl( \Exp{-2\phi}\,\cE_A{}^I\bigr) = \Exp{\Delta} (d-9)\,\xi_A\,,\qquad \text{with} \qquad  D_A\Delta = \xi_A\,,
	\intertext{and $D_A\equiv E_A{}^I\,\partial_I$. Furthermore, we define the generalized fluxes for $E_A{}^I$ as before by}\label{eqn:defcF3}
	F_{ABC} & \equiv - 3 E_{[A}{}^I \partial_I E_B{}^J E_{C]J}\,, \qquad\text{and}\qquad
	\hat{F}_{ABC} \equiv F_{ABC} + \Sigma_{ABC}\,.
\end{align}
Using these quantities, we make the Scherk-Schwarz ansatz \cite{Aldazabal:2011nj,Geissbuhler:2011mx}
\begin{equation}\label{eq:gSS}
    \begin{aligned}
    	e_{\bat}^\mu & = \Exp{\Delta(x^i)}\,\hat{e}_{\bat}^\mu(x^\mu)\,,       \\
    	B_{\mu\nu}   & = \Exp{-2\Delta(x^i)}\,\hat{B}_{\mu\nu}(x^\mu)\,,       \\
    	A_{\mu}{}^I  & = \Exp{-\Delta(x^i)} A_\mu{}^A(x^\mu)\,\cE_A{}^I(x^i)\,,                     \\
    	\cV_A{}^{I}  & = \hat{\cV}_A{}^B(x^\mu)\,\cE_B{}^I(x^i)\,,             \\
    	-2\,\phi     & = -2\,\hat{\phi}(x^\mu) + (8-d)\,\Delta(x^i) + \ln v\,,
    \end{aligned}
\end{equation}
where $v\equiv \det(v_i^a)$ is the left-invariant one-form which we will define later. We, then, obtain
\begin{equation}\label{eqn:scEfluxtoscVfluxes}
	\cF_{ABC} = \Exp{\Delta} \hat{\cV}_A{}^D \hat{\cV}_B{}^E \hat{\cV}_C{}^F F_{DEF}\,,\qquad
	\cF_A = \hat{\cV}_A{}^B F_B\,,
\end{equation}
and their hatted counterparts that are required to evaluate the contributions \eqref{eqn:ingr-DFT-action-d} to the reduced action. Under the above ansatz, the quantities that appear in the action \eqref{eq:DFT-action-d} become
\begin{equation}
    \begin{aligned}
    	\cF_2^I        & = \Exp{-\Delta}\,\bigl(\rmd A^A + \tfrac{1}{2}\, \hat{F}_{BC}{}^A \, A^B\wedge A^C + \xi_B\,A^B\wedge A^A -2\,\xi^A\,\hat{B}_2\bigr)\,\cE_A{}^I \,,
    	\\
    	\cH_3          & = \Exp{-2\Delta}\,\bigl(\rmd \hat{B}_2 + 2\,A^A\,\xi_A\wedge \hat{B}_2
    	- \tfrac{1}{2}\, A^A\wedge \rmd A_A -\tfrac{1}{3!}\,\hat{F}_{ABC}\,A^A\wedge A^B\wedge A^C \bigr)\,,
    	\\
    	D_\mu \phi     & = \partial_\mu \hat{\phi} + \tfrac{8-d}{2}\,A_\mu{}^A \xi_A \,,
    	\\
    	D_\mu \mathbb{H}_{IJ} & = \partial_\mu \mathbb{H}_{IJ} -2\,A_\mu{}^A\,\hat{F}_{A(I}{}^K\,\mathbb{H}_{J)K}-2\,A_{\mu(I}\,\mathbb{H}_{J)K}\,\xi^K + 2\,A_\mu{}^A\,\xi_{(I}\,\mathbb{H}_{J)A}\,,
    \end{aligned}
\end{equation}
where $\cE_A{}^I$ is used to relate flat indices with curved ones. This only leaves the potential $V$ to be computed -- a thing we will deal with in Subsection~\ref{sec:reltogSS}. Note that the scale factor $\Delta$ completely drops out from the action, and we have successfully reproduced the action of a half-maximal gauged supergravity up to a boundary term.

\section{Half-maximal gSUGRAs in more than three dimensions}\label{sec:halfmaxgSUGRAs}
We now come back to the main challenge of this article, namely, the construction of the frames that satisfy the algebra \eqref{eqn:framealg} with constant $X_{AB}{}^C$. They are central in both the generalized Scherk-Schwarz reduction already discussed in Subsection~\ref{sec:gSSansatz} and the definition of generalized dualities with which we will deal later. Our approach to tackle it consists of the following steps: First, we need to clarify the algebraic structure underlying the Lie derivative given in \eqref{eqn:genLieDef} and its extension to larger duality groups, as they appear in \eqref{eqn:halfmaxGduality}. This will be done in Subsection~\ref{sec:genLieder}. Then, in Subsection~\ref{sec:embeddingtensor} we will deal with the analysis of the embedding tensor in each dimension $d\leq 7$, and, finally, in Subsection~\ref{sec:reltogSS} we come back to the Scherk-Schwarz reduction and identify conditions on upliftable gaugings. After this, we will have all the necessary tools to construct, in Section~\ref{sec:frames}, the frames.

\subsection{Generalized Lie derivative}\label{sec:genLieder}
In \eqref{eqn:genLieDef} we have already encountered the generalized Lie derivative in DFT. To make contact with half-maximal gSUGRAs, we furthermore need to understand how it can be extended to their duality groups given in \eqref{eqn:halfmaxGduality}. For this purpose, the more algebraic formulation \cite{Ciceri:2016hup,Sakatani:2021eqo}
\begin{align}\label{eqn:genLieAlgebraic}
	\gLie_V W^{\Ih} \equiv V^{\Jh}\,\partial_{\Jh} W^{\Ih} + (t^{\dot{\bm{\alpha}}})_{\Lh}{}^{\Kh}\,\partial_{\Kh}\,V^{\Lh} \, (t_{\dot{\bm{\alpha}}})_{\Jh}{}^{\Ih}\, W^{\Jh} + w \,(\partial_{\Kh} V^{\Kh}) \, W^{\Ih}
\end{align}
is better suited. Note that this Lie derivative is different form its DFT counterpart and we will see later in Subsection~\ref{sec:reltogSS} how they can be related. The definition \eqref{eqn:genLieAlgebraic} relies on the generators $(t_{\dot{\bm{\alpha}}})_{\Ih}{}^{\Jh}$ of the respective duality groups; we will deal with them in full detail in the next paragraph. In contrast to \eqref{eqn:genLieDef}, generalized vectors $W^{\Ih}$ with arbitrary weight $w$ are here considered; for example, parameters of generalized diffeomorphisms have a natural weight $w=\beta$, where
\begin{equation}
	\beta\equiv \frac{1}{8-d}\,.
\end{equation}
It is customary to define the $Y$-tensor
\begin{align}
	Y^{\Ih\Jh}_{\Kh\Lh} \equiv \delta^{\Ih}_{\Kh}\,\delta^{\Jh}_{\Lh} + (t^{\dot{\bm{\alpha}}})_{\Kh}{}^{\Jh}\, (t_{\dot{\bm{\alpha}}})_{\Lh}{}^{\Ih} + \beta \,\delta^{\Ih}_{\Lh}\,\delta^{\Jh}_{\Kh}\,
	\label{eq:Y-tensor-c-def}
\end{align}
to express the section condition required for this Lie derivative to close as
\begin{equation}
	Y^{\Ih\Jh}_{\Kh\Lh} \partial_{\Ih} \, \cdot \, \partial_{\Jh} \, \cdot \, = 0\,.
\end{equation}
At the same time it allows to write the generalized Lie derivative \eqref{eqn:genLieAlgebraic} in the more compact form
\begin{align}\label{Ygenlieder}
	\gLie_{V} W^{\Ih} & \equiv V^{\Jh}\,\partial_{\Jh} W^{\Ih} - W^{\Jh}\,\partial_{\Jh} V^{\Ih} + Y^{\Ih\Kh}_{\Lh\Jh}\,\partial_{\Kh} V^{\Lh}\, W^{\Jh} + (w-\beta)\,(\partial_{\Kh} V^{\Kh})\,W^{\Ih}\,.
\end{align}

Coming back to the generators of the duality group, note that we distinguish between them in the curved indices, denoted by $(t_{\dot{\bm{\alpha}}})_{\Ih}{}^{\Jh}$, and their flat counterparts $(t_{\bm{\alpha}})_{\Ah}{}^{\Bh}$. However, they can always be chosen to have the same components and, therefore, we just give more details on the latter. As the duality group decomposes into two factors, namely $\Gdualityt$ and $\OO(d,\mathfrak{n})$, we choose the generators accordingly,
\begin{align}
	\{t_{\bm{\alpha}}\} = \{\tilde{t}_{\tilde{\bm{\alpha}}},\, t_{\hat{\bm{\alpha}}}\}\,.
\end{align}

Let us start with the dimension independent $\OO(d,\mathfrak{n})$ factor and its generators $t_{\hat{\bm{\alpha}}}$. Since we want, eventually, to take into account solutions of the section condition, like \eqref{eqn:solSC}, it is convenient to decompose them following the branching
\begin{equation}\label{eqn:OdntoGldxOn}
	\OO(d,\mathfrak{n})\rightarrow\GL(d)\times \OO(n)\,.
\end{equation}
For example, the adjoint representation branches as
\begin{equation}\label{eqn:decompAdj}\arraycolsep=2pt\def\arraystretch{1.8}
	\begin{array}{ccccccccccccccc}

		\mathrm{adj} & \rightarrow & (\raisebox{0.25em}{\ydiagram{1,1}},\mathbf{1}) & \oplus & (\raisebox{-0.25em}{\ydiagram{1}}, \raisebox{-0.25em}{\ydiagram{1}}) & \oplus & (\mathrm{adj},\mathbf{1}) & \oplus & (\mathbf{1},\raisebox{0.25em}{\ydiagram{1,1}}) & \oplus & (\overline{\raisebox{-0.25em}{\ydiagram{1}}}, \raisebox{-0.25em}{\ydiagram{1}}) & \oplus & (\overline{\raisebox{0.25em}{\ydiagram{1,1}}}, \mathbf{1})\,, \\
		             &             & -2                                             &        & -1                                       &        & 0                         &        & 0                                              &        & 1                                         &        & 2
	\end{array}
\end{equation}
where the numbers below each summand indicate a natural grading which arises by the decomposition. Following this prescription, we denote the generators of $\OO(d,\mathfrak{n})$ by
\begin{align}\label{generators}
	\{t_{\hat{\bm{\alpha}}}\} \equiv \bigl\{R_{a_1a_2},\,R_a^{\gA},\,K^{a_1}{}_{a_2},\, R_{\gA_1\gA_2} ,\,R^a_{\gA},\, R^{a_1a_2} \bigr\}\,,
\end{align}
where $a=1,\dotsc,d$ and $\gA=\bm{1},\dotsc,\bm{n}$\,, and the indices of $R_{a_1 a_2}$, $R_{\cA_1\cA_2}$ and $R^{ab}$ are antisymmetric (according to the first, fourth and last term in \eqref{eqn:decompAdj}). Their commutators and matrix realizations in the fundamental representation are given in Appendix~\ref{app:dualityalgebra}. To obtain the latter, one also needs the branching of the fundamental under \eqref{eqn:OdntoGldxOn}. This reads
\begin{equation}
	\begin{array}{ccccccc}\arraycolsep=2pt\def\arraystretch{1.8}
		\raisebox{-0.25em}{\ydiagram{1}} & \rightarrow & (\raisebox{-0.25em}{\ydiagram{1}},\mathbf{1}) & \oplus & (\mathbf{1},\raisebox{-0.25em}{\ydiagram{1}}) & \oplus & (\overline{\raisebox{-0.25em}{\ydiagram{1}}},\mathbf{1}) \\
		              &             & -1                         &        & 0                          &        & 1
	\end{array}
\end{equation}
and, therefore, we compose generalized vectors $Z_A$'s as
\begin{equation}\label{eqn:decompFundOdn}
	Z_A = \begin{pmatrix} Z_a & Z_{\cA} & Z^a \end{pmatrix}\,.
\end{equation}
Now, let us come to $\Gdualityt$, which depends on the internal dimension of the reduction $d$,
\begin{equation}
	\Gdualityt = \begin{cases}
		\mathbb{R}^+ & d\leq 5 \\
		\SL(2)       & d=6\,.
	\end{cases}
\end{equation}
Their generators will be denoted as
\begin{equation}\label{eqn:defGtgen}
	\tilde{t}_{\tilde{\bm{\alpha}}} \equiv \begin{cases}
		R_*                   & d\leq 5 \\
		R^{\bba_1}{}_{\bba_2} & d=6\,,
	\end{cases}
\end{equation}
where $\bba=+,-$ and $R^{\bba}{}_{\bba}=0$\,. The generators $t_{\hat{\bm{\alpha}}}$ and $\tilde{t}_{\tilde{\bm{\alpha}}}$ commute with each other and $R^{\bba}{}_{\bbb}$ satisfies the sl($2$) algebra
\begin{align}
	[R^{\bba}{}_{\bbb},\,R^{\bbc}{}_{\bbd}] = \delta_{\bbb}^{\bbc}\,R^{\bba}{}_{\bbd} -  \delta_{\bbd}^{\bba}\,R^{\bbc}{}_{\bbb}\,.
\end{align}
We define, also, the dual generators
\begin{equation}\label{eqn:splitGenGdual}
	\{t^{\bm{\alpha}}\} \equiv  \{\tilde{t}^{\tilde{\bm{\alpha}}},\, t^{\hat{\bm{\alpha}}}\}\,.
\end{equation}
They are dual in the sense that $\tr_{R_1}(t^{\bm{\alpha}}\,t_{\bm{\beta}})$ is diagonal\footnote{Concretely, we find
	\begin{align*}
		\begin{alignedat}{2}
			\tr_{R_1}(\tilde{t}^{\tilde{\bm{\alpha}}}\,\tilde{t}_{\tilde{\bm{\beta}}}) &= -\beta\,(c_d+n)\,\delta^{\tilde{\bm{\alpha}}}_{\tilde{\bm{\beta}}}\,,\qquad &
			\tr_{R_1}(t^{\hat{\bm{\alpha}}}\,t_{\hat{\bm{\beta}}}) &= -2\,\delta^{\hat{\bm{\alpha}}}_{\hat{\bm{\beta}}} \qquad d\le 5\,,
			\\
			\tr_{R_1}(\tilde{t}^{\tilde{\bm{\alpha}}}\,\tilde{t}_{\tilde{\bm{\beta}}}) &= -(12+n)\,\delta^{\tilde{\bm{\alpha}}}_{\tilde{\bm{\beta}}}\,,\qquad &
			\tr_{R_1}(t^{\hat{\bm{\alpha}}}\,t_{\hat{\bm{\beta}}}) &= -4\,\delta^{\hat{\bm{\alpha}}}_{\hat{\bm{\beta}}} \qquad d=6\,,
		\end{alignedat}
	\end{align*}
	where $c_d=2,2,6,8,14$ for $d=1,2,3,4,5$\,.}, where $R_1$ is the vector representation of $\Gduality$. As expected, the grading of the dual generators is the opposite and we find again decompositions that mimic \eqref{generators} and \eqref{eqn:defGtgen}; to be specific
\begin{align}
	\{t^{\hat{\bm{\alpha}}}\}       & \equiv \{R^{a_1a_2} ,\,R^a_{\gA},\,K_{a_1}{}^{a_2},\, R^{\gA_1\gA_2} ,\,R_a^{\gA},\, R_{a_1a_2}\}
	\intertext{and}\label{eqn:namesGenGt}
	\tilde{t}^{\tilde{\bm{\alpha}}} & \equiv \begin{cases}
		                                         R^*                   & d\leq 5 \\
		                                         R_{\bba_1}{}^{\bba_2} & d=6\,,
	                                         \end{cases}
\end{align}
where
\begin{align}
	R^* \equiv - \beta^{-1}\,R_*\,,\qquad R_{\bba}{}^{\bbb} \equiv - R^{\bbb}{}_{\bba} \,,\qquad K_a{}^b \equiv - K^b{}_a\,.
\end{align}

Finally, we join these two parts together to obtain the full matrices $(t_{\bm{\alpha}})_{\Bh}{}^{\Ch}$ that are needed in the definition \eqref{eqn:genLieAlgebraic} of the generalized Lie derivative. At this point, one has to fix how the hatted indices decompose. For $d<5$, this is straightforward, because it reduces to the decomposition \eqref{eqn:decompFundOdn}; for larger $d$, however, the situation becomes more subtle since the ten-dimensional theory \eqref{eqn:hmaxSUGRA10d} contains a three-form which is dual to a six-form, describing an NS5-brane. This has one time direction and five spacial ones, and all of the latter have to be in the compact $d$-dimensional space in the reduction. That can only happen for $d\ge 5$. If this bound is saturated, the 5-form is dual to a scalar field that we call $Z_* = \tfrac{1}{5!} \epsilon^{a_1\dots a_5} Z_{a_1\dots a_5}$ . In $d=6$ the same quantity is dual to the vector $Z^{-a} = \tfrac{1}{5!} \epsilon^{a b_1\dots b_5} Z_{b_1\dots b_5}$ which is related by the $\SL(2)$ S-duality to $Z^{+a}$. Therefore, we choose the following decompositions:
\begin{align}\label{eqn:decompIndices}
	Z_{\Ah} = \begin{cases}
		          Z_{A} = \begin{pmatrix} Z_a & Z_{\gA} & Z^a\end{pmatrix}                                        & d\leq 4 \\
		          \begin{pmatrix} Z_A & Z_*\end{pmatrix} = \begin{pmatrix} Z_a & Z_{\gA} & Z^a & Z_*\end{pmatrix} & d=5     \\
		          Z_{\bba A} = \begin{pmatrix} Z_{\bba a} & Z_{\bba \gA} & Z_{\bba}{}^a\end{pmatrix}  \quad       & d=6\,,
	          \end{cases}
\end{align}
where $A=1,\dotsc,d + \mathfrak{n}$\,.  All $\OO(d,\mathfrak{n})$ generators, which are presented in full detail in the Appendix~\ref{app:dualityalgebra}, act from the left on $Z_A$. $\Gdualityt$'s action is more involved: First we have $R_*$ in dimensions $d\le 5$; this is realized by the matrices,
\begin{equation}
	\begin{array}{ll}
		\text{for $d\le 4$:}                            & \text{for $d=5$:} \\
		(R_*)_A{}^B \equiv \beta
		\begin{pmatrix}
			\delta_a^b & 0                  & 0          \\
			0          & \delta_{\gA}^{\gB} & 0          \\
			0          & 0                  & \delta^a_b \\
		\end{pmatrix}\,, \qquad &
		(R_*)_{\Ah}{}^{\Bh} \equiv \beta
		\begin{pmatrix}
			\delta_a^b & 0                  & 0          & 0  \\
			0          & \delta_{\gA}^{\gB} & 0          & 0  \\
			0          & 0                  & \delta^a_b & 0  \\
			0          & 0                  & 0          & -2
		\end{pmatrix}\,,
	\end{array}
\end{equation}
depending on the value of $d$, as explained before. Consequently, $(R_*)_{\Ah}{}^{\Bh}$ is proportional to the identity matrix in $d\leq 4$.
In $d=6$\,, we have to additionally implement the $\SL(2)$-factor; this is achieved by defining
\begin{align}
	(R^{\bbc}{}_{\bbd})_{\Ah}{}^{\Bh} \equiv (R^{\bbc}{}_{\bbd})_{\alpha A}{}^{\beta B} = \bigl(\delta_{\bba}^{\bbc}\,\delta^{\bbb}_{\bbd}-\tfrac{1}{2}\delta_{\bba}^{\bbb}\delta^{\bbc}_{\bbd}\bigr)\,\delta_A^B\,,\qquad
	(t_{\hat{\bm{\alpha}}})_{\Ah}{}^{\Bh} \equiv (t_{\hat{\bm{\alpha}}})_{\alpha A}{}^{\beta B} = \delta_{\bba}^{\bbb}\,(t_{\hat{\bm{\alpha}}})_{A}{}^{B}\,,
\end{align}
where $(t_{\hat{\bm{\alpha}}})_{A}{}^{B}$ take the same form as those in $d\leq 4$\,.

\subsection{Embedding tensor}\label{sec:embeddingtensor}
After having settled the explicit details of the generalized Lie derivative, we come back to obtaining the generalized frames satisfying the algebra
\begin{align}\label{eq:gen-flux}
	\gLie_{E_{\Ah}} E_{\Bh} = - X_{\Ah\Bh}{}^{\Ch}\,E_{\Ch}\,.
\end{align}
Note that this is still not the algebra \eqref{eqn:framealg} we based the generalized Scherk-Schwarz reduction on in the last section, since the frames here are valued in the duality groups given in \eqref{eqn:halfmaxGduality} and not in $\OO(d,\mathfrak{n})$. Later on, it will become clear how these two algebras are related; for the moment, we just look at how the embedding tensor $X_{\Ah\Bh}{}^{\Ch}$ arises from the frame. To this end, it is convenient to introduce the Weitzenböck connection
\begin{align}
	\Omega_{\Ah\Bh}{}^{\Ch} \equiv D_{\Ah} E_{\Jh}{}^{\Ch}\, E_{\Bh}{}^{\Jh} = - D_{\Ah} E_{\Bh}{}^{\Jh}\,E_{\Jh}{}^{\Ch}\,,
\end{align}
with $D_{\Ah}\equiv E_{\Ah}{}^{\Ih}\,\partial_{\Ih}$. Employing the definition of the generalized Lie derivative \eqref{Ygenlieder} from the last subsection, we can express the embedding tensor as
\begin{align}\label{eqn:XOmegaMap}
	X_{\Ah\Bh}{}^{\Ch} = \Omega_{\Ah\Bh}{}^{\Ch} + (t^{\bm{\alpha}})_{\Dh}{}^{\Eh}\,\Omega_{\Eh\Ah}{}^{\Dh}\,(t_{\bm{\alpha}})_{\Bh}{}^{\Ch} - \beta\, \Omega_{\Dh\Ah}{}^{\Dh}\,(t_0)_{\Bh}{}^{\Ch}\,,
\end{align}
where $(t_0)_{\Ah}{}^{\Bh}=-\delta_{\Ah}^{\Bh}$ can be understood as a generator of an $\mathbb{R}^+$ symmetry (associated to the trombone symmetry). To see how this additional symmetry relates to the duality group, we distinguish two different cases depending on the value of $d$.\begin{enumerate}
	\item
	      In $d\leq 5$\,, we make a general decomposition
	      \begin{align}
		      \Omega_{\Ah\Bh}{}^{\Ch} = \Omega_{\Ah}{}^{\bm{\alpha}}\,(t_{\bm{\alpha}})_{\Bh}{}^{\Ch} = \Omega_{\Ah}{}^{\hat{\bm{\alpha}}}\,(t_{\hat{\bm{\alpha}}})_{\Bh}{}^{\Ch} + \Omega_{\Ah}{}^*\,(R_*)_{\Bh}{}^{\Ch} + \Omega_{\Ah}{}^0\,(t_0)_{\Bh}{}^{\Ch} \,,
	      \end{align}
	      where we use the splitting of generators \eqref{eqn:splitGenGdual} and \eqref{eqn:namesGenGt}, and we set $\Omega_{\Ah}{}^*=0$ in $d\leq 4$ because $R_*$ and $t_0$ are not independent. To see how the different components of the Weitzenböck connection contribute to the embedding tensor, we first define the invariant tensors
	      \begin{align}
		      \mathbb{I}_{\Ah}^{\Bh}                                         & \equiv (R_*)_{\Ah}{}^{\Bh} + (\beta-1)\,(t_0)_{\Ah}{}^{\Bh} = \begin{pmatrix} \delta_A^B & 0 \\ 0 & 0 \end{pmatrix}\,,
		      \\
		      \mathbb{J}_{\Ah}^{\Bh}                                         & \equiv -(\beta\,t_0 + R_*)_{\Ah}{}^{\Bh} = \begin{pmatrix} 0 & 0 \\ 0 & 1 \end{pmatrix}\,,
		      \intertext{and}
		      \mathbb{P}_{\Ah}{}^{\hat{\bm{\alpha}}\Bh}{}_{\hat{\bm{\beta}}} & \equiv \mathbb{I}_{\Ah}^{\Bh}\,\delta_{\hat{\bm{\beta}}}^{\hat{\bm{\alpha}}} + (t_{\hat{\bm{\beta}}}\cdot t^{\hat{\bm{\alpha}}})_{\Ah}{}^{\Bh}\,.
	      \end{align}
	      While the first two can be easily seen to commute with all the generators due to their block form, the last is slightly more complicated. It can be understood as a projector onto the totally antisymmetric representation in the decomposition of the tensor product of the fundamental with the adjoint,
	      \begin{equation}\label{eqn:decompFundxAdj}
		      \raisebox{-0.2em}{\ydiagram{1}} \otimes \raisebox{0.25em}{\ydiagram{1, 1}} \rightarrow  \raisebox{0.25em}{\ydiagram{2,1}} \oplus \raisebox{0.75em}{\ydiagram{1,1,1}}\,,
	      \end{equation}
	      because it satisfies
	      \begin{align}
		      \mathbb{P}_{\Ah}{}^{\hat{\bm{\alpha}}\Bh}{}_{\hat{\bm{\beta}}}\,(t^{\hat{\bm{\beta}}})_{\Bh}{}^{\Ch}=0\,.
	      \end{align}
	      Note that the projector has to be rescaled by $\tfrac{1}{3}$ to have the correct normalization.

	      Using these three invariants and remembering $(t^{\hat{\bm{\alpha}}})_*{}^{\Bh}=(t^{\hat{\bm{\alpha}}})_{\Ah}{}^*=0$\,, we find
	      \begin{align}
		      X_{\Ah\Bh}{}^{\Ch}
		       & = \mathbb{P}_{\Ah}{}^{\hat{\bm{\alpha}}\Dh}{}_{\hat{\bm{\beta}}}\,\Omega_{\Dh}{}^{\hat{\bm{\beta}}}\,(t_{\hat{\bm{\alpha}}})_{\Bh}{}^{\Ch}
		      + (t^{\hat{\bm{\alpha}}})_{\Ah}{}^{\Dh}\,\xi_{\Dh}\,(t_{\hat{\bm{\alpha}}})_{\Bh}{}^{\Ch}
		      + \beta^{-1}\,\xi_{\Ah}\, (R_*)_{\Bh}{}^{\Ch}
		      \nn                                                                                                                                           \\
		       & \quad + \mathbb{J}_{\Ah}^{\Dh}\,\Omega_{\Dh}{}^{\hat{\bm{\alpha}}}\,(t_{\hat{\bm{\alpha}}})_{\Bh}{}^{\Ch}
		      - \beta^{-1}\,\mathbb{I}_{\Ah}^{\Dh}\,\Vtheta_{\Dh} \,\mathbb{J}_{\Bh}^{\Ch}
		      + \tfrac{1}{\beta-1} \,\mathbb{J}_{\Ah}^{\Dh}\,\Vtheta_{\Dh} \,\mathbb{I}_{\Bh}^{\Ch} \,,
		      \label{eq:X-general}
	      \end{align}
	      where
	      \begin{align}
		      \xi_{\Ah} \equiv \mathbb{I}_{\Ah}^{\Bh}\,\bigl(\beta\,\Omega_{\Bh}{}^{*} - \Omega_{\Bh}{}^{0}\bigr) \,,\qquad
		      \Vtheta_{\Ah} \equiv \Omega_{\Ah}{}^0 - \beta\, \Omega_{\Dh\Ah}{}^{\Dh} \,.
	      \end{align}
	      Therefore, the only non-vanishing components of $X_{\Ah\Bh}{}^{\Ch}$ are
	      \begin{equation}\label{eqn:Xabchatdleq5}
		      \begin{aligned}
			      X_{AB}{}^{C} & = F_{AB}{}^{C} + (t^{\hat{\bm{\alpha}}})_{A}{}^{D}\,\xi_{D}\,(t_{\hat{\bm{\alpha}}})_{B}{}^{C} + \xi_{A}\, \delta_{B}^{C} \,, \\
			      X_{*B}{}^{C} & = \xi_B{}^C + \tfrac{1}{\beta-1} \, \Vtheta_{*} \,\delta_B^C \,,                                                            \\
			      X_{A*}{}^{*} & = -2\,\xi_{A} - \beta^{-1}\, \Vtheta_{A}  \,,
		      \end{aligned}
	      \end{equation}
	      where $X_{*B}{}^{C}$ and $X_{A*}{}^{*}$ are not present in $d\leq 4$\,, and we have defined
	      \begin{align}
		      F_{AB}{}^C = \mathbb{P}_{A}{}^{\hat{\bm{\alpha}}D}{}_{\hat{\bm{\beta}}}\,\Omega_{D}{}^{\hat{\bm{\beta}}}\,(t_{\hat{\bm{\alpha}}})_{B}{}^{C}\,,\qquad
		      \xi_A{}^B \equiv \Omega_{*}{}^{\hat{\bm{\alpha}}}\,(t_{\hat{\bm{\alpha}}})_{B}{}^{C}\,.
	      \end{align}
	      Since, as explained above, $\mathbb{P}_{\Ah}{}^{\hat{\bm{\alpha}}\Bh}{}_{\hat{\bm{\beta}}}$ projects on the totally antisymmetric representation, we find that $F_{ABC} = F_{AB}{}^{D}\,\eta_{DC}$ is totally antisymmetric. Moreover, from the antisymmetricity in $AB$ of $(t_{\hat{\bm{\alpha}}})_{AB} = (t_{\hat{\bm{\alpha}}})_{A}{}^{C}\,\eta_{CB}$, $\xi_{AB} = \xi_A{}^C\,\eta_{CB}$ is also antisymmetric with respect to these two indices.
	      By applying the identity
	      \begin{align}
		      (t^{\hat{\bm{\alpha}}})_{A}{}^{D}\,(t_{\hat{\bm{\alpha}}})_{B}{}^{C} = \eta_{AB}\,\eta^{CD} - \delta_{A}^{C}\,\delta_{B}^{D}\,,
	      \end{align}
	      $X_{AB}{}^{C}$ simplifies to
	      \begin{align}
		      X_{AB}{}^{C} & = F_{AB}{}^{C} + \eta_{AB}\,\xi^C + \xi_{A}\, \delta_{B}^{C} - \xi_B\,\delta_A^C\,.
	      \end{align}
	      As a crosscheck of our results so far, we can take $d=5$ and truncate $\Vtheta_{\Ah}$. After the redefinitions $\xi_{\Ah}\to -\frac{1}{2}\,\xi_{\Ah}$\,, $\xi_{AB}\to -\xi_{AB}$\,, and $F_{ABC}\to -f_{ABC}$, our expression \eqref{eqn:Xabchatdleq5} for $X_{\Ah\Bh}{}^{\Ch}$ reproduces the known embedding tensor given in equation~(3.6) of \cite{Schon:2006kz}.

	\item
	      In $d=6$\,, we follow conceptually the same steps and begin with the decomposition
	      \begin{align}
		      \Omega_{\Ah\Bh}{}^{\Ch} = \Omega_{\Ah}{}^{\bm{\alpha}}\,(t_{\bm{\alpha}})_{\Bh}{}^{\Ch} = \Omega_{\Ah}{}^{\hat{\bm{\alpha}}}\,(t_{\hat{\bm{\alpha}}})_{\Bh}{}^{\Ch} + \Omega_{\Ah\bba}{}^{\bbb}\,(R^{\bba}{}_{\bbb})_{\Bh}{}^{\Ch} + \Omega_{\Ah}{}^0\,(t_0)_{\Bh}{}^{\Ch} \,.
	      \end{align}
	      Similarly to the case $d\leq 5$\,, we define
	      \begin{align}
		      \mathbb{P}_{\Ah}{}^{\hat{\bm{\alpha}}\Bh}{}_{\hat{\bm{\beta}}} \equiv \delta_{\bba}^{\bbb}\,\delta_{A}^{B}\,\delta_{\hat{\bm{\beta}}}^{\hat{\bm{\alpha}}} + (t_{\hat{\bm{\beta}}}\,t^{\hat{\bm{\alpha}}})_{\Ah}{}^{\Bh}\,,
	      \end{align}
	      which again satisfies
	      \begin{align}
		      \mathbb{P}_{\Ah}{}^{\hat{\bm{\alpha}}\Bh}{}_{\hat{\bm{\beta}}}\,(t^{\hat{\bm{\beta}}})_{\Bh}{}^{\Ch}=0\,,
	      \end{align}
	      and after a rescaling by $\tfrac{1}{3}$ it projects on the totally anti-symmetric representation in \eqref{eqn:decompFundxAdj}, which is now  a doublet of $\SL(2)$.
	      Accordingly, we define
	      \begin{align}
		      F_{\Ah\Bh}{}^{\Ch} \equiv \mathbb{P}_{\Ah}{}^{\hat{\bm{\alpha}}\Dh}{}_{\hat{\bm{\beta}}}\,\Omega_{\Dh}{}^{\hat{\bm{\beta}}}\,(t_{\hat{\bm{\alpha}}})_{\Bh}{}^{\Ch} \equiv F_{\bba AB}{}^C\,\delta_{\bbb}^{\bbc}\,,
	      \end{align}
	      and then $F_{\bba ABC}\equiv F_{\bba AB}{}^D\,\eta_{DC}$ is totally antisymmetric in $ABC$\,.
	      Consequently, we have
	      \begin{align}
		      \begin{split}
			      X_{\Ah\Bh}{}^{\Ch} &= F_{\bba AB}{}^{C}\,\delta_{\bbb}^{\bbc}
			      + (t^{\hat{\bm{\alpha}}})_{A}{}^{D}\,\xi_{\bba D}\,(t_{\hat{\bm{\alpha}}})_{B}{}^C\,\delta_{\bbb}^{\bbc}
			      + 2\,\xi_{\bbb A}\,\delta_{\bba}^{\bbc}\,\delta_B^C
			      - \xi_{\bba A}\,\delta_{\bbb}^{\bbc}\,\delta_B^C
			      \\
			      &\quad + 2\,\bigl(\Vtheta_{\bbb A}\,\delta_{\bba}^{\bbc} - \Vtheta_{\bba A}\,\delta_{\bbb}^{\bbc}\bigr)\,\delta_{B}^C\,,
		      \end{split}
		      \label{eq:X-4d-1}
	      \end{align}
	      after the decomposition $(t_{\hat{\bm{\alpha}}})_{\Bh}{}^{\Ch}=(t_{\hat{\bm{\alpha}}})_{B}{}^{C}\,\delta_{\bbb}^{\bbc}$, and having defined
	      \begin{align}
		      \xi_{\bba A} \equiv \Omega_{\bbb \bba A}{}^{\bbb} -\Omega_{\bba A}{}^0\,,\qquad
		      \Vtheta_{\bba A} \equiv \Omega_{\bba A}{}^0 - \beta\,\Omega_{\Bh\bba A}{}^{\Bh}\,,
		      \label{eq:X-4d-2}
	      \end{align}
	      which mimic their counterparts in the discussion of $d\leq 5$.
	      Similarly, with
	      \begin{align}
		      (t^{\hat{\bm{\alpha}}})_{A}{}^{D}\,(t_{\hat{\bm{\alpha}}})_{B}{}^{C} = \eta_{AB}\,\eta^{CD} - \delta_{A}^{C}\,\delta_{B}^{D}\,,
	      \end{align}
	      we further simplify it to
	      \begin{align}
		      X_{\Ah\Bh}{}^{\Ch} & = F_{\bba AB}{}^{C}\,\delta_{\bbb}^{\bbc}
		      + \eta_{AB}\,\eta^{DC}\,\xi_{\bba D}\,\delta_{\bbb}^{\bbc}
		      - \xi_{\bba B}\,\delta_A^C\,\delta_{\bbb}^{\bbc}
		      + \xi_{\bbb A}\,\delta_{\bba}^{\bbc}\,\delta_B^C
		      - \epsilon_{\bba\bbb}\,\epsilon^{\bbd\bbc}\,\xi_{\bbd A}\,\delta_B^C \nn                                                                         \\
		                         & \quad + 2\,\bigl(\Vtheta_{\bbb A}\,\delta_{\bba}^{\bbc} - \Vtheta_{\bba A}\,\delta_{\bbb}^{\bbc}\bigr)\,\delta_{B}^C\,.
	      \end{align}
	      Again a crosscheck with the existing literature is in order. If we redefine the fluxes as $F_{\bba ABC}\to -f_{\bba ABC}$ and $\xi_{\alpha A}\to -\tfrac{1}{2}\,\xi_{\alpha A} - \Vtheta_{\alpha A}$, then the expression for $X_{\hat{A}\hat{B}}{}^{\hat{C}}$ matches equation (4.5) of~\cite{Ciceri:2016hup}. If we further truncate $\Vtheta_{\bba A}$, this reproduces equation (2.24) of~\cite{Schon:2006kz}.
\end{enumerate}

\noindent To sum up, $X_{\Ah\Bh}{}^{\Ch}$ is decomposed into
\begin{equation}
	\begin{aligned}
		 & \{F_{ABC},\,\xi_A\}                                               & d\leq 4\,, \\
		 & \{F_{ABC},\,\xi_A,\,\Vtheta_A,\,\xi_{AB},\,\Vtheta_*\} \qquad & d=5\,,     \\
		 & \{F_{\bba ABC},\,\xi_{\bba A},\,\Vtheta_{\bba A}\}              & d=6\,.
	\end{aligned}
\end{equation}
Here we note that $X_{\Ah\Bh}{}^{\Ch}$ is supposed to be constant, and indeed in Section~\ref{sec:frames} we construct the frames $E_{\Ah}$ such that $X_{\Ah\Bh}{}^{\Ch}$ becomes constant. Accordingly, its constituents, like $F_{AB}{}^C$ and $\xi_{\Ah}$, are constant. However, $\Vtheta_{\Ah}$ is non-trivial. In $d=5,6$, it is contained in $X_{\Ah\Bh}{}^{\Ch}$ and reduces to a constant $\vartheta_{\Ah}$, which is known as the trombone gauging, but in $d\leq 4$ it does not appear in $X_{\Ah\Bh}{}^{\Ch}$ and is not necessarily constant.
We also note that if the trombone gauging in $d=5,6$ is truncated the embedding tensor $X_{\Ah\Bh}{}^{\Ch}$ is constructed only from the components $X_{AB}{}^C$.

\subsection{Scherk-Schwarz ansatz revisited}\label{sec:reltogSS}
We now are in the position to identify what we call half-maximal geometric gaugings as those that admit an uplift to the ten-dimensional action \eqref{eqn:hmaxSUGRA10d} we discussed in the Section~\ref{sec:genSSred}. As before, we have to distinguish between the two cases depending on $d$.
\begin{enumerate}
	\item In $d \leq 5$, the standard solution of the section condition \eqref{eqn:solSC} has to be supplemented with $\partial_*=0$ in $d=5$ \cite{Malek:2017njj,Sakatani:2021eqo}. Consequently, one finds that $\Omega_{*}{}^{\hat{\bm{\alpha}}}=0$ and this, in turn, implies
	      \begin{equation}
		      \xi_{AB}=\vartheta_*=0\,,
	      \end{equation}
	      leading, eventually, to  $X_{AB}{}^* = 0$. We can use this observation to restrict the frame algebra for the duality group $\mathbb{R}^+ \times \OO(d,\mathfrak{n})$ to the second factor. This is done in two stages: First, the frame algebra \eqref{eq:gen-flux} becomes
	      \begin{equation}
		      \gLie_{E_{A}} E_{B}{}^{\Ih} = - X_{AB}{}^{C}\,E_{C}{}^{\Ih}
	      \end{equation}
	      after removing the $*$ component from $\Ah$ and $\Bh$.
	      Then, we note that the generalized Lie derivative with index $\Ih$ restricted to $I$ is the same as that of double field theory, and we find
	      \begin{equation}
		      \gLie_{E_{A}} E_{B}{}^{I} = \gLieDFT_{E_{A}} E_{B}{}^{I} = - X_{AB}{}^{C}\,E_{C}{}^{I}\,,
		      \label{eq:gen-flux-DFT}
	      \end{equation}
	      where $X_{AB}{}^{C}$ is a restriction of $X_{\Ah\Bh}{}^{\Ch}$. To compare this setting with the generalized Scherk-Schwarz reduction discussed in Subsection~\ref{sec:gSSansatz}, we need to parametrize the frame in terms of an $\OO(d,\mathfrak{n})$ frame, $\cE_A{}^I$, and two additional fields. A suitable parametrization is
	      \begin{equation}\label{eqn:paramEaIdle5}
		      E_{\Ah}{}^{\Ih} = \exp\left[\Delta\, t_0 - \bar{\lambda}\,(R_* + \beta\,t_0)\right]{}_{\Ah}{}^{\Bh}\, \cE_{\Bh}{}^{\Ih}\,,
	      \end{equation}
	      with the new fields being $\Delta$ and $\bar{\lambda}$ (which is absent in $d\leq 4$ because of $R_* + \beta\,t_0=0$).
	\item  Turning to the case $d=6$, we have to impose $\partial_{-I}=0$ to supplement the section condition, and we find
	      \begin{equation}
		      F_{-ABC}=0\,.
	      \end{equation}
	      A direct consequence is that $X_{+A+B}{}^{-C}=0$, and thus, replicating the discussion for the previous case, we find that the relation \eqref{eq:gen-flux-DFT} still holds for the submatrix $E_A{}^I\equiv E_{+A}{}^{+I}$.
	      In the following we will denote $+I$ and $+A$ simply as $I$ and $A$ respectively, as the plus components appear much more often than their minus counterparts. In this case the parametrization of the frame
	      \begin{equation}\label{eqn:paramEaId6}
		      E_{\Ah}{}^{\Ih} = \Bigl[\exp(\Delta\, t_0) \exp(-\gamma\,R^-{}_+) \exp[- \bar{\lambda}\,(R^+{}_+ + \beta\,t_0)]\Bigr]{}_{\Ah}{}^{\Bh}\, \cE_{\Bh}{}^{\Ih}
	      \end{equation}
	      is made in terms of an additional field $\gamma$.
\end{enumerate}
Combining the two results, we see that in any dimensions $d\leq 6$ \eqref{eq:gen-flux-DFT} is satisfied with
\begin{align}\label{eqn:Xdecomp}
	X_{AB}{}^{C} & = F_{AB}{}^{C} + \eta_{AB}\,\eta^{DC}\,\xi_{D} + \xi_{A}\, \delta_B^C - \xi_{B}\,\delta_A^C\,.
\end{align}
Now we have all that is needed to eventually relate the embedding tensor identified in this section to the generalized Scherk-Schwarz reduction from the last section. Restricting the parametrizations \eqref{eqn:paramEaIdle5}  and \eqref{eqn:paramEaId6} for $E_{\Ah}{}^{\Ih}$ to the $\OO(d,\mathfrak{n})$ subsector one reproduces \eqref{eqn:cEfromE}.
Furthermore, plugging the frame $\cE_A{}^I$ in the generalized fluxes \eqref{eqn:defcF1} and \eqref{eqn:defcF3} gives rise to
\begin{align}\label{eqn:F1andF3fromParam}
	F_A = \Exp{\Delta}\,\bigl(2\,D_A \phi - D_A \Delta - \partial_I E_A{}^I\bigr)
	\,,\qquad
	F_{ABC} = X_{[ABC]}\,.
\end{align}
To compute the scalar potential in \eqref{eqn:ingr-DFT-action-d}, let us first focus on the case $D=4,5$. 
Here we take the ansatz
\begin{align}
	\Exp{-2\phi} =\Exp{-2\hat{\phi}(x^\mu)}\Exp{\bar{\lambda}(x^i)}
\label{eq:dilaton-ansatz}
\end{align}
for $\phi$ to obtain
\begin{align}
	F_A = -\Exp{\Delta}\,\bigl(D_A \Delta + D_A\bar{\lambda} + \partial_I E_A{}^I\bigr)\,.
\end{align}
Moreover, we find
\begin{equation}
	\xi_A = D_A\Delta
\end{equation}
and, by imposing the heterotic section\footnote{We call the heterotic section the partial fixing of the section $\partial_{\Ih}=\begin{pmatrix} \partial_I & 0 \end{pmatrix}$ for $d\le 5$ and $\partial_{\Ih}=\begin{pmatrix} \partial_{+ I } & 0 \end{pmatrix}$ for $d=6$.}, the trombone gauging $\Vtheta_A=\vartheta_A$ reduces to
\begin{align}
	\vartheta_A = \Omega_A{}^0 + \beta\,\partial_{\Ih} E_A{}^{\Ih} = - D_A\Delta + \beta\,D_A\bar{\lambda} + \beta\,\partial_{I} E_A{}^{I} \,,
\end{align}
after taking into account
\begin{align}
	\Omega_A{}^0 = \beta\,D_A\bar{\lambda} - D_A\Delta
\end{align}
as a result of the above parametrization. At end, we combine them to obtain
\begin{align}
	F_A = -\Exp{\Delta}\,\bigl(1+\beta^{-1}\bigr)\,\xi_A - \beta^{-1}\,\Exp{\Delta}\, \vartheta_A \,.
\end{align}
From the above relations we compute the scalar potential as\footnote{Note that in \eqref{eqn:potential}, we reintroduced hats over the generalized fluxes $F_{ABC}$, although in our discussion we dropped $\Sigma_{ABC}$ for the moment. Therefore strictly, we are dealing with unhatted quantities. However, after reintroducing $\Sigma_{ABC}$ as we will do later, it is easy to see that the general expression with the hats holds.}
\begin{align}\label{eqn:potential}
	\begin{split}
		V =& -\Exp{2\Delta}\,\Bigl[
			-\tfrac{1}{12}\,\mathbb{H}^{AD}\,\mathbb{H}^{BE}\,\mathbb{H}_{CF}\,\hat{F}_{AB}{}^{C}\,\hat{F}_{DE}{}^{F}
			- \tfrac{1}{4}\,\mathbb{H}^{AB}\, \hat{F}_{BD}{}^{C}\,\hat{F}_{AC}{}^{D}+
			\\
			&- (9-d)\,\mathbb{H}^{AB}\, \xi_{A} \, \xi_{B}
			- 2\,\beta^{-1}\, \mathbb{H}^{AB}\, \xi_{A} \, \vartheta_{B}
			+ \beta^{-2} \,\mathbb{H}^{AB}\, \vartheta_{A} \, \vartheta_{B}\Bigr]\,,
	\end{split}
\end{align}
where we see explicitly why the last two terms in \eqref{eqn:ingr-DFT-action-d} are important to reproduce the known scalar potential for non-unimodular gaugings (i.e., gaugings with $\xi_A\neq 0$). 
In $D=4$ and $D=5$, this is consistent with (2.11) and (3.16) of \cite{Schon:2006kz} respectively.
Then the ten-dimensional action depends on the internal coordinates $x^i$ only through the overall factor
\begin{align}
	\Exp{-\beta^{-1}\,(\Delta-\beta\,\bar{\lambda})}\,,
\end{align}
and if it reduces to a scalar density $v$ on the internal space that induces a left-invariant integration measure, then the full action integral \eqref{eq:DFT-action-d} splits into
\begin{align}\label{actionwithleft}
	S = \int\rmd^{D} x\,\bigl(\text{$x^i$-independent terms}\bigr) \times \int\rmd^d x\,v\,.
\end{align}
In this case, the dimensional reduction of the full ten-dimensional action to $D$ dimensions is possible. However, this only works when the trombone gauging vanishes and the section condition holds. Then, we have
\begin{equation}
	-\beta^{-1}\,D_A (\Delta-\beta\,\bar{\lambda}) = - \partial_{I} E_A{}^{I}= D_A \ln v\,,
\end{equation}
where $v$ is identified with the determinant of the Maurer-Cartan 1-form $v_i^a$, to be defined later in Subsection~\ref{sec:constgenframes}. 
Then, the only thing left to be verified is that $v$ gives rise to a left-invariant measure on the coset $E_A{}^I$ is defined on. This is, indeed, the case, as we discuss in details in Appendix \ref{leftinv}.
When the trombone gauging does not vanish, we find 
\begin{equation}
	-\beta^{-1}\,D_A (\Delta-\beta\,\bar{\lambda}) = D_A \ln v + \beta^{-1}\,E_A{}^i\,\vartheta_i \,,
\end{equation}
where we have decomposed the trombone gauging as $\vartheta_A = \begin{pmatrix} \vartheta_a & \vartheta_{\tilde{\alpha}} \end{pmatrix}$ and defined $\vartheta_i\equiv \delta_i^a\,\vartheta_a$\,. 
Then the volume factor becomes
\begin{align}\label{eqn:volD4,5}
	\Exp{-\beta^{-1}\,(\Delta-\beta\,\bar{\lambda})} = v\,\Exp{\beta^{-1}\,\vartheta_i\,x^i}\,.
\end{align}
As we show in Appendix \ref{leftinv}, if $\vartheta_{\tilde{\alpha}}$ vanishes, $v$ gives a left-invariant measure, but the additional factor $\Exp{\beta^{-1}\,\vartheta_i\,x^i}$ breaks the left invariance, and the action \eqref{actionwithleft} is not a well-defined $D$-dimensional action.
Of course, since the equations of motion depend only on the $D$-dimensional fields and the constant embedding tensor $X_{\Ah\Bh}{}^{\Ch}$, they are well-defined in $D$-dimensions and define a class of gauged supergravity. 
This is consistent with the known fact that the gauged supergravity with trombone gauging does not admit an action. 

Now, we consider the case $D\geq 6$. 
Here we have $\bar{\lambda}=0$ and we need to modify the ansatz \eqref{eq:dilaton-ansatz}. 
In $D=4,5$, the ansatz \eqref{eq:dilaton-ansatz} for $\phi$ can be expressed with the aid of~\eqref{eqn:volD4,5} as
\begin{align}\label{eq:dilaton-ansatz-D45}
	\Exp{-2\phi} =\Exp{-2\hat{\phi}(x^\mu)}\Exp{\beta^{-1} \Delta(x^i) + \beta^{-1}\,\vartheta_i\,x^i + \ln v(x^i)}\,,
\end{align}
but in $D\geq 6$, the embedding tensor $X_{\Ah\Bh}{}^{\Ch}$ contains only $F_{AB}{}^C$ and $\xi_A$ and the definition of $\vartheta_i$ is not clear in principle. 
To find a way to naturally define the trombone gauging in $D\geq 6$, we start recalling the notion of trombone symmetry. 
The $D$-dimensional theory given by the action \eqref{eq:DFT-action-d} has a global symmetry given by
\begin{equation}\label{eqn:Dsymm}
\begin{aligned}
 e'^\mu_{\bat} &= \Exp{\alpha_1}\,\hat{e}_{\bat}^\mu \,,\\
 B'_{\mu\nu} &= \Exp{-2\alpha_1}\,\hat{B}_{\mu\nu} \,,\\
 A'_{\mu}{}^I &= \Exp{-\alpha_1} A_\mu{}^I \,,\\
 \cV'_A{}^{I} &= \hat{\cV}_A{}^I\,, 
\\
 -2\,\phi' &= -2\,\hat{\phi} + \beta^{-1}\,\bigl(\alpha_1 + \alpha_2\bigr) \,,
\end{aligned}
\end{equation}
where $\alpha_1$ and $\alpha_2$ are constant. 
Setting $\alpha_2=0$ we obtain a symmetry under which the action is invariant while $\alpha_2\neq 0$ induces an overall rescaling of the action and therefore it is only a symmetry of the equations of motion. 
The $D$-dimensional trombone symmetry is defined by setting $(\alpha_1,\,\alpha_2)=(\alpha,\,-\alpha)$ in~\eqref{eqn:Dsymm} so that $\phi$ is invariant. 
Gauging these symmetries means promoting the transformation parameters to functions that depend on the internal coordinates $x^i$ such that the $D$-dimensional equations of motion do not depend on $x^i$. 
Recalling our Scherk-Schwarz ansatz \eqref{eq:gSS} and the ansatz \eqref{eq:dilaton-ansatz-D45} for the dilaton in $D=4,5$, we can understand that $\alpha_1$ and $\alpha_2$ are promoted to functions $\Delta$ and $\vartheta_i\,x^i$, respectively. 
This suggests us to make an ansatz for the dilaton in $D\geq 6$ as
\begin{align}\label{eq:dilaton-ansatz-D6}
 \Exp{-2\phi} =\Exp{-2\hat{\phi}(x^\mu)}\Exp{\beta^{-1} \Delta(x^i) + \beta^{-1}\,\sigma(x^i) +\ln v(x^i)}\,.
\end{align}
We then find
\begin{align}
	F_A = -\Exp{\Delta}\,\bigl(1+\beta^{-1}\bigr)\,\xi_A - \beta^{-1}\,\Exp{\Delta}\, \tilde{\Vtheta}_A \,,
\label{eq:FA-d<=4}
\end{align}
where
\begin{align}
 \tilde{\Vtheta}_A \equiv \Vtheta_A + \xi_A + D_A\sigma + \beta\,D_A\ln v\,.
\end{align}
An important difference from the case $D=4,5$ is that $\tilde{\Vtheta}_A$ is not always constant. 
If we use the matrix $M_A{}^B$ and $A_a{}^B$, to be defined later in \eqref{eqn:paramM} and \eqref{eqn:defAaBh}, this can be expressed as
\begin{align}
 \tilde{\Vtheta}_A = D_A\sigma + \beta\,\bigl(M_A{}^{\tilde{\beta}} - M_A{}^c\,A_c{}^{\tilde{\beta}}\bigr)\,X_{\tilde{\beta}d}{}^d\,.
\end{align}
If we can choose a parametrization of $M_A{}^B$ and $E_A{}^I$ such that $\tilde{\Vtheta}_A$ is constant, $\tilde{\Vtheta}_A(x^i)=\vartheta_A$, we can evaluate its value as
\begin{align}\label{eq:trombone-D6}
 \vartheta_A = \begin{pmatrix} \vartheta_a & \beta\,X_{\tilde{\alpha}b}{}^b\end{pmatrix} ,
\end{align}
where $\vartheta_a\equiv \delta_a^i\,\partial_i\sigma$ and we have exploited the fact that there is a point $x^i=0$ at which the matrix $M_A{}^B$ reduces to $\delta_A^B$ and $E_A{}^I$ can be parameterized such that $E_A{}^I\rvert_{x^i=0} = \delta_A^I$. 
The constant $\vartheta_A$ plays the role of the trombone gauging in $D\geq 6$, and the scalar potential takes the same from as \eqref{eqn:potential}. 
In particular, if $\vartheta_A=0$ this scalar potential in $D=6$ reproduces (3.7) of \cite{Dibitetto:2019odu}. 
As we show in Appendix \ref{leftinv}, a sufficient condition for $v$ to be left-invariant is that $X_{\tilde{\alpha}b}{}^b=0$, and the integration measure $v\,\Exp{\beta^{-1}\,\sigma(x^i)}$ is left-invariant only by further requiring $\vartheta_a=0$\,. 
Namely, similar to the case $D=4,5$, the $D$-dimensional action is well-defined if the trombone gauging is absent. 
We also note that half-maximal gauged supergravites in $D\geq 6$ are not characterized only by the embedding tensor $X_{\Ah\Bh}{}^{\Ch}$ (which determines the gauge group $G$) but it depends on the choice of the trombone gauging. The top components $\vartheta_a$ are introduced by hand, and the dual components $\vartheta_{\tilde{\alpha}}=\beta\,X_{\tilde{\alpha}b}{}^b$ depend on the choice of the subgroup $H$ (which will be defined later). 
An example of a gauged supergravity with a trombone gauging has been studied in \cite{Kerimo:2003am,Kerimo:2004md} (based on \cite{Lavrinenko:1997qa}), where $d=4$ and $n=0$, and the only non-zero parameters are $\xi_4=m_1-m_2$ and $\vartheta_4=m_2$\,.

\section{Construction of the frames}\label{sec:frames}
In the last section we identified necessary constraints for half-maximal gSUGRAs to admit an uplift to the ten-dimensional action \eqref{eq:DFT-action-d}. However, we did not yet construct the explicit ans\"atze for the uplifts; this is what we are going to do now and we will see that further constraints arise if we require the existence of the $D$-dimensional action. All the relevant fields can be easily read-off from the extended frame $E_{\Ah}{}^{\Ih}$; thus, we are left with getting those of them that satisfy the section condition for heterotic/type I supergravity, namely that they only depend on the coordinates $x^i$. Their embedding tensors $X_{\Ah\Bh}{}^{\Ch}$ are called geometric and satisfy a Leibniz algebra,
\begin{align}
	T_{\Ah}\circ T_{\Bh} = X_{\Ah\Bh}{}^{\Ch}\,T_{\Ch}\,,
\end{align}
to be called heterotic/type I geometric algebra.

In Subsection~\ref{sec:geomalgebras} we discuss these geometric algebras, identifying all the resulting geometric gaugings in $d\leq 7$. Afterwards, frames are constructed from basic elements of group theory
in Subsection~\ref{sec:constgenframes}.
In Subsection~\ref{sec:twist}, we discuss an alternative way to
introduce the torsion term $\Sigma_{\cI\cJ}{}^\cK$ of \eqref{eqn:torsiondef} through an
additional twist of the frame. In this way the non-abelian structure of the theory will be made explicitly manifest.

\subsection{Heterotic/type I geometric algebras}\label{sec:geomalgebras}
Assume that the frame $E_{\Ah}{}^{\Ih}$ just depends on the coordinates $x^i$. Then, we can always find a global transformations such that it becomes the identity element of the respective duality group at the distinguished point $x^i = 0$. We define a geometric gauging by the requirement that the Weitzenb\"ock connection evaluated at this point,
\begin{equation}\label{eqn:constantW}
	W_{\Ah\Bh}{}^{\Ch} =\Omega_{\Ah\Bh}{}^{\Ch}|_{x^i=0}= W_{\Ah}{}^{\bm\delta} (t_{\bm\delta})_{\Bh}{}^{\Ch}\,,
\end{equation}
has only contributions from $W_a{}^{\bm{\beta}}$, while all the others are removed imposing the section condition. At this point, one computes the components of the embedding tensor, which are coordinate-independent. As before, the details depend on the dimensions $d$. One thing they all have in common are the contributions coming from $\OO(d,\mathfrak{n})$ leading to the constants in $W_A{}^{\hat{\bm{\beta}}}$
\begin{align}
	W_a{}^{\hat{\bm{\beta}}} = \begin{pmatrix} f_a{}^{b_1b_2} & f_a{}^b{}_{\gA} & f_{a,b_1}{}^{b_2} & f_a{}^{\gB_1\gB_2} & f_{a,b}{}^{\gA} & f_{a,b_1b_2}\end{pmatrix}\,.
\end{align}
Keep in mind that we used here the decomposition of the generators already encountered in \eqref{generators}. However, not all of these components appear in $X_{\Ah\Bh}{}^{\Ch}$, but only those that survive after antisymmetrizing the lower indices, namely
\begin{align}
	\begin{pmatrix} f_a{}^{bc} & f_a{}^b{}_{\gA} & f_{[a,b]}{}^{c} & f_a{}^{\gB\gC} & f_{[a,b]}{}^{\gC} & f_{[a,bc]} \end{pmatrix}\,.
\end{align}
\begin{enumerate}
	\item Additionally, for $d\le 5$, there are two abelian generators $(R_*+\beta\,t_0)$ and $t_0$\,, with the corresponding structure constants denoted by
	      \begin{align}
		      \begin{pmatrix} f_a & Z_a \end{pmatrix}\,.
	      \end{align}
	      Note that the generator $(R_*+\beta\,t_0)$ vanishes in $d\leq 4$\,, and therefore the structure constants $f_a$ appear only in $d=5$\,. After making suitable redefinitions, we parametrize all geometric gaugings by
	      \begin{align}\label{eqn:Fcomponents}
		      \begin{alignedat}{3}
			      F_{abc}&=h_{abc}\,,\hspace{0.5cm}&
			      F_{ab}{}^c&=f_{ab}{}^c + \delta_a^c\,Z_b - \delta_b^c\,Z_a\,,\hspace{0.5cm}&
			      F_{a}{}^{bc}&=f_{a}{}^{bc}\,,
			      \\
			      F_{ab\gC}&=h_{ab\gC}\,,\hspace{0.5cm}&
			      F_{a\gB\gC}&=f_{a\gB\gC}\,,\hspace{0.5cm}&
			      F_{\gA\gB\gC}&=0 \,,
			      \\
			      F^{abc}&=0\,,
		      \end{alignedat}
	      \end{align}
	      and
	      \begin{equation}
		      \begin{aligned}
			      \xi_A       & = \begin{pmatrix} Z_a&0&0\end{pmatrix} ,\hspace{0.5cm}                                        &
			      \vartheta_A & = \begin{pmatrix} \beta\,f_a-Z_a & \beta\,f_b{}^b{}_{\gA} & \beta\,f_b{}^{ba} \end{pmatrix} ,                   \\
			      \xi_{AB}    & =0\,,\hspace{0.5cm}                                                                           & \vartheta_* & =0\,,
		      \end{aligned}
	      \end{equation}
	      where $\xi_{AB}$, $\vartheta_A$, and $\vartheta_*$ are defined only in $d=5$, but the trombone gauging \eqref{eq:trombone-D6} in $d\leq 4$ takes a similar form $\vartheta_A = \begin{pmatrix} \vartheta_a & \beta\, f_b{}^b{}_{\gA} & \beta\,f_b{}^{ba} \end{pmatrix}$.

	\item In $d=6$, one has to take into account four more generators, $R^\alpha{}_\beta$ of sl$(2)$, and $t_0$. We arrange these as
	      \begin{align}
		      \begin{pmatrix} (R^+{}_++\beta\,t_0) & R^+{}_- & R^-{}_+ & t_0\end{pmatrix}
	      \end{align}
	      to obtain
	      \begin{align}
		      \begin{pmatrix} W_{+a}{}^{\tilde{\bm{\beta}}} & W_{+a}{}^0 \end{pmatrix} = \begin{pmatrix} f_a & f_{a+}{}^- & f_{a-}{}^+ & Z_a \end{pmatrix}\,.
	      \end{align}
	      Again not all of these components will be part of $X_{\Ah\Bh}{}^{\Ch}$; indeed $f_{a+}{}^-$ disappears completely. The reason for this can be understood by looking at the general expressions \eqref{eq:X-4d-1} and \eqref{eq:X-4d-2} from where it is clear that  $f_{a+}{}^-$ can appear only through $\xi_{\alpha A}$ and $\vartheta_{\alpha A}$\,. However, under $W_{-a}{}^{\bm{\alpha}}=W_{-a}{}^0=0$\,, we easily see that $W_{-a+b}{}^{-c}$ cannot appear in $\xi_{\alpha A}$ and $\vartheta_{\alpha A}$\,. Consequentially, geometric gaugings for $d=6$ are given by
	      \begin{align}
		      \begin{alignedat}{2}
			      F_{+ABC}&=F_{ABC}\,,& F_{-ABC}&=0\,,
			      \\
			      \xi_{+A}&=\begin{pmatrix} Z_a & 0 & 0 \end{pmatrix}, & \xi_{-A}&=\begin{pmatrix} f_{a-}{}^+ & 0 & 0 \end{pmatrix},
			      \\
			      \vartheta_{+A}&= \begin{pmatrix} \beta\,f_a - Z_a & \beta\,f_{b}{}^b{}_{\gA} & \beta\,f_b{}^{ba} \end{pmatrix},\qquad & \vartheta_{-A}&=\begin{pmatrix} -\beta\,f_{a-}{}^+ & 0 & 0 \end{pmatrix},
		      \end{alignedat}
	      \end{align}
	      where $F_{ABC}$ is the same as in \eqref{eqn:Fcomponents}.
\end{enumerate}
Some of the gaugings we discovered here were already known in the context of extended Drinfel'd algebras \cite{Sakatani:2021eqo}. The new ones that have not been discussed before, originate from $h_{abc}$ and $h_{ab\gC}$. It is possible to write the universal expression
\begin{equation}\label{eqn:GeomAlgX}
	\begin{aligned}
		X_{a}     & =
		\tfrac{1}{2!}\, h_{abc}\,R^{bc}
		+ h_{ab}{}^{\gA}\,R^{b}_{\gA}
		+ f_{ab}{}^c\,K^b{}_c
		+ \tfrac{1}{2!}\, f_a{}^{bc}\,R_{bc}
		+ f_a{}^b{}_{\gC}\,R_b{}^{\gC}
		+ \tfrac{1}{2!}\, f_a{}^{\gA\gB}\,R_{\gA\gB}                                                           \\
		          & \quad + f_a\,\bigl(R_*+\beta\,t_0\bigr)
		+ f_{a-}{}^{+}\,R^{-}{}_{+}
		- Z_a\,(K + t_0) \,,                                                                                   \\
		X_{\gA}   & = f_{a}{}^b{}_{\gA}\, K^a{}_b
		- f_{a\gA}{}^{\gB}\, R^a{}_{\gB}
		+ \tfrac{1}{2!}\, h_{ab\gA}\, R^{ab}
		- Z_a\,R^a{}_{\gA}
		+ f_a{}^a{}_{\gA}\,\bigl(R_*+\beta\,t_0\bigr) \,,                                                      \\
		X^a       & = f_b{}^{ca}\,K^b{}_c
		- f_b{}^{a\gC} \,R^b{}_{\gC}
		+ \bigl(\tfrac{1}{2}\,f_{bc}{}^a - 2\,Z_{[b}\,\delta_{c]}^a \bigr)\,R^{bc}
		+ f_b{}^{ba}\,\bigl(R_*+\beta\,t_0\bigr)\,,                                                            \\
		X_*       & =0\,,                                                                                      \\
		X_{-a}    & = -f_{b-}{}^{+}\,K^b{}_a - f_{a-}{}^{+}\,\bigl(R_*+\beta\,t_0\bigr) + f_a\,R^{+}{}_{-} \,, \\
		X_{-\gA}  & = f_a{}^a{}_{\gA}\,R^{+}{}_{-} - f_{a-}{}^{+}\,R^a{}_{\gA}\,,                              \\
		X_{-}{}^a & = f_b{}^{ba}\,R^{+}{}_{-} + f_{b-}{}^{+}\,R^{ab}\,,
	\end{aligned}
\end{equation}
for all geometric gaugings in $d\leq 6$, where $K\equiv K^a{}_a$ and $X_*$ is defined only in $d=5$ while $X_{-A}$ is defined only in $d=6$. Moreover, we denote $X_{+A}$ as $X_A$ and $R_*$ as $R^+{}_+$ in the latter case. The full structure of the geometric algebras is presented in Appendix~\ref{geomalg}.

\subsection{Generalized frame fields}\label{sec:constgenframes}
We have now completed all the algebraic considerations we needed and we can finally start with the construction of the generalized frame that realize the geometric algebras. For several other duality groups this construction has been already executed successfully. A common pattern that emerges is that one should start from the parametrization
\begin{align}\label{eqn:ansatzGenFrame}
	E_{\Ah}{}^{\Ih} = M_{\Ah}{}^{\Bh}\,\hat{V}_{\Bh}{}^{\Ih}\,,
\end{align}
where
\begin{align}\label{eqn:vHat}
	\hat{V}_{\Ah}{}^{\Ih} = V_{\Ah}{}^{\Jh}\,N_{\Jh}{}^{\Ih}
\end{align}
decomposes further into the frame $V_{\Ah}{}^{\Jh}$ and $N_{\Ih}{}^{\Jh}$ given by
\begin{align}\label{eqn:leftinvMC}
	N_{\Ih}{}^{\Jh} \equiv \bigl[\exp(-\tfrac{1}{2!}\,\mathfrak{b}_{ij}\,R^{ij})\,\exp(-\mathfrak{a}_k^{\gI}\,R^{k}_{\gI})\bigr]_{\Ih}{}^{\Jh}\,.
\end{align}
The matrix $M_{\Ah}{}^{\Bh}$ mediates the adjoint action of an underlying Lie group and is therefore constrained by
\begin{align}
	(M^{-1})_{\Ah}{}^{\Ch}\,\rmd M_{\Ch}{}^{\Bh} = -v^{\Ch}\,X_{\Ch\Ah}{}^{\Bh}\,,
\end{align}
where $v^{\Ah}$ is an extension of the ordinary Maurer-Cartan 1-form, and satisfies the modified Maurer-Cartan equation \cite{Grutzmann:2014hkn}
\begin{align}\label{maurercartan}
	\rmd v^{\Ah} = - \tfrac{1}{2}\,X_{\Bh\Ch}{}^{\Ah}\,v^{\Bh}\wedge v^{\Ch} -w^{\Ah}\,,
\end{align}
with the 2-form $w^{\Ah}$. The exterior derivative of this object encodes the violation of the Jacobi identity for $X_{[\Bh \Ch]}{}^{\Ah}$ \cite{Hassler:2022egz}, but still it is irrelevant for the construction of the generalized frames, since
\begin{equation}
	w^{\Ah} X_{\Ah \Bh}{}^{\Ch}=0
\end{equation}
holds. All these objects are defined on the coset $G/H$, where $G$ is the gauge group of the gSUGRA in $D$ dimensions, which is completely fixed by the embedding tensor. $H$ is a subgroup of $G$ that we specify next. To this end, we first introduce the matrices
\begin{equation}
    (T_{\Ah})_{\Bh}{}^{\Ch} = - X_{\Ah\Bh}{}^{\Ch}
\end{equation}
which satisfy
\begin{equation}
    [ T_{\Ah}, T_{\Bh} ] = X_{\Ah\Bh}{}^{\Ch} T_{\Ch}
\end{equation}
under normal matrix multiplication due to the Leibniz identity
\begin{equation}\label{eqn:LeibnizId}
    (T_{\Ah} \circ T_{\Bh}) \circ T_{\Ch} + T_{\Bh} \circ (T_{\Ah} \circ T_{\Ch}) = T_{\Ah} \circ (T_{\Bh} \circ T_{\Ch})\,.
\end{equation}
They decompose into three distinguished parts,
\begin{equation}\label{eqn:decompTgen}
    T_{\Ah} = \begin{pmatrix} 
        T_a & T_{\grave{\alpha}} & T_{\acute{\alpha}}
    \end{pmatrix}  = \begin{pmatrix}
        T_a & T_{\tilde{\alpha}}
    \end{pmatrix} = \begin{pmatrix}
        T_{\grave{a}} & T_{\acute{\alpha}}
    \end{pmatrix}\,.
\end{equation}
We already know the index $a$ from \eqref{eqn:decompIndices}, while the two additional conditions,
\begin{equation}
    X_{(\Bh\Ch)}{}^a = 0 \qquad \text{and} \qquad
    X_{(\Bh\Ch)}{}^{\grave{\alpha}} = 0\,,
\end{equation}
fix the full decomposition completely. The gauge group $G$ is generated by $T_{\grave{a}}$, while the Lie algebra of its subgroup $H$ is spanned by $T_{\grave{\alpha}}$. Revisiting \eqref{eqn:leftinvMC} in this light, it is instructive to rewrite it as
\begin{equation}\label{eqn:MinvdM}
    M^{-1} \dd M = v^{\Ah} T_{\Ah}\,,
\end{equation}
to see that it just defines the left-invariant Maurer-Cartan form for a coset representative $M\in G/H$. For example, one might choose this representative as
\begin{equation}\label{eqn:paramM}
    M_{\Ah}{}^{\Bh} = \exp\left( x^c T_c \right){}_{\Ah}{}^{\Bh}\,.
\end{equation}

Finally, we look at $V_{\Ah}{}^{\Ih}$, which depends on the dimension $d$ and reads
\begin{align}
	V_{\Ah}{}^{\Ih} \equiv \begin{cases}
		                       \begin{pmatrix} V_A{}^I & 0 \\ 0 & v
		\end{pmatrix} & d\leq 5
		                       \\
		                       \begin{pmatrix}
			1 & 0 \\ 0 & v
		\end{pmatrix} \otimes  V_A{}{}^I     & d=6\,,
	                       \end{cases}
\end{align}
where
\begin{align}\label{eqn:vAIframe}
	V_A{}^I \equiv \begin{pmatrix} v_a^i & 0                 & 0     \\
                0     & \delta_{\gA}{}^{\gI} & 0     \\
                0     & 0                 & v^a_i
	               \end{pmatrix}\,.
\end{align}
Here, $v^a_i$ originates from $v_i{}^{\Ah} \dd x^i$ after adopting the parametrization
\begin{align}
	v_i{}^{\Ah} = \begin{cases}
		              \begin{pmatrix} v_i^a & v_i^{\gA} & v_{ia} & v_i^* \end{pmatrix}                     & d\leq 5
		              \\
		              \begin{pmatrix} v_i{}^{\bba a} & v_i{}^{\bba \gA} & v_{i}{}^{\bba}_{a} \end{pmatrix} & d=6\,.
	              \end{cases}
\end{align}
Last but not least, we have $v_a^i$, that represents the dual vector fields to the one-forms $v^a_i$.

With the frame fixed, we can now start to compute the Weitzenb\"ock connection
\begin{equation}
    \begin{aligned}\label{eqn:OmegaAhBhCh}
    	\Omega_{\Ah\Bh}{}^{\Ch} & = M_{\Ah}{}^{\Dh}\,M_{\Bh}{}^{\Eh}\,(M^{-1})_{\Fh}{}^{\Ch}\,\bigl[\hat{\Omega}_{\Dh\Eh}{}^{\Fh} - \hat{V}_{\Dh}{}^{\Ih}\,(M^{-1}\,\partial_IM)_{\Eh}{}^{\Fh}\bigr] \\
    	                        & = M_{\Ah}{}^d\,M_{\Bh}{}^{\Eh}\,(M^{-1})_{\Fh}{}^{\Ch}\,\bigl(\hat{\Omega}_{d\Eh}{}^{\Fh} + A_d{}^{\Gh}\,X_{\Gh\Eh}{}^{\Fh} \bigr),
    \end{aligned}
\end{equation}
where we have introduced
\begin{align}\label{eqn:defAaBh}
	\hat{\Omega}_{a\Bh}{}^{\Ch} = v_a^i\,\hat{V}_{\Bh}{}^{\Jh}\,\partial_i \hat{V}_{\Jh}{}^{\Ch}\,,\qquad
	A_a{}^{\Bh} \equiv v_a^i\,v_i{}^{\Bh}\,,
\end{align}
to keep the equations more compact. In order to evaluate $\hat{\Omega}_{a\Bh}{}^{\Ch}$, we need three more ingredients. First, the Maurer-Cartan equation \eqref{maurercartan}, which results in
\begin{align}
	\rmd v^a & = - \tfrac{1}{2}\,X_{\hat{B} \hat{C}}{}^a\,v^{\hat{B}} \wedge v^{\hat{C}}
	\nn                                                                                                                                                                 \\
	         & = -\bigl(\tfrac{1}{2}\,f_{bc}{}^a\,v^b\wedge v^c - f_b{}^{ac}\,v^b\wedge v_{c} - f_b{}^a{}_{\gC}\,v^b\wedge v^{\gC} + f_{b-}{}^+\,v^b\wedge v^{-a}\bigr)\,,
\end{align}
and is needed to cope with the derivatives of the one-forms in \eqref{eqn:vAIframe}. Next, we need to deal with their dual vector fields. For them, after using $\rmd v^c(v_a,v_b)=-v^c([v_a,v_b])$, we get
\begin{align}
	v^c([v_a,v_b]) = - \rmd v^c(v_a,v_b) = f_{ab}{}^c - 2\,f_{[a}{}^{cd}\, A_{b]d} - 2\,f_{[a|}{}^c{}_{\gC}\, A_{|b]}{}^{\gC} + 2\,f_{[a|-}{}^+\,A_{|b]}{}^{-c}\,.
	\label{eq:va-vb}
\end{align}
The last contribution comes from the flat derivative
\begin{align}
	\mathfrak{D}_a N_{\hat{J}}{}^{\hat{L}} \, (N^{-1})_{\hat{L}}{}^{\hat{K}} = -\mathfrak{D}_a \mathfrak{a}^{\gI}_i\,(R^i_{\gI})_{\hat{J}}{}^{\hat{K}} - \tfrac{1}{2!}\,\bigl(\mathfrak{D}_a \mathfrak{b}_{ij} - \kappa_{\gI\gJ}\,\mathfrak{D}_a\mathfrak{a}_{i}^{\gI}\,\mathfrak{a}_j^{\gJ}\bigr)\,(R^{ij})_{\hat{J}}{}^{\hat{K}} \,,
\end{align}
where $\mathfrak{D}_a\equiv v_a^i\, \partial_i$\,. Putting all of them together gives rise to
\begin{equation}
    \begin{aligned}\label{eqn:Omegahat}
    	\hat{\Omega}_{a\Bh}{}^{\Ch} & = -\mathfrak{D}_a V_{\Bh}{}^{\Jh}\,V_{\Jh}{}^{\Ch}
    	- V_{\Bh}{}^{\Jh}\, V_{\Kh}{}^{\Ch}\, \mathfrak{D}_a N_{\Jh}{}^{\Lh} \, (N^{-1})_{\Lh}{}^{\Kh}
    	                                                                                         \\
    	                            & = k_{ab}{}^c\,(K^b{}_c)_{\Bh}{}^{\Ch}
    	+ \mathfrak{D}_a\ln v\,\bigl(R_*+\beta\,t_0\bigr)_{\Bh}{}^{\Ch}
    	- V_{\Bh}{}^{\Jh}\, V_{\Kh}{}^{\Ch}\, \mathfrak{D}_a N_{\Jh}{}^{\Lh}\, (N^{-1})_{\Lh}{}^{\Kh}\,,
    \end{aligned}
\end{equation}
where we have defined
\begin{align}\label{eqn:fandhflux}
	k_{ab}{}^c \equiv v_a^j\,v_b^i\,\partial_i v_j^c\,, \qquad
	\mathfrak{f}_{ij}{}^{\gI} \equiv 2\,\partial_{[i} \mathfrak{a}_{j]}^{\gI} \,,\qquad \text{and} \qquad
	\mathfrak{h}_{ijk} \equiv 3\,\bigl(\partial_{[i} \mathfrak{b}_{jk]} - \kappa_{\gI\gJ}\,\partial_{[i}\mathfrak{a}_{j}^{\gI}\,\mathfrak{a}_{k]}^{\gJ}\bigr)\,.
\end{align}

To further proceed, it is important to keep in mind that only certain projections of $\hat{\Omega}_{a\Bh}{}^{\Ch}$ will enter the embedding tensor $X_{\Ah\Bh}{}^{\Ch}$. To take this fact into account quantitatively, we define the equivalence relation $\sim$ which allow to neglect terms from the Weitzenb\"ock connection that will not contribute. This identifies the following three classes of terms
\begin{itemize}
	\item terms of the form $S_{ab}{}^c\,(K^b{}_c)_{\Bh}{}^{\Ch}+S_{ba}{}^b\,\bigl(R^*+\beta\,t_0\bigr)_{\Bh}{}^{\Ch}$ with $S_{[ab]}{}^c=0$,
	\item terms of the form $S_{ade} \,(R^{de})_{\Bh}{}^{\Ch}$ with $S_{[ade]}=0$,
	\item terms of the form $S_{a}\,(R^+{}_-)_{\Bh}{}^{\Ch}$,
\end{itemize}
which can appear in $\hat{\Omega}_{a\Bh}{}^{\Ch}$ but not in $X_{\Ah\Bh}{}^{\Ch}$. Applying it to \eqref{eqn:Omegahat} leads to the simplifications
\begin{equation}
    \begin{aligned}
    	\hat{\Omega}_{a\Bh}{}^{\Ch} & \sim \bigl[-\bigl(\tfrac{1}{2}\,f_{ab}{}^c - f_{[a}{}^{cd}\, A_{b]d} - f_{[a|}{}^c{}_{\gC}\, A_{|b]}{}^{\gC} + f_{[a|-}{}^+\,A_{|b]}{}^{-c}\bigr) \,K^b{}_c
    	                                                                                                                                                                                       \\
    	                            & \quad  + k_{[ba]}{}^b\,\bigl(R_*+\beta\,t_0\bigr)
    	+ \tfrac{1}{2!}\,\mathfrak{f}_{ab}{}^{\gA}\,R^b_{\gA} + \tfrac{1}{3!}\,\mathfrak{h}_{abc}\, R^{bc} \bigr]_{\Bh}{}^{\Ch}\,,
    \end{aligned}
\end{equation}
after taking into account \eqref{eq:va-vb} and $\mathfrak{D}_a\ln v = v_b^i\,\mathfrak{D}_a v_i^b = k_{ba}{}^b$. From here on, we suppress the last two indices of the Weitzenb\"ock connection and rather use $\hat{\Omega}_a$ that arises from $\hat{\Omega}_{a \Bh}{}^{\Ch} = \hat{\Omega}_a{}^{\bm{\beta}} (t_{\bm{\beta}})_{\Bh}{}^{\Ch}$.

Eventually, we want to verify that our ansatz \eqref{eqn:ansatzGenFrame} results in the embedding tensor \eqref{eqn:GeomAlgX} for the geometric gaugings from the previous subsection. To this end, we need to check that $\Omega_{\Ah\Bh}{}^{\Ch}$ in \eqref{eqn:OmegaAhBhCh} gives rise to the correct $X_{\Ah\Bh}{}^{\Ch}$. Since the embedding tensor is invariant under the adjoint action mediated by $M_{\Ah}{}^{\Bh}$, it is sufficient to consider the two terms in the bracket on the right hand side of \eqref{eqn:OmegaAhBhCh}\footnote{To come to this conclusion, keep in mind that the map from the Weitzenb\"ock connection and the embedding tensor \eqref{eqn:XOmegaMap} is equivariant under the adjoint action of the duality group.}. They give rise to
\begin{equation}
    \begin{aligned}\label{eqn:OmegahatplusAX}
    	\hat{\Omega}_{a} + A_a{}^{\hat{B}}\,X_{\hat{B}}
    	 & \sim X_a + \tfrac{1}{2}\,f_{ba}{}^c\,K^b{}_c + \tfrac{1}{2}\,f_{ba}{}^b\,\bigl(R_*+\beta\,t_0\bigr)\\
    	 & \quad +\bigl(\tfrac{1}{2}\,\mathfrak{f}_{ac}{}^{\gC} - f_{c\gB}{}^{\gC}\,A_a{}^{\gB} - A_a{}^{\gC}\,Z_{c} - f_c{}^{d\gC}\,A_{ad} - f_{c-}{}^+\,A_a{}^{-\gC}\bigr)\,R^c_{\gC}\\
    	 & \quad +\bigl(\tfrac{1}{3!}\,\mathfrak{h}_{abc} + \tfrac{1}{2}\,h_{bc\gA}\,A_a{}^{\gA} + \tfrac{1}{2}\,f_{bc}{}^d\,A_{ad} - 2\,Z_{b}\,A_{ac} - f_{b-}{}^+\,A_a{}^{-}{}_c \bigr)\,R^{bc} \,,
    \end{aligned}
\end{equation}
after taking into account \eqref{eqn:GeomAlgX} and the definition of $A_a{}^B$ \eqref{eqn:defAaBh}. Note again that $R_*$ is $R^+{}_+$ in $d=6$.
The right hand side has to match
\begin{equation}\label{eq:Omega-a}
	W_a \sim X_a + \tfrac{1}{2}\,f_{ba}{}^c\,K^b{}_c + \tfrac{1}{2}\,f_{ba}{}^b\,\bigl(R_*+\beta\,t_0\bigr)
	- \tfrac{1}{3}\,h_{abc}\,R^{bc} - \tfrac{1}{2}\,h_{ab}{}^{\gC}\,R^b_{\gC} \,.
\end{equation}
One can understand this relation as the inverse of the map \eqref{eqn:XOmegaMap}, evaluated at the point $x^i=0$ defined in \eqref{eqn:constantW}. At this point, it is also clear why we need the equivalence relation. As the map has a non-trivial kernel, we can only specify the Weitzenb\"ock connection up to elements from this kernel. It is straightforward to check that this choice for $\Omega_a$ will indeed result in the embedding tensor in \eqref{eqn:ansatzGenFrame}. The right hand sides of \eqref{eqn:OmegahatplusAX} and \eqref{eq:Omega-a} only match if we impose
\begin{equation}
    \begin{aligned}
    	\mathfrak{f}_{ab}{}^{\gC} & = -h_{ab}{}^{\gC} + 2\,A_{[a}{}^{\gB}\, f_{b]\gB}{}^{\gC} + 2\,A_{[a}{}^{\gC}\, Z_{b]} + 2\,A_{[a|d|}\, f_{b]}{}^{d\gC} + 2\,A_{[a}{}^{-\gC}\, f_{b]-}{}^+\,,
    	\\
    	\mathfrak{h}_{abc}        & = -2\,h_{abc} - 3\,A_{[a}{}^{\gA}\,h_{bc]\gA} - 3\,A_{[a|d}\,f_{bc]}{}^d - 12\,A_{[ab}\,Z_{c]} - 6\,A_{[a}{}^{-}{}_{b}\, f_{c]-}{}^+\,,
    \end{aligned}
\end{equation}
which can also be expressed more compactly as
\begin{equation}\label{eqn:fieldstrength}
    \begin{aligned}
    	\mathfrak{f}_2{}^{\gA} & = h_2{}^{\gA} - X_{\Bh c}{}^{\gA}\,v^{\Bh}\wedge v^c \,,
    	\\
    	\mathfrak{h}_3         & = h_3 - \tfrac{1}{2}\,X_{\Ah b c}\,v^{\Ah}\wedge v^b\wedge v^c\,.
    \end{aligned}
\end{equation}
From the definition \eqref{eqn:fandhflux}, we see that this only constrains the derivatives of the potentials $\mathfrak{a}_i^{\cJ}$ and $\mathfrak{b}_{ij}$, required to fully fix the generalized frame. To obtain the potentials, one has to integrate, which requires that certain integrability conditions are satisfied. These are represented by Bianchi identities, which we now discuss.
\begin{enumerate}
	\item In $d\leq 5$, we find
	      \begin{equation}
                \begin{aligned}
    		      \rmd \mathfrak{f}_2{}^{\gA} & = -\tfrac{1}{2}\,\bigl(f_{eb}{}^{\gA}\,F_{cd}{}^e - X_{\Eh b}{}^{\gA}\,F_{cd}{}^{\Eh} + X_{b e}{}^{\gA}\,F_{cd}{}^e\bigr)\,v^b\wedge v^c\wedge v^d                                                                                           \\
    		                                  & \quad + \bigl(f_{ec}{}^{\gA}\,F_{\tilde{\beta}d}{}^e - X_{\Eh c}{}^{\gA}\,F_{\tilde{\beta}d}{}^{\Eh} + X_{ce}{}^{\gA}\,F_{\tilde{\beta}d}{}^e-\tfrac{1}{2}\,X_{\tilde{\beta}e}{}^{\gA}\,F_{cd}{}^e\bigr)\,v^{\tilde{\beta}}\wedge v^c\wedge v^d
    		                                                                                                                                                                                                                                                                                 \\
    		                                  & \quad + \bigl(\tfrac{1}{2}\,X_{\Eh d}{}^{\gA}\,F_{\tilde{\beta}\tilde{\gamma}}{}^{\Eh}-X_{[\tilde{\beta}|e}{}^{\gA}\,F_{|\tilde{\gamma}]d}{}^e\bigr)\,v^{\tilde{\beta}}\wedge v^{\tilde{\gamma}}\wedge v^d
    		                                                                                                                                                                                                                                                                                 \\
    		                                  & = \tfrac{1}{3!}\,L_{bcd}{}^{\gA}\,v^b\wedge v^c\wedge v^d
    		      + \tfrac{1}{2}\,L_{c\tilde{\beta}d}{}^{\gA}\,v^{\tilde{\beta}}\wedge v^c\wedge v^d
    		      + \tfrac{1}{2}\,L_{\tilde{\beta}\tilde{\gamma}d}{}^{\gA} \,v^{\tilde{\beta}}\wedge v^{\tilde{\gamma}}\wedge v^d\,.
    	      \end{aligned}
	      \end{equation}
	      Note that here we encounter the indices $\tilde{\alpha}$ which appear in the decomposition \eqref{eqn:decompTgen} of the Leibniz algebra's generators, and we 
	      have defined the Leibniz identity as
	      \begin{align}
		      L_{\Ah\Bh\Ch}{}^{\Dh} \equiv X_{\Ah\Ch}{}^{\Eh}\,X_{\Bh\Eh}{}^{\Dh} - X_{\Bh\Ch}{}^{\Eh}\,X_{\Ah\Eh}{}^{\Dh} + X_{\Ah\Bh}{}^{\Eh}\,X_{\Eh\Ch}{}^{\Dh}\,.
	      \end{align}
	      Of course this tensor vanishes due to \eqref{eqn:LeibnizId}. Similarly we find
	      \begin{align}
		      \rmd \mathfrak{h}_3 + \tfrac{1}{2}\,\kappa_{\gA\gB}\,\mathfrak{f}_2{}^{\gA}\wedge \mathfrak{f}_2{}^{\gB}
		       & = \tfrac{1}{8}\,L_{abcd}\,v^a\wedge v^b\wedge v^c\wedge v^d
		      + \tfrac{1}{12}\,\bigl(L_{\tilde{\alpha}bcd}-3\,L_{b\tilde{\alpha}cd}\bigr)\,v^{\tilde{\alpha}}\wedge v^b\wedge v^c\wedge v^d
		      \nn                                                                                                                           \\
		       & \quad + \tfrac{1}{4}\,L_{\tilde{\alpha}\tilde{\beta}cd}\,v^{\tilde{\alpha}}\wedge v^{\tilde{\beta}}\wedge v^c\wedge v^d\,.
	      \end{align}
	\item In $d=6$\,, by defining
	      \begin{align}
		      \Delta_{\Ah\Bh\Ch}{}^{\Dh} \equiv - \tfrac{1}{4}\,\epsilon_{\bba\bbb}\,\epsilon^{\bbc\bbd}\,\bigl(L_{\bbc A,\bbd B,\Ch}{}^{\Dh} + L_{\bbd B,\bbc A,\Ch}{}^{\Dh}\bigr)\,,
	      \end{align}
	      we find
	      \begin{align}
    		      \rmd \mathfrak{f}_2{}^{\gA} & = \tfrac{1}{3!}\,L_{bcd}{}^{\gA}\,v^b\wedge v^c\wedge v^d
    		      + \tfrac{1}{2}\,L_{c\tilde{\beta}d}{}^{\gA}\,v^{\tilde{\beta}}\wedge v^c\wedge v^d
    		      \nn                                                                                                                                                               \\
    		                                  & \quad + \tfrac{1}{2}\,(L+\Delta)_{\tilde{\beta}\tilde{\gamma}d}{}^{\gA}\, v^{\tilde{\beta}}\wedge v^{\tilde{\gamma}}\wedge v^d
    		      \,,
    		      \intertext{and}
    		      \rmd \mathfrak{h}_3 + \tfrac{1}{2}\,\kappa_{\gA\gB}\,\mathfrak{f}_2{}^{\gA}\wedge \mathfrak{f}_2{}^{\gB}
    		                                  & = \tfrac{1}{8}\,L_{abcd}\,v^a\wedge v^b\wedge v^c\wedge v^d
    		      + \tfrac{1}{12}\,\bigl(L_{\tilde{\alpha}bcd}-3\,L_{b\tilde{\alpha}cd}\bigr)\,v^{\tilde{\alpha}}\wedge v^b\wedge v^c\wedge v^d
    		      \nn                                                                                                                                                               \\
    		                                  & \quad + \tfrac{1}{4}\,(L+\Delta)_{\tilde{\alpha}\tilde{\beta}cd}\,v^{\tilde{\alpha}}\wedge v^{\tilde{\beta}}\wedge v^c\wedge v^d\,.
    	      \end{align}
\end{enumerate}
Thus the Bianchi identities are ensured by the Leibniz identity.

\subsection{Non-abelian twist}\label{sec:twist}
Up to now, we have introduced by hand the structure constants $f_{\cI\cJ}{}^{\cK}$ of the Lie algebra of $\cG$ in the ten-dimensional theory through the term $\Sigma_{\Iht\Jht}{}^{\Kht}$ in \eqref{eqn:10dhatF}. This can be seen as a deformation of the ten-dimensional generalized Lie derivative. Here, we want to take an alternative approach where $f_{\cI\cJ}{}^{\cK}$ is introduced through a twist in the generalized frames $E_{\Ah}{}^{\Ih}$\,. In this way the non-abelian nature of the gauge potentials in~\eqref{eqn:leftinvMC} is made manifest.

We start defining the right-twist
\begin{equation}\label{eqn:LucaTwist}
	\hat{E}_{\Ah}{}^{\Ih}=E_{\Ah}{}^{\Jh}\, \cU_{\Jh}{}^{\Ih},
\end{equation}
with
\begin{align}
	\cU_{\Jh}{}^{\Ih} \equiv \begin{cases}
		                       \begin{pmatrix} \cU_J{}^I & 0 \\ 0 & 1
		\end{pmatrix} & d\leq 5
		                       \\
		                       \begin{pmatrix} 1 & 0 \\ 0 & 1
		\end{pmatrix} \otimes  \cU_J{}^I     & d=6\,,
	                       \end{cases}
\end{align}
and
\begin{equation}
	\cU_J{}^I=\begin{pmatrix}
		\delta_j{}^i & 0               & 0            \\
		0            & u_{\cJ}{}^{\cI} & 0            \\
		0            & 0               & \delta^j{}_i
	\end{pmatrix}.
\end{equation}
Here, the matrix $u_{\cI}{}^{\cJ}$ is an element in the adjoint representation of the gauge group, and its inverse $u^{\cJ}{}_{\cI}$ is defined such that $u_{\cI}{}^{\cK} u^{\cJ}{}_{\cK} = \delta_{\cI}^{\cJ}$. 
The twist matrix $u_{\cJ}{}^\cI$ is chosen such that
\begin{equation}
	\Sigma_{\mathcal{I}\mathcal{J}}{}^{\mathcal{K}} = f_{\cI\cJ}{}^{\cK} = -2 u_{[\cI}{}^\cL \partial_\cL u_{\cJ]}{}^\cM u^{\cK}{}_{\cM}\,
\end{equation}
holds. In $d=5,6$, we additionally assume the unimodularity condition $\partial_\cI u_{\cJ}{}^\cI=0$, in order to avoid new undesired terms arising from the symmetric part of $u_{\cI}{}^\cL \partial_\cL u_{\cJ}{}^\cM u^{\cK}{}_{\cM}$. Note that for a non-vanishing $f_{\cI\cJ\cK}$, $u_{\cI}{}^{\cJ}$ has to depend on the coordinates $x^{\cI}$, and this breaks the heterotic section. But this is not an inconsistency because the gauge algebra still closes. Due to the additional twist, the generalized fluxes are modified as
\begin{align}
 X_{\Ah\Bh}{}^{\Ch} \to \hat{X}_{\Ah\Bh}{}^{\Ch} \equiv X_{\Ah\Bh}{}^{\Ch} + E_{\Ah}{}^{\Ih}\, E_{\Bh}{}^{\Jh}\, E_{\Jh}{}^{\Ch}\, \Sigma_{\Ih\Jh}{}^{\Kh}\,,
\label{eq:modify}
\end{align}
and the twisted frames $\hat{E}_{\Ah}{}^{\Ih}$ satisfy the desired relation
\begin{align}
	\gLie_{\hat{E}_{\Ah}} \hat{E}_{\Bh} = - \hat{X}_{\Ah\Bh}{}^{\Ch}\,\hat{E}_{\Ch}\,,
\end{align}
a modification of \eqref{eq:gen-flux}.
The $X_{\Ah\Bh}{}^{\Ch}$ part of the modified flux is constructed from the Maurer-Cartan form precisely in the same way as explained in Section~\ref{sec:constgenframes}. 
Therefore, the expressions \eqref{eqn:fieldstrength} of the field strengths $\mathfrak{f}_2{}^{\gA}$ and $\mathfrak{h}_3$ are not modified and the Bianchi identities for the field strengths are still satisfied. 

In terms of the structure constants, the modification \eqref{eq:modify} can be also expressed as
\begin{alignat}{2}
 f_{a}{}^{\gA\gB} \ &\to&\ \hat{f}_{a}{}^{\gA\gB} &\equiv f_{a}{}^{\gA\gB} - \mathfrak{a}_a{}^{\gC}\,f_{\gC}{}^{\gA\gB}\,,
\\
 h_{ab}{}^{\gA} \ &\to&\ \hat{h}_{ab}{}^{\gA} &\equiv h_{ab}{}^{\gA} + \mathfrak{a}_a{}^{\gB}\,\mathfrak{a}_b{}^{\gC}\,f_{\gB\gC}{}^{\gA}\,,
\\
 h_{abc} \ &\to&\ \hat{h}_{abc} &\equiv h_{abc} -\mathfrak{a}_a{}^{\gA}\,\mathfrak{a}_b{}^{\gB}\,\mathfrak{a}_c{}^{\gC}\,f_{\gA\gB\gC}\,,
\end{alignat}
as well as a constant shift
\begin{align}
 X_{\gA\gB}{}^{\gC} = 0 \to \hat{X}_{\gA\gB}{}^{\gC} = f_{\gA\gB}{}^{\gC}\,.
\end{align}
Then, if we prefer to express \eqref{eqn:fieldstrength} by using the modified flux $\hat{X}_{\Ah\Bh}{}^{\Ch}$, we find that
\begin{align}
    \begin{aligned}
    	\hat{\mathfrak{f}}_2{}^{\gA} & = \hat{h}_2{}^{\gA} - \hat{X}_{\Bh c}{}^{\gA}\,v^{\Bh}\wedge v^c \,,
    	\\
    	\hat{\mathfrak{h}}_3         & = \hat{h}_3 - \tfrac{1}{2}\,\hat{X}_{\Ah b c}\,v^{\Ah}\wedge v^b\wedge v^c\,,
    \end{aligned}
\end{align}
where the field strengths \eqref{eqn:fandhflux} are modified as follows:
\begin{equation}\label{eqn:nonAbelianStrengths}
    \begin{aligned}
    	\hat{\mathfrak{f}}_{ij}{}^{\gI} & \equiv 2\,\left(\partial_{[i} \mathfrak{a}_{j]}^{\gI}+\frac12 \mathfrak{a}_{[i}^{\gJ}\mathfrak{a}_{j]}^{\gK} f_{\gJ\gK}{}^{\gI}\right) \,,                                                                                               \\
    	\hat{\mathfrak{h}}_{ijk}        & \equiv 3\,\left(\partial_{[i} \mathfrak{b}_{jk]} - \kappa_{\gI\gJ}\,\partial_{[i}\mathfrak{a}_{j}^{\gI}\,\mathfrak{a}_{k]}^{\gJ}-\frac13 \mathfrak{a}_{[i}^{\gI} \mathfrak{a}_{j}^{\gJ} \mathfrak{a}_{k]}^{\gK} f_{\gI\gJ\gK} \right)\,.
    \end{aligned}
\end{equation}
Since the undeformed field strengths satisfy the Bianchi identity with $f_{\gJ\gK}{}^{\gI}=0$\,, we can easily check that in this case \eqref{eqn:nonAbelianStrengths} satisfy a non-abelian version of the Bianchi identities, that is,
\begin{align}\label{eqn:nonAbBianchi}
 \rmd \hat{\mathfrak{f}}_2{}^{\cI}-f_{\cJ\cK}{}^{\cI}\,\hat{\mathfrak{f}}_2{}^{\cJ}\wedge \mathfrak{a}^{\cK}&=0\,,
\\
 \rmd \hat{\mathfrak{h}}_3 + \tfrac{1}{2}\,\kappa_{\gI\gJ}\,\hat{\mathfrak{f}}_2{}^{\gI}\wedge \hat{\mathfrak{f}}_2{}^{\gJ}&=0\,.
\end{align}

\subsection{Generalized dualities}\label{sec:gendualities}
After settling all the technical aspects of consistent truncations and the corresponding uplifts of half-maximal gSUGRAs, we are ready to look at the results from the point of view of generalized dualities. Conceptually the situation is not different from the one for the bosonic string, type II strings or M-theory. The key points that are important for those cases, and for our as well, are:
\begin{itemize}
    \item All the relevant information about the internal space is encoded in the generalized frames we constructed in the last section. The generalized Scherk-Schwarz ansatz presented in Section~\ref{sec:genSSred}, makes this information accessible and allows to obtain the metric and all the relevant gauge potentials.
    \item All of the physics in the resulting $D$-dimensional gauged supergravity is just governed by the embedding tensor. The explicit realization in terms of a generalized frame, as long as it exists, is irrelevant.
    \item While the gauge group $G$ is completely fixed by the embedding tensor, the subgroup $H$ required to construct the coset $G/H$ for the frame is not. Of course it can not be chosen completely arbitrarily and it has to be compatible with the embedding tensor. More specifically, after the transformation with an appropriate element of the respective duality group, the embedding has to be in the form of \eqref{eqn:GeomAlgX}. 
\end{itemize}
Combining these points, it is obvious that when there are different admissible choices for the subgroup $H$ there are also different possible uplifts to the ten-dimensional low-energy effective action of the heterotic/type I string. Each of them might have very different configurations for the metric and the gauge potentials, but still all of them result exactly in the same physics. This is the definition of a duality, or when there are more than two choices for $H$, a plurality.

More formally, we define $p$-plural geometric algebras spanned by the generators $T_{\Ah}^{(i)}$, $i=1,\dots,p$ as those that can be related by transformations $\cO^{(ij)}_{\Ah}{}^{\Bh}$ in $\Gduality$ acting as
\begin{equation}
    T^{(i)}_{\Ah}{}^{\Bh} = \cO^{(ij)}_{\Ah}{}^{\Bh} T^{(j)}_{\Bh}\,.
\end{equation}
From this data, one constructs all the plural generalized frames and, therewith, the corresponding metrics and gauge potentials.

At this point, the trombone gauging $\vartheta_{\Ah}$ requires additional attention. In $D=5$, it is given by
\begin{equation}
    X_{\Ah\Bh}{}^{\Bh} + X_{\Bh\Ah}{}^{\Bh} = - 3\,\vartheta_{\Ah}\,,
\end{equation} while in $D=4$, we have
\begin{equation}
    X_{\Ah\Bh}{}^{\Bh} = - 2\,(2d+n)\,\vartheta_{\Ah}\,.
\end{equation}
Under the generalized dualities, the structure constants $X_{\Ah\Bh}{}^{\Ch}$ transform covariantly, and with them the trombone gauging $\vartheta_{\Ah}$ too. Therefore, if we require the absence of a trombone gauging, this can not be reintroduced by a generalized duality transformation. However, the situation is different in $D\geq 6$\,, where $X_{\Ah\Bh}{}^{\Ch} =  X_{AB}{}^{C}$ contains only $F_{AB}{}^C$ and $\xi_A$. 
In this case, the trombone gauging $\vartheta_A$ defined in \eqref{eq:trombone-D6} is not a covariant object. 
Its absence corresponds to the situation of having $\vartheta_a=0$, $f_b{}^b{}_{\gA}=0$, and $f_b{}^{ba}=0$, but these objects are not covariant under generalized dualities, and a (possibly non-constant) $\vartheta_A$ can arise after a generalized duality. 

As an example, take the algebra whose only only non-vanishing structure constants are $f_{12}{}^2=1$ and $f_{13}{}^3=1$, and with $\vartheta_A=0$. In $D\geq 6$, by performing a generalized duality which exchanges $T_a$ with $T^a$, the algebra is mapped to one with $f_2{}^{12}=1$ and $f_3{}^{13}=1$. In this case a non-vanishing $\vartheta^a$ arises due to the presence of $f_b{}^{ba}$, and the dual background becomes the configuration of a different gauged supergravity. In a conservative approach, when $\vartheta_A$ shows up under a duality transformation, we prohibit this duality. This approach was taken in non-abelian T-duality for isometry algebras with traceful structure constants. Another approach is to choose the function $\sigma$ in \eqref{eq:dilaton-ansatz-D6} such that $\vartheta_A$ does not appear after the duality. However, to eliminate the dual components $\vartheta^a$, we need to allow the function $\sigma$ to depend on the dual coordinates $x^I$, and the dilaton $\phi$ gets the dual-coordinate dependence. Then the resulting configuration does not uplift to the standard ten-dimensional supergravity. In $D=4$ or $D=5$, if we apply the same duality map ($T_a\leftrightarrow T^a$), the resulting algebra goes beyond geometric algebras. This means that, under this duality, the generalized frames get the dependence on the dual coordinates (without breaking the section condition), and if we stick to the standard section, this duality should be prohibited. In the context of Poisson-Lie T-duality, the issue of the upliftability in the presence of $f_b{}^{ba}$ has been discussed in \cite{Demulder:2018lmj,Sakatani:2019jgu}, and the appearance of $\vartheta^a$ or the need to change the section can be naturally understood in the framework of generalized supergravity \cite{Wulff:2016tju,Sakatani:2016fvh}.

\section{Examples}\label{sec:examples}
The space of admissible embedding tensors is very large and exploring it systematically is an extremely hard problem inherited by the geometric gaugings introduced in the last section. Most challenging is to solve the Leibniz identity. By brute force, we identified two, more or less random, solutions which are geometric in $d=6$ with $\xi_A\neq 0$. The first one permits to construct two backgrounds which are related by a generalized T-duality, while the second one is based on a more complicated gauging and emphasizes that our approach is applicable to any gauging that can be brought into the form \eqref{eqn:GeomAlgX}.
In both the examples the duality group is $\SL(2)\times \mathrm{O}(6,6+n)$, with $n\geq 2$. The fundamental representation of the Leibniz algebra, then, is $24+2n$ dimensional, while the heterotic/type I gauge group $\cG$ can be chosen to be $\mathrm{U}(1)^n$. The number of vector multiplets of these theories, therefore, is $\mathfrak{n}=6+n$. 

\subsection{A generalized duality}
Let us start choosing the gaugings $f_{ab}{}^c$, $f_a{}^{bc}$, $h_{abc}$, $Z_a$, $f_a$, $f_a{}^b{}_{\cA}$, $f_a{}^{\cA\cB}$, and $h_{ab}{}^\cA$ as
\begin{align}	\label{eq:Ex1}
	\begin{alignedat}{6}
		f_{12}{}^3 &= 1\,,&\qquad
		f_{45}{}^6 &= 1\,,&\qquad
		f_1{}^{23} &= 1\,,&\qquad
		h_{123} &= 1 \,,&\qquad
		h_{456} &= -\tfrac{1}{2}\,,&\qquad
		Z_1 &= 1\,,
		\\
		f_1 &= 2\,,&\qquad
		f_1{}^6{}_{\bm{2}} &= -1\,,&\qquad
		f_1{}^{\bm{1}\bm{2}} &= 1\,,&\qquad
		h_{16}{}^{\bm{1}} &= 1\,,&\qquad
		h_{16}{}^{\bm{2}} &= \tfrac{1}{2}\,,&\qquad
		h_{45}{}^{\bm{1}} &= -1\,,
	\end{alignedat}
\end{align}
such that the Leibniz identity holds \eqref{eqn:LeibnizId}. Here the trombone gauging $\vartheta_{\pm A}$ vanishes and $\xi_{\pm A}$ has the only non-zero component $\xi_{+1}=1$\,. We also have $F_{-ABC}=0$, and $F_{+ABC}$ can be expressed by using the structure constants given in the first line of \eqref{eq:Ex1}.
Here the Lie algebra of the gauge group $G$ can be identified as a subalgebra of the Leibniz algebra, as explained in~\ref{sec:constgenframes}. 
The procedure to do it is outlined for instance in~\cite{Hassler:2022egz}, and relies on the idea of bringing the Leibniz algebra in a canonical representation (called the ``Leibniz representation''), that can be seen as the intertwining of the adjoint representation of the Lie algebra of $G$ with a second representation of the same algebra. 
In our case, then, the Leibniz representation is $24+2n$ dimensional, while the adjoint representation of Lie($G$) is $13+n$ dimensional and can be explicitly computed taking the generators $T_{\grave{c}}$ of the Leibniz algebra for which $X_{(\Ah \Bh)}{}^{\grave{c}}=0$. 
In particular, if we focus on the case $n=2$, Lie($G$) is a direct sum of a $14$-dimensional solvable Lie algebra and a $1$-dimensional abelian algebra. 
The former is generated by
\begin{align}\label{eq:G-generators}
\begin{alignedat}{5}
 \mathfrak{t}_a &= T_{a} &\qquad 
 (a=&1,\dots,6)\,,&\qquad 
 \mathfrak{t}_7 &= T_{-1}\,,&\qquad 
 \mathfrak{t}_8 &= T_{+\bm{1}}\,,&\qquad 
 \mathfrak{t}_9 &= T_{+\bm{2}}\,,
\\
 \mathfrak{t}_{10} &= T_+{}^2\,,&\qquad 
 \mathfrak{t}_{11} &= T_+{}^3\,,&\qquad 
 \mathfrak{t}_{12} &= T_+{}^4\,,&\qquad 
 \mathfrak{t}_{13} &= T_+{}^5\,,&\qquad 
 \mathfrak{t}_{14} &= T_+{}^6\,, 
\end{alignedat}
\end{align}
and in this basis the structure constants are
\begin{align}
\begin{alignedat}{6}
\mathfrak{f}_{1, 2}{}^3 &= 1\,,&\quad
\mathfrak{f}_{1, 2}{}^{11} &= 1\,,&\quad 
\mathfrak{f}_{1, 3}{}^{10} &= -1\,,&\quad 
\mathfrak{f}_{1, 6}{}^{8} &= 1\,,&\quad 
\mathfrak{f}_{1, 6}{}^{9} &= \tfrac{1}{2}\,, 
\\
\mathfrak{f}_{1, 7}{}^{7} &= -2\,,&\quad 
\mathfrak{f}_{1, 8}{}^{8} &= 1\,,&\quad
\mathfrak{f}_{1, 8}{}^{9} &= 1\,,&\quad 
\mathfrak{f}_{1, 8}{}^{14} &= -1\,,&\quad 
\mathfrak{f}_{1, 9}{}^{6} &= 1\,, 
\\
\mathfrak{f}_{1, 9}{}^{8} &= -1\,,&\quad 
\mathfrak{f}_{1, 9}{}^{9} &= 1\,,&\quad 
\mathfrak{f}_{1, 9}{}^{14} &= -\tfrac{1}{2}\,,&\quad 
\mathfrak{f}_{1, 10}{}^{3} &= 1\,,&\quad
\mathfrak{f}_{1, 10}{}^{10} &= 2\,,
\\
\mathfrak{f}_{1, 11}{}^{2} &= -1\,,&\quad 
\mathfrak{f}_{1, 11}{}^{10} &= -1\,,&\quad 
\mathfrak{f}_{1, 11}{}^{11} &= 2\,,&\quad 
\mathfrak{f}_{1, 12}{}^{12} &= 2\,,&\quad 
\mathfrak{f}_{1, 13}{}^{13} &= 2\,,
\\
\mathfrak{f}_{1, 14}{}^{9} &= -1\,,&\quad 
\mathfrak{f}_{1, 14}{}^{14} &= 2\,,&\quad 
\mathfrak{f}_{4, 5}{}^{6} &= 1\,,&\quad 
\mathfrak{f}_{4, 5}{}^{8} &= -1\,,&\quad
\mathfrak{f}_{4, 5}{}^{14} &= -\tfrac{1}{2}\,,
\\
\mathfrak{f}_{4, 6}{}^{13} &= \tfrac{1}{2}\,,&\quad 
\mathfrak{f}_{4, 8}{}^{13} &= 1\,,&\quad
\mathfrak{f}_{4, 14}{}^{13} &= -1\,,&\quad 
\mathfrak{f}_{5, 6}{}^{12} &= -\tfrac{1}{2}\,,&\quad 
\mathfrak{f}_{5, 8}{}^{12} &= -1\,,
\\
\mathfrak{f}_{5, 14}{}^{12} &= 1 \,.
\end{alignedat}
\end{align}

In order to construct the frames,  following the general procedure outlined in Subsection~\ref{sec:constgenframes}, we first have to fix the coset representative $M$. 
Here the subgroup $H$ generated by $7+n$ generators $T_{\grave{\alpha}}$ is abelian, and the coset $G/H$ can be parameterized by six generators $T_a$. For the representative, a parametrization different from \eqref{eqn:paramM}, namely
\begin{align}
	M = \Exp{x\,T_1}\Exp{y\,T_2}\Exp{z\,T_3}\Exp{u\,T_4}\Exp{v\,T_5}\Exp{w\,T_6}\,,
\end{align}
turns out the be convenient. Of course, \eqref{eqn:paramM} would also work, but it would result in more complicated expressions related by a diffeomorphism to the ones given here. Note that $M_{\Ah}{}^{\Bh}$ is regular everywhere and has unit determinant.
Next, we compute the one-form fields $v^{\Ah}$ as
\begin{align}
	v_i{}^{+A} = \bigl(v_i{}^{a},\,v_i{}^{\cA},\,v_{ia}\bigr)\,,\qquad
	v_i{}^{-A} = 0\,,
\end{align}
where
    \begin{align}
    	v_i{}^{a} & = \begin{pmatrix}
    		              1 & 0 & y & 0 & 0 & 0 \\
    		              0 & 1 & 0 & 0 & 0 & 0 \\
    		              0 & 0 & 1 & 0 & 0 & 0 \\
    		              0 & 0 & 0 & 1 & 0 & v \\
    		              0 & 0 & 0 & 0 & 1 & 0 \\
    		              0 & 0 & 0 & 0 & 0 & 1
    	              \end{pmatrix},\qquad &
    	v_i{}^{\cA} & =\begin{pmatrix}
    		           w  & \frac{w}{2} & 0 & \cdots \\
    		           0  & 0           & 0 & \cdots \\
    		           0  & 0           & 0 & \cdots \\
    		           -v & 0           & 0 & \cdots \\
    		           0  & 0           & 0 & \cdots \\
    		           0  & 0           & 0 & \cdots
    	           \end{pmatrix},\nonumber
    	\\
    	v_i{}^{a} & = \begin{pmatrix}
    		              -\frac{5 w^2}{8}-y^2 & -z & y & 0            & 0           & 0            \\
    		              z                    & 0  & 0 & 0            & 0           & 0            \\
    		              0                    & 0  & 0 & 0            & 0           & 0            \\
    		              v w                  & 0  & 0 & 0            & \frac{w}{2} & -\frac{v}{2} \\
    		              0                    & 0  & 0 & -\frac{w}{2} & 0           & 0            \\
    		              0                    & 0  & 0 & 0            & 0           & 0
    	              \end{pmatrix}.&
    \end{align}            
From the matrix $v_i{}^{a}$, we obtain the matrix $V_{\Ah}{}^{\Ih}$ of \eqref{eqn:vAIframe}, which has unit determinant too.
Moreover, combining the one-form fields $v^{\Ah}$ and the structure constants $X_{\Ah\Bh}{}^{\Ch}$ gives rise to the field strengths
\begin{equation}
    \begin{aligned}
    	\mathfrak{f}_2^{\bm{1}} & = -\rmd x\wedge\rmd w \,,\\
    	\mathfrak{f}_2^{\bm{2}} & = -\tfrac{1}{2}\,\rmd u\wedge\rmd v\,,\\
    	\mathfrak{f}_2^{\cA} & = 0,\qquad \cA\geq \bm{3}\,,                                                                                             \\
    	\mathfrak{h}_3          & = -2 \rmd x\wedge \rmd y\wedge \rmd z - w\,\rmd x\wedge \rmd y\wedge \rmd v + \rmd u\wedge\rmd v\wedge\rmd w\,,
    \end{aligned}
\end{equation}
required to fix the matrix $N_{\Ih}{}^{\Jh}$ in \eqref{eqn:leftinvMC} up to gauge transformations. Since the Bianchi identities are satisfied, it is possible to integrate them, resulting in the associated potentials
\begin{equation}
\begin{aligned}
	\mathfrak{a}_1^{\bm{1}} & = u\,\rmd v - x\,\rmd w \,,\\
	\mathfrak{a}_1^{\bm{2}} & = - \tfrac{x}{2}\,\rmd w \,,
	\\
	\mathfrak{b}_2          & = u\,w\,\rmd x\wedge \rmd v -2\,x\,\rmd y\wedge \rmd z + u\,\bigl(1-\tfrac{x}{2}\bigr)\,\rmd v\wedge\rmd w\,.
\end{aligned}
\end{equation}
Combining these results, we obtain the matrix $E_{\Ah}{}^{\Ih}$ which is regular and has unit determinant. In this way, one systematically constructs the generalized frame fields for any geometric algebra.

A generalized T-duality arises in this example by an $\OO(6,6+n)$ transformation
\begin{equation}
\begin{alignedat}{2}
	 T'_{\pm a} & = T_{\pm a}\,, \qquad a=1,3,4\,, \hspace{4cm} & T'_{\pm a} & = T_{\pm}{}^a\,, \qquad a=2,5\,,\\
	 T'_{\pm 6} & = T_{\pm 6}- \tfrac{1}{2}\,T_{\pm}{}^6 - T_{\pm}{}^{\bm{1}}\,, \quad & T'_{\pm}{}^{\bm{1}} & = T_{\pm}{}^{\bm{1}} + T_{\pm}{}^{6} \,,\\
	 T'_{\pm}{}^{\bm{2}} & = T_{\pm}{}^{\bm{2}} - T_{\pm}{}^{\bm{1}} + T_{\pm 6} - \tfrac{1}{2}\,T_{\pm}{}^6\,,\quad &
	T'_{\pm}{}^{\cA} & = T_{\pm}{}^{\cA}\,, \qquad \cA\geq \bm{3}\,,
	\\
	 T'_{\pm}{}^{a} & = T_{\pm}{}^{a}\,, \qquad a=1,3,4\,, \quad & T'_{\pm}{}^a & = T_{\pm a}\,, \qquad a=2,5\,,\\
	 T'_{\pm}{}^6 & = \tfrac{5}{4}\,T_{\pm}{}^6 - \tfrac{1}{2}\,T_{\pm 6} + \tfrac{1}{2}\,T_{\pm}{}^{\bm{1}} - T_{\pm}{}^{\bm{2}}\,.
\end{alignedat}
\end{equation}
As we demanded in Subsection~\ref{sec:gendualities}, it results in a new geometric algebra with the structure constants\footnote{Note that, consistently with our claims made in Section~\ref{sec:gendualities}, a trombone gauging does not appear after the duality transformation since we are working in $d=6$.}
\begin{align}
	\begin{alignedat}{4}
		f'_{12}{}^2 &= 2\,, & \qquad
		f'_{12}{}^3 & = 1\,, & \qquad
		f'_{13}{}^2 & = -1\,, & \qquad
		f'_{15}{}^5 & = 2\,,
		\\
		f'_1{}^{23} &= 1\,, & \qquad
		f'_4{}^{56} & = 1\,, & \qquad
		Z'_1 & = 1\,, & \qquad
		f'_1 & = 2\,.
	\end{alignedat}
\end{align}
For consistency, finding again the subalgebra Lie($G$), one checks that it is still $13+n$-dimensional.

Now the generators $T'_a$ form a Lie subalgebra, and the computation becomes easier. Again, we have to first parametrize the coset representative
\begin{align}
	M = \Exp{x\,T'_1}\Exp{y\,T'_2}\Exp{z\,T'_3}\Exp{u\,T'_4}\Exp{v\,T'_5}\Exp{w\,T'_6}\,,
\end{align}
and then obtain
\begin{align}
	v_i{}^{+A} = \bigl(v_i{}^{a},\,v_i{}^{\cA},\,v_{ia}\bigr)= \bigl(v_i{}^{a},\,0,\,0\bigr)\,,\qquad
	v_i{}^{-A} = 0\,,
\end{align}
where
\begin{align}
	v_i{}^{a} & = \begin{pmatrix}
		              1 & 2 y-z & y & 0 & 2 v & 0 \\
		              0 & 1     & 0 & 0 & 0   & 0 \\
		              0 & 0     & 1 & 0 & 0   & 0 \\
		              0 & 0     & 0 & 1 & 0   & 0 \\
		              0 & 0     & 0 & 0 & 1   & 0 \\
		              0 & 0     & 0 & 0 & 0   & 1
	              \end{pmatrix}.
\end{align}
Since here $\mathfrak{f}_2^{\cA} = 0$ and $\mathfrak{h}_3=0$\,, we choose $\mathfrak{a}_1^{\cA}=0$ and $\mathfrak{b}_2=0$\,, resulting in the simple generalized frame $E_{\Ah}{}^{\Ih}=M_{\Ah}{}^{\Bh}\,\hat{V}_{\Bh}{}^{\Ih}$\,, which still has unit determinant. 
Note, also, that in this case, being $T'_a$ a subalgebra of Lie($G$), the manifold obtained doing the coset $G/H$ is a group manifold.

\subsection{Proof of concept}
To show that our approach works in full generality, we switch now to a more difficult gauging. Specifically, we take $f_{ab}{}^c$, $f_a{}^{bc}$, $h_{abc}$, $Z_a$, $f_a$, $f_a{}^b{}_{\cA}$, $f_a{}^{\cA\cB}$, and $h_{ab}{}^\cA$ as
\begin{align}
	\begin{alignedat}{6}
		f_{23}{}^{3} &= 1\,, & \quad
		f_{24}{}^{4}  & = 1\,, & \quad
		f_{25}{}^{5}  & = -1\,, & \quad
		f_1{}^{34}  & = -1\,, & \quad
		f_1{}^{56}  & = 2\,, & \quad
		f_2{}^{34}  & = 1\,,\\
		f_2{}^{56}  & = 1\,, & \quad
		h_{123} &= 1\,, & \quad
		h_{134}  & = -1\,, & \quad
		h_{156}  & = \tfrac{1}{2}\,, & \quad
		h_{234}  & = 1\,, & \quad
		h_{256}  & = \tfrac{1}{4}\,,\\
		Z_1  & = 1\,, & \quad
		f_1  & = 2 \,,& \quad
		f_1{}^{3}{}_{\bm{1}} &= 1\,, & \quad
		f_1{}^{4}{}_{\bm{2}}  & = 1\,, & \quad
		f_2{}^{3}{}_{\bm{1}}  & = -1\,, & \quad
		f_2{}^{4}{}_{\bm{2}}  & = -1\,,\\
		f_{1}{}^{\bm{1}\bm{2}} &  = 1\,, &\quad
		f_{2}{}^{\bm{1}\bm{2}}  & = -1\,,& \quad
		h_{13}{}^{\bm{1}} &= 1\,, & \quad
		h_{14}{}^{\bm{2}}  & = 1\,, & \quad
		h_{23}{}^{\bm{1}}  & = -1\,, & \quad
		h_{24}{}^{\bm{2}}  & = -1\,.
	\end{alignedat}
	\label{eq:Ex2}
\end{align}
While the Leibniz representation is still $24+2n$-dimensional, this time we find a $12+n$-dimensional solvable Lie algebra of $G$.
In particular for $n=2$, $G$ is generated by the generators given in~\eqref{eq:G-generators} and the structure constants are
\begin{align}
\begin{alignedat}{5}
\mathfrak{f}_{1, 2}{}^{11} &= 1\,,&\quad 
\mathfrak{f}_{1, 3}{}^{8} &= 1\,,&\quad 
\mathfrak{f}_{1, 3}{}^{10} &= -1\,,&\quad 
\mathfrak{f}_{1, 3}{}^{12} &= -1\,,&\quad 
\mathfrak{f}_{1, 4}{}^{9} &= 1\,,\ 
\\
\mathfrak{f}_{1, 4}{}^{11} &= 1\,,&\quad 
\mathfrak{f}_{1, 5}{}^{14} &= \tfrac{1}{2}\,,&\quad
\mathfrak{f}_{1, 6}{}^{13} &= -\tfrac{1}{2}\,,&\quad
\mathfrak{f}_{1, 7}{}^{7} &= -2\,,&\quad 
\mathfrak{f}_{1, 8}{}^{3} &= -1\,,\quad 
\\
\mathfrak{f}_{1, 8}{}^{8} &= 1\,,&\quad 
\mathfrak{f}_{1, 8}{}^{9} &= 1\,,&\quad
\mathfrak{f}_{1, 8}{}^{11} &= -1\,,&\quad 
\mathfrak{f}_{1, 9}{}^{4} &= -1\,,&\quad
\mathfrak{f}_{1, 9}{}^{8} &= -1\,,\ 
\\
\mathfrak{f}_{1, 9}{}^{9} &= 1\,,&\quad 
\mathfrak{f}_{1, 9}{}^{12} &= -1\,,&\quad 
\mathfrak{f}_{1, 10}{}^{10} &= 2\,,&\quad 
\mathfrak{f}_{1, 11}{}^{4} &= -1\,,&\quad 
\mathfrak{f}_{1, 11}{}^{8} &= 1\,,\ 
\\
\mathfrak{f}_{1, 11}{}^{11} &= 2\,,&\quad
\mathfrak{f}_{1, 12}{}^{3} &= 1\,,&\quad 
\mathfrak{f}_{1, 12}{}^{9} &= 1\,,&\quad 
\mathfrak{f}_{1, 12}{}^{12} &= 2\,,&\quad 
\mathfrak{f}_{1, 13}{}^{6} &= 2\,,\ 
\\
\mathfrak{f}_{1, 13}{}^{13} &= 2\,,&\quad 
\mathfrak{f}_{1, 14}{}^{5} &= -2\,,&\quad 
\mathfrak{f}_{1, 14}{}^{14} &= 2\,,&\quad
\mathfrak{f}_{2, 3}{}^{3} &= 1\,,&\quad 
\mathfrak{f}_{2, 3}{}^{8} &= -1\,,\ 
\\
\mathfrak{f}_{2, 3}{}^{12} &= 1\,,&\quad 
\mathfrak{f}_{2, 4}{}^{4} &= 1\,,&\quad 
\mathfrak{f}_{2, 4}{}^{9} &= -1\,,&\quad 
\mathfrak{f}_{2, 4}{}^{11} &= -1\,,&\quad 
\mathfrak{f}_{2, 5}{}^{5} &= -1\,,
\\
\mathfrak{f}_{2, 5}{}^{14} &= \tfrac{1}{4}\,,&\quad 
\mathfrak{f}_{2, 6}{}^{13} &= -\tfrac{1}{4}\,,&\quad 
\mathfrak{f}_{2, 8}{}^{3} &= 1\,,&\quad 
\mathfrak{f}_{2, 8}{}^{9} &= -1\,,&\quad 
\mathfrak{f}_{2, 8}{}^{11} &= 1\,,\ 
\\
\mathfrak{f}_{2, 9}{}^{4} &= 1\,,&\quad 
\mathfrak{f}_{2, 9}{}^{8} &= 1\,,&\quad
\mathfrak{f}_{2, 9}{}^{12} &= 1\,,&\quad 
\mathfrak{f}_{2, 11}{}^{4} &= 1\,,&\quad 
\mathfrak{f}_{2, 11}{}^{8} &= -1\,,\ 
\\
\mathfrak{f}_{2, 11}{}^{11} &= -1\,,&\quad 
\mathfrak{f}_{2, 12}{}^{3} &= -1\,,&\quad 
\mathfrak{f}_{2, 12}{}^{9} &= -1\,,&\quad
\mathfrak{f}_{2, 12}{}^{12} &= -1\,,&\quad 
\mathfrak{f}_{2, 13}{}^{6} &= 1\,,\ 
\\
\mathfrak{f}_{2, 13}{}^{13} &= 1\,,&\quad 
\mathfrak{f}_{2, 14}{}^{5} &= -1\,,&\quad 
\mathfrak{f}_{3, 4}{}^{10} &= 1\,,&\quad 
\mathfrak{f}_{3, 8}{}^{10} &= -1\,,&\quad 
\mathfrak{f}_{3, 11}{}^{10} &= 1\,,\ 
\\
\mathfrak{f}_{4, 9}{}^{10} &= -1\,,&\quad 
\mathfrak{f}_{4, 12}{}^{10} &= 1\,,&\quad 
\mathfrak{f}_{5, 6}{}^{10} &= \tfrac{1}{4}\,,&\quad 
\mathfrak{f}_{5, 13}{}^{10} &= -1\,,&\quad 
\mathfrak{f}_{8, 9}{}^{10} &= -1\,,\ 
\\
\mathfrak{f}_{8, 11}{}^{10} &= 1\,,&\quad 
\mathfrak{f}_{9, 12}{}^{10} &= 1\,,&\quad 
\mathfrak{f}_{11, 12}{}^{10} &= 1\,,&\quad 
\mathfrak{f}_{13, 14}{}^{10} &= 1\,.
\end{alignedat}
\end{align}
In this case, the subgroup $H$ is an $8$-dimensional (non-abelian) solvable Lie algebra. 

Parametrizing the coset representative $M$ by
\begin{align}
	M = \Exp{x\,T_1}\Exp{y\,T_2}\Exp{z\,T_3}\Exp{u\,T_4}\Exp{v\,T_5}\Exp{w\,T_6}\,,
\end{align}
we find that $M_{\Ah}{}^{\Bh}$ is regular everywhere and has unit determinant.
Next, we compute the one-form fields $v^{\Ah}$ as
\begin{align}
	v_i{}^{+A} = \bigl(v_i{}^{a},\,v_i{}^{\cA},\,v_{ia}\bigr)\,,\qquad
	v_i{}^{-A} = 0\,,
\end{align}
where
\begin{align}
	v_i{}^{a} & = \begin{pmatrix}
		              1 & 0 & \frac{\sin (\sqrt{2} y)}{\sqrt{2}}-\sin y & \cos y -1 & 0  & 0 \\
		              0 & 1 & z                                         & u         & -v & 0 \\
		              0 & 0 & 1                                         & 0         & 0  & 0 \\
		              0 & 0 & 0                                         & 1         & 0  & 0 \\
		              0 & 0 & 0                                         & 0         & 1  & 0 \\
		              0 & 0 & 0                                         & 0         & 0  & 1
	              \end{pmatrix},\nonumber
	\\
	v_i{}^{\cA} & =\begin{pmatrix}
		             \sin y -\frac{\sin(\sqrt{2} y)}{\sqrt{2}}+\frac{1-\cos (\sqrt{2} y)}{2} +z & u+\frac{1-2 \cos y +\cos (\sqrt{2} y)}{2} & 0 & \cdots \\
		             -z                                                                         & -u                                        & 0 & \cdots \\
		             0                                                                          & 0                                         & 0 & \cdots \\
		             0                                                                          & 0                                         & 0 & \cdots \\
		             0                                                                          & 0                                         & 0 & \cdots \\
		             0                                                                          & 0                                         & 0 & \cdots
	             \end{pmatrix},
	\\
	v_i{}^{a} & = \begin{pmatrix}
		              c_1                       & c_2                        & c_3 & c_4 & -\frac{w}{2} & \frac{v}{2} \\
		              \frac{(u-z)^2-v w+2 z}{2} & -\frac{2 u^2+v w+2 z^2}{4} & -u  & z   & -\frac{w}{4} & \frac{v}{4} \\
		              -u                        & u                          & 0   & 0   & 0            & 0           \\
		              0                         & 0                          & 0   & 0   & 0            & 0           \\
		              \frac{w}{2}               & \frac{w}{4}                & 0   & 0   & 0            & 0           \\
		              0                         & 0                          & 0   & 0   & 0            & 0
	              \end{pmatrix},\nonumber
\end{align}
and
\begin{equation}
    \begin{aligned}
	c_1       & \equiv {\textstyle\frac{(u-z) [2 \sin y -\sqrt{2} \sin (\sqrt{2} y)] +2 (u+z-1) \cos y +(1-u+z) \cos (\sqrt{2} y)-u (u+1)-z (z+3)+1}{2}}\,,
	\\
	c_2       & \equiv {\textstyle\frac{\sqrt{2} (u-z) \sin(\sqrt{2} y)-2 u \sin y +(u+z)^2-2 z \cos y}{2}}\,,
	\\
	c_3       & \equiv {\textstyle \frac{1-\cos (\sqrt{2} y)}{2}} + u + \sin  y\,,\\
	c_4       & \equiv {\textstyle \frac{1 +\cos (\sqrt{2} y)}{2}} - z - \cos y \,.
    \end{aligned}
\end{equation}
Like in the last subsection, we also need the field strengths
\begin{equation}
    \begin{aligned}
    	\mathfrak{f}_2^{\bm{1}} & = \bigl[\cos (\sqrt{2} y)-\cos y-\tfrac{\sin(\sqrt{2} y)}{\sqrt{2}}\bigr]\,\rmd x\wedge \rmd y + (\rmd y-\rmd x)\wedge \rmd z \,,
    	\\
    	\mathfrak{f}_2^{\bm{2}} & = \bigl[\tfrac{\sin (\sqrt{2} y)}{\sqrt{2}}-\sin y \bigr] \,\rmd x\wedge \rmd y + (\rmd y-\rmd x)\wedge \rmd u\,, \\
    	\mathfrak{f}_2^{\cA} & = 0\,,\qquad \cA\geq \bm{3}\,,                                                                                                                                         \\
    	\mathfrak{h}_3          & = \bigl[u-\tfrac{\sin \left(\sqrt{2} y\right)}{\sqrt{2}}-2 \cos y\bigr]\,\rmd x\wedge \rmd y\wedge \rmd z
    	                                                                                                                                                                                   \\
    	                        & \quad + \rmd x\wedge \rmd y\wedge \bigl\{ \bigl[z - 2 \sin y +\sqrt{2} \sin (\sqrt{2} y)\bigr]\,\rmd u + \tfrac{w}{2}\,\rmd v + \tfrac{v}{2}\,\rmd w\bigr\}
    	                                                                                                                                                                                   \\
    	                        & \quad + 2\,(\rmd x-\rmd y)\wedge \rmd z\wedge \rmd u
    	- \bigl(\rmd x+\tfrac{1}{2}\,\rmd y\bigr)\wedge \rmd v\wedge \rmd w \,,
    \end{aligned}
\end{equation}
and the associated potentials
\begin{equation}
    \begin{aligned}
    	\mathfrak{a}_1^{\bm{1}}  &= \bigl[\sin y -\tfrac{\sin (\sqrt{2} y)}{\sqrt{2}}-\tfrac{\cos (\sqrt{2} y)}{2} \bigr]\,\rmd x +(y-x)\,\rmd z \,,
    	\\
    	\mathfrak{a}_1^{\bm{2}} &= \tfrac{1}{2} \bigl[\cos (\sqrt{2} y)-2 \cos y \bigr]\,\rmd x +(y-x)\,\rmd u \,,
    	\\
    	\mathfrak{b}_2          &=b_{12} \,\rmd x\wedge \rmd y
    	+ 2(x-y)\,\rmd z\wedge \rmd u -\bigl(x+\tfrac{y}{2}\bigr)\,\rmd v\wedge \rmd w\,,
    \end{aligned}
\end{equation}
where we have defined
\begin{equation}
    \begin{aligned}
    	b_{12}                  &\equiv \tfrac{1}{4}\,\bigl\{
    	u (x-y-4) [2\sin y -\sqrt{2} \sin (\sqrt{2} y)]-2 u \cos y +u \cos (\sqrt{2} y)+4 u z+2 v w
    	                                                                                                                                          \\
    	                        & \quad  + \bigl[\sqrt{2} x \sin (\sqrt{2} y)-(8-2 x+2 y) \cos y - (1+2 x-2 y) \cos (\sqrt{2} y)
    	                                                                                                                                          \\
    	                        & \quad +2 \sin y -\sqrt{2} y \sin (\sqrt{2} y)-3 \sqrt{2} \sin (\sqrt{2} y)\bigr]\,z\bigr\} \,,
    \end{aligned}
\end{equation}
allowing us to obtain the matrix $N_{\Ih}{}^{\Jh}$\,.
Combining these results, we obtain again the generalized frame $E_{\Ah}{}^{\Ih}$, which is regular and has unit determinant.

\section{Conclusions}\label{sec:conclusion}
The main goal of this paper was to fill a gap in the literature and to define generalised T-dualities for heterotic/type I string theories. In order to do this, we leveraged half-maximal gSUGRAs and their possible uplifts to ten dimensions. The starting point for our discussion is heterotic DFT with the duality group $\OO(10,10+n)$. After a generalized Scherk-Schwarz reduction, this gives rise to half-maximal gSUGRAs in various dimensions; in our discussion we considered the case $D\geq 4$. Although conceptually not very different from other reductions of this type, we are not aware of previous results of this kind starting from heterotic DFT. Therefore, we developed the required ans\"atze in three steps:
\begin{itemize}
    \item Subsection~\ref{sec:hetDFT} introduced heterotic DFT in the frame formalism and related its action to the one of ten-dimensional half-maximal SUGRA;
    \item Next, Subsection~\ref{sec:splitHS} set the ground for dimensional reductions by introducing a splitting of the ten-dimensional target space into a $d$-dimensional internal part and a $D$-dimensional external one in the spirit of the Hohm-Samtleben prescription;
    \item Based on this decomposition, Subsection~\ref{sec:gSSansatz} introduced the appropriate Scherk-Schwarz ansatz and discussed the resulting gSUGRAs.
\end{itemize}

Note that our discussion held for arbitrary non-abelian gauge groups $\cG$ in ten dimensions, characterized by structure constants $\Sigma_{\cI\cJ\cK}$. To identify upliftable gSUGRAs, we analyzed their embedding tensors and expressed them in terms of a generalized Lie derivative for the respective duality groups $\Gduality$ in Subsection~\ref{sec:genLieder}. The details for this depend on the dimension $d$ of the internal space. For example, $d=6$ is more cumbersome to work out as the factor $\Gdualityt$ picked up an additional $\SL(2)$ instead of the usual $\mathbb{R}^+$. Moreover, $d=5$ is more complicated than $d\leq 4$, due to the presence of an additional coordinate on the extended space originating from NS5-branes. Subsequently, we turned to the relation between the generalized fluxes, that arose in the generalized Scherk-Schwarz reduction ansatz of Section~\ref{sec:genSSred}, and the embedding tensor. A frame in $\Gduality$ can always be used as an alternative way to encode the latter and we showed how this works in Subsection~\ref{sec:embeddingtensor}, comparing our results with the known results for the embedding tensor in the literature. Finally Subsection~\ref{sec:reltogSS} presented upliftability conditions which restrict certain components of the embedding tensor and that are required to make contact with the previously obtained Scherk-Schwarz ansatz.

Section~\ref{sec:frames} dealt with the explicit construction of generalized frames that satisfy those conditions. First, we found that the section condition, which is required for the closure of the generalized Lie derivative, imposes additional constraints. They give rise to what we call heterotic/type I geometric algebras in Subsection~\ref{sec:geomalgebras}. Imposing a well-established ansatz, generalized frames were constructed explicitly for all geometric gauge algebras in Subsection~\ref{sec:constgenframes}. In particular, we found \eqref{eqn:fieldstrength}, relating the field strength of the potentials in the ansatz with the gaugings. In Subsection~\ref{sec:twist}, we employed a further right-twisting of the generalized frames to make the non-abelian structure of the field strengths manifest. Central consistency checks in our construction were the Bianchi identities for the field strengths given in \eqref{eqn:fieldstrength} and \eqref{eqn:nonAbelianStrengths}. They were verified at the end of Subsections~\ref{sec:constgenframes} and \ref{sec:twist}, for the abelian and non-abelian cases respectively. With all these tools in place, Subsection~\ref{sec:gendualities} showed that generalized dualities follow exactly the same rules as for the bosonic string or in M-theory. Finally, several examples demonstrated the utility and generality of the presented results in Section~\ref{sec:examples}.

\subsection*{Acknowledgements}
We would like to thank David Osten for helpful discussions and comments on the draft. The work by YS is supported by JSPS KAKENHI Grant Number JP23K03391. FH and LS are supported by the SONATA BIS grant 2021/42/E/ST2/00304 from the National Science Centre (NCN), Poland. LS acknowledges financial support from the doctoral school of the University of Wrocław.

\appendix
\section{Indexology}\label{app:index}
In the main text, we use many different indices. We did our best to keep things as simple as possible, but at some point alphabets run out and one has to resort to further decorations which in turn get more and more exotic. To assist the reader, we provide here a summary of the most important indices we are using and the groups they describe. When two indices appears in the second column they refers, respectively, to curved and flat ones; also, to make the table lighter, when we did not use explicitly, in the main text, the splitting of curved or flat indices, we introduced a -- in their place.
\begin{center}
\begin{tabular}{|c|c|c|} 
\hline
Group & Indices &  Splitting \\ 
\hline\hline
$\OO(10,10+n)$ & $\Iht\,, \Aht$ & $\begin{matrix}
\begin{pmatrix} {}_m & {}_{\cI} & {}^m \end{pmatrix}\,, \begin{pmatrix} {}_{\hat{a}} & {}_{\cA} & {}^{\hat{a}} \end{pmatrix}\\
\begin{pmatrix} {}_\mu & {}_{I} & {}^\mu \end{pmatrix}\,, \begin{pmatrix} {}_{\bat} & {}_{A} & {}^{\bat} \end{pmatrix}
\end{matrix}$\\ 
\hline
$\OO(d,d+n)$ & $I\,, A$ & $\begin{pmatrix} {}_i & {}_{\cI} & {}^i\end{pmatrix}\,, \begin{pmatrix} {}_a & {}_{\cA} & {}^a \end{pmatrix}$  \\
\hline
$\cG$ & $\cI\,, \cA$ &  \\
\hline
$\GL(10)$ & $\begin{pmatrix} {}_m & {}^m \end{pmatrix}$, $\begin{pmatrix} {}_{\hat{a}} & {}^{\hat{a}} \end{pmatrix}$ & \\
\hline
$\GL(d)$ & $\begin{pmatrix} {}_i & {}^i \end{pmatrix}\,, \begin{pmatrix} {}_a & {}^a \end{pmatrix}$ & \\
\hline
$\GL(D)$ & $\begin{pmatrix} {}_\mu & {}^\mu \end{pmatrix}\,,\begin{pmatrix} {}_{\bat} & {}^{\bat} \end{pmatrix} $ & \\
\hline
$\Gduality$ & $\Ih\,,\Ah$ & --, $\begin{cases}
		          {}_{A} = \begin{pmatrix} {}_a & {}_{\gA} & {}^a\end{pmatrix}                                        & d\leq 4 \\
		          \begin{pmatrix} {}_A & {}_*\end{pmatrix} = \begin{pmatrix} {}_a & {}_{\gA} & {}^a & {}_*\end{pmatrix} & d=5     \\
		          {}_{\bba A} = \begin{pmatrix} {}_{\bba a} & {}_{\bba \gA} & {}_{\bba}{}^a\end{pmatrix}  \quad       & d=6
	          \end{cases}$ \\
\hline
$\SL(2)$ & $\alpha=+,-$ &\\
\hline
$\mathbb{R}^+$ & $*$&\\
\hline
\end{tabular}
\end{center}
It is also helpful to keep in mind the generators and their respective decompositions for all the relevant groups.
\begin{center}
\begin{tabular}{|c|c|c|} 
\hline
Group & Generators & Decomposition \\ 
\hline\hline
$\Gduality$ & $T_{\Ah}$ & $\{ 
        T_a\,, T_{\grave{\alpha}}\,, T_{\acute{\alpha}}
    \}  = \{
        T_a\,, T_{\tilde{\alpha}}
    \} = \{  T_{\grave{a}}\,, T_{\acute{\alpha}}
    \} $\\
\hline
$\Gduality$ (adj) & $\{t_{\dot{\bm{\alpha}}}\}\,, \{t_{\bm{\alpha}}\}$ & --, $\{\tilde{t}_{\tilde{\bm{\alpha}}},\, t_{\hat{\bm{\alpha}}}\}$\\
\hline
$G$ & $\{ 
        T_a\,, T_{\grave{\alpha}}\}=\{  T_{\grave{a}}\}$ &\\
\hline
$H$ & $\{T_{\grave{\alpha}}\}$ &\\
\hline
$\OO(d,d+n)$ (adj) & $\{t^{\hat{\bm{\alpha}}}\}$ & $\{R^{a_1a_2} ,\,R^a_{\gA},\,K_{a_1}{}^{a_2},\, R^{\gA_1\gA_2} ,\,R_a^{\gA},\, R_{a_1a_2}\}$\\
\hline
$\Gdualityt$ & $\{\tilde{t}_{\tilde{\bm{\alpha}}}\}$ & $\begin{cases}
		                                         R^*                   & d\leq 5 \\
		                                         R_{\bba_1}{}^{\bba_2} & d=6\,,
	                                         \end{cases}$\\
\hline
\end{tabular}
\end{center}
\section{Duality algebra \texorpdfstring{$\OO(d,\mathfrak{n})$}{O(d,n)}}\label{app:dualityalgebra}
For completeness, here we write all the details of the Lie algebras $\OO(d,\mathfrak{n})$. Its generators are decomposed according to \eqref{generators} and satisfy the following commutation relations:
\begin{align}
	\begin{alignedat}{2}
		[K^a{}_b,\,K^{c}{}_d] &= \delta^c_b\,K^a{}_d-\delta^a_d\,K^c{}_b\,,&
		[K^a{}_b,\,R_{\gC\gD}]&=0\,,\\
		[K^a{}_b,\,R^{c}_{\gC}] &= \delta^c_b\,R^{a}_{\gC}\,, &
		[K^a{}_b,\,R_c^{\gC}] &= -\delta_c^a\,R_b^{\gC}\,,\\
		[K^a{}_b,\,R^{cd}] &= 2\,\delta^{cd}_{be}\,R^{ae}\,,&
		[K^a{}_b,\,R_{cd}] &= -2\,\delta_{cd}^{ae}\,R_{be}\,,\\
		[R_{\gA\gB},\,R_{\gC\gD}] &=-2\,\bigl(\delta_{\gC[\gA}\,R_{\gB]\gD}-\delta_{\gD[\gA}\,R_{\gB]\gC}\bigr)\,, \qquad &
		[R_{\gA\gB},\,R^{cd}] &= 0\,,\\
		[R_{\gA\gB},\,R_{cd}] &= 0 \,,&
		[R_{\gA\gB},\,R^c_{\gC}] &= - 2\,\kappa_{\gC[\gA}\,\kappa_{\gB]}^{\gD}\,R^c_{\gD} \,,\\
		[R_{\gA\gB},\,R_c^{\gC}]  &= - 2\,\kappa^{\gC}_{[\gA}\,\kappa_{\gB]\gD} \,R_c^{\gD} \,, &
		[R^{ab},\,R^{cd}] &= 0\,,\\
		[R^{ab},\,R_{cd}] &= -4\,\delta^{[a}_{[c}\,K^{b]}{}_{d]} \,,&
		[R^{ab},\,R^{c}_{\gC}] &= 0 \,,\\
		[R^{ab},\,R_{c}^{\gC}] &= -2\,\kappa^{\gC\gD}\,\delta^{[a}_c\,R^{b]}_{\gD}\,,&
		[R_{ab},\,R_{cd}] &= 0\,,\\
		[R_{ab},\,R_{c}^{\gC}] &= 0 \,,&
		[R_{ab},\,R^{c}_{\gC}] &= -2\,\kappa_{\gC\gD}\,\delta_{[a}^c\,R_{b]}^{\gD}\,, \\
		[R^a_{\gA},\,R^b_{\gB}] &= - \kappa_{\gA\gB}\,R^{ab}\,,&
		[R^a_{\gA},\,R_b^{\gB}] &= -\kappa_{\gA}^{\gB}\,K^a{}_b - \delta^a_b\,\kappa^{\gB\gC}\,R_{\gA\gC}\,,\\
		[R_a^{\gA},\,R_b^{\gB}] &= - \kappa^{\gA\gB}\,R_{ab}\,.
	\end{alignedat}
\end{align}
Their fundamental representation can be expressed in term of the matrices
\begin{equation}
	\begin{aligned}
		(R_{\gC\gD})_A{}^B                                                                                   & \equiv
		\begin{pmatrix} 0 & 0                                                                         & 0 \\
                0 & \kappa_{\gC\gA}\,\delta_{\gD}^{\gB} - \kappa_{\gD\gA}\,\delta_{\gC}^{\gB} & 0 \\
                0 & 0                                                                         & 0
		\end{pmatrix}\,, \quad &
		(K^c{}_d)_A{}^B                                                                                      & \equiv
		\begin{pmatrix}
			\delta_a^c\,\delta_d^b & 0 & 0                        \\
			0                      & 0 & 0                        \\
			0                      & 0 & - \delta_d^a\,\delta_b^c \\
		\end{pmatrix}\,,                                                                                                                            \\
		(R_{c_1c_2})_A{}^B                                                                                   & \equiv
		\begin{pmatrix}
			0                       & 0 & 0 \\
			0                       & 0 & 0 \\
			2\,\delta_{c_1c_2}^{ab} & 0 & 0
		\end{pmatrix}\,,                                                                   &
		(R^{c_1c_2})_A{}^B                                                                                   & \equiv
		\begin{pmatrix}
			0 & 0 & 2\, \delta^{c_1c_2}_{ab} \\
			0 & 0 & 0                        \\
			0 & 0 & 0                        \\
		\end{pmatrix}\,,                                                                                                                                                 \\
		(R_c^{\gC})_A{}^B                                                                                    & \equiv \begin{pmatrix}
			                                                                                                              0                                & 0                           & 0 \\
			                                                                                                              - \delta_c^b\,\delta^{\gC}_{\gA} & 0                           & 0 \\
			                                                                                                              0                                & \kappa^{\gC\gB}\,\delta^a_c & 0 \\
		                                                                                                              \end{pmatrix}\,, &
		(R^c_{\gC})_A{}^B                                                                                    & \equiv
		\begin{pmatrix}
			0 & \delta^c_a\,\delta_{\gC}^{\gB} & 0                            \\
			0 & 0                              & -\kappa_{\gC\gA}\,\delta_b^c \\
			0 & 0                              & 0                            \\
		\end{pmatrix}\,.
	\end{aligned}
\end{equation}
As expected, one can check that they leave the following metric invariant:
\begin{equation}
	\eta_{AB} = \begin{pmatrix}
		0          & 0               & \delta_a^b \\
		0          & \kappa_{\cA\cB} & 0          \\
		\delta_b^a & 0               & 0
	\end{pmatrix}.
\end{equation}

\section{On the left invariance of the measure}\label{leftinv}
The second integral in \eqref{actionwithleft} is well-defined only if $v$ defines a good measure. To check when this is the case, we introduce a matrix representation of the generators of the gauge group $G$ as
\begin{align}
	(T_{\grave{a}})_{\Ah}{}^{\Bh} = -X_{\grave{a}\Ah}{}^{\Bh}\,.
\end{align}
Following \eqref{eqn:decompTgen}, they decompose into the generators $T_{\grave{\alpha}}$ of $H$ and the remaining $T_a$ that span the coset $G/H$ with dimension $d$. The matrix $M_{\Ah}{}^{\Bh}$, which is used in \eqref{eqn:ansatzGenFrame} to construct the generalized frame fields, is here be expressed as suggested in \eqref{eqn:paramM}. Moreover, we decompose the left hand side of \eqref{eqn:MinvdM} according to 
\begin{align}
	M^{-1}\,\rmd M = v^{a}\, T_{a} + \omega^{\grave{\alpha}}\, T_{\grave{\alpha}} \,.
\end{align}

Under left multiplication by a constant element of $g \in G$, the coset representative $M_{\Ah}{}^{\Bh}$ transforms as
\begin{align}
    M_{\Ah}{}^{\Bh}(x) \to \left[g\,M(x)\right]_{\Ah}{}^{\Bh}\,.
\end{align}
Alternatively, this can be reexpressed as a transformation of the coordinates $x^i \rightarrow x'^i$ accompanied by a compensating $H$-transformation from the right given by the element $h(x', x)\in H$. Under this action, we find that $v^{a}(x) =v_i{}^a(x)\,\rmd x^i $ transforms as
\begin{align}
	v^a(x') = (\text{Ad}_h)_b{}^a(x',\,x)\,v^b(x) \,,
\end{align}
where
\begin{align}
	h\,T_a\,h^{-1} \equiv (\text{Ad}_h)_a{}^b\,T_b + (\text{Ad}_h)_a{}^{\tilde{\beta}}\,T_{\tilde{\beta}}\,.
\end{align}
Without loss of generality one can take $x^i=0$ and relabel $x'$ as $x$ to find
\begin{align}
	v^a(x) = (\text{Ad}_h)_b{}^a(x,\,0)\,v^b(0)\,.
\end{align}
Form the transformation of the frame, one sees that the volume element changes according to
\begin{align}
	\rmd \mu(x) \equiv \tfrac{1}{d!}\epsilon_{a_1\cdots a_{d}}\,v^{a_1}(x)\wedge \cdots \wedge v^{a_{d}}(x)
	= \det\bigl(\text{Ad}_h(x,\,0)\bigr)\,\rmd \mu(0)\,.
\end{align}
Next, we want to show that the determinant of the adjoint action is 1 and therefore the measure is left-invariant. One way to do so is to parametrize $\text{Ad}_h(x,0)$ in terms of functions $\gamma^{\grave\alpha}$ as
\begin{equation}
    (\text{Ad}_h)_b{}^a(x,\,0)=\exp\bigl(\gamma^{\grave{\alpha}}(x)\,T_{\grave{\alpha}}\bigr)_b{}^a\,.
\end{equation}
From the form of the geometric algebra we already discussed, we conclude
\begin{align}
	\det\bigl(\text{Ad}_h(x,\,0)\bigr) =\exp\bigl(- \gamma^{\grave{\alpha}}(x)\,X_{\grave{\alpha}b}{}^b\bigr)\,,
\end{align}
and furthermore we see that
\begin{align}
	X_{\grave{\alpha}b}{}^{b} = \beta^{-1}\,\vartheta_{\grave{\alpha}}\,.
\end{align}
If the $\vartheta_{\grave{\alpha}}$ components of the trombone gauging are absent then the measure becomes left invariant, namely
\begin{align}
	\rmd \mu(x) = \rmd^d x\,v(x) = \rmd \mu(0)
	    \qquad \text{with} \qquad 
    v\equiv \det(v_i^a)\,.
\end{align}

\section{Geometric algebra}\label{geomalg}
We present, here, the full structure of the studied geometric algebras.
In $d\leq 5$\,, this can be expressed as follows (note that $T_*$ is absent in $d\leq 4$)
\begin{align}
	T_{a}\circ T_{b} &= f_{ab}{}^c\,T_{c} + f_{ab}{}^{\gA}\,T_{\gA} + f_{abc}\,T^c\,,
	\nn \\
	T_{a}\circ T_{\gB} &= -f_{a}{}^c{}_{\gB} \,T_{c} + f_{a\gB}{}^{\gC}\,T_{\gC} + Z_a\,T_{\gB} - f_{ac\gB}\,T^c \,,
	\nn \\
	T_{a}\circ T^b &= f_a{}^{bc}\,T_{c} + f_a{}^{b\gC}\,T_{\gC} - f_{ac}{}^b\,T^c + 2\,Z_a\,T^b\,,
	\nn \\
	T_a \circ T_* &= (Z_a - f_a)\,T_*\,,
	\nn \\
	T_{\gA} \circ T_{b} &= f_b{}^{c}{}_{\gA}\,T_{c} - f_b{}_{\gA}{}^{\gB}\,T_{\gB} -Z_b\,T_{\gA} + f_{bc\gA}\,T^c\,,
	\nn \\
	T_{\gA}\circ T_{\gB} &= f_{c\gA\gB}\,T^c + \delta_{\gA\gB}\,Z_c\,T^c\,,
	\nn \\
	T_{\gA}\circ T^b &= -f_c{}^b{}_{\gA}\,T^c\,,
	\nn \\
	T_{\gA}\circ T_* &= - f_c{}^c{}_{\gA}\,T_*\,,
	\nn \\
	T^a\circ T_{b} &= -f_b{}^{ac}\,T_{c} - f_b{}^{a\gC}\,T_{\gC} + \bigl(f_{bc}{}^a+2\,\delta^a_b\,Z_c-2\,\delta^a_c\,Z_b\bigr)\,T^c \,,
	\nn \\
	T^a\circ T_{\gB} &= f_c{}^{a}{}_{\gB}\,T^c \,,
	\nn \\
	T^a\circ T^b &= f_c{}^{ab}\,T^c \,,
	\nn \\
	T^a \circ T_* &= - f_c{}^{ca}\,T_*\,,
	\nn \\
	T_*\circ T_b &=0\,,
	\nn \\
	T_*\circ T_{\gB} &=0\,,
	\nn \\
	T_*\circ T^b &=0\,,
	\nn \\
	T_*\circ T_* &=0\,.
\end{align}
In $d=6$\,, instead, we find
\begin{align}
	T_{+a}\circ T_{+b} &= f_{ab}{}^c\,T_{+c} + f_{ab}{}^{\gA}\,T_{\gA} + f_{abc}\,T_{-}{}^c\,,
	\nn \\
	T_{+a}\circ T_{-b} &= f_{a-}{}^+\,T_{+b} + \bigl(f_{ab}{}^c-f_a\,\delta_b^c \bigr)\,T_{-c} + f_{ab}{}^{\gC}\,T_{-\gC} + f_{abc}\,T_{-}{}^c\,,
	\nn \\
	T_{+a}\circ T_{+\gB} &= -f_{a}{}^c{}_{\gB} \,T_{+c} + f_{a\gB}{}^{\gC}\,T_{+\gC} + Z_a\,T_{+\gB} - f_{ac\gB}\,T^c\,,
	\nn \\
	T_{+a}\circ T_{-\gB} &= -f_{a}{}^c{}_{\gB} \,T_{-c} + f_{a-}{}^+\,T_{+\gB} + f_{a\gB}{}^{\gC}\,T_{-\gC} +(Z_a-f_a)\,T_{-\gB} - f_{ac\gB}\,T_{-}{}^c\,,
	\nn \\
	T_{+a}\circ T_{+}{}^b &= f_a{}^{bc}\,T_{+c} + f_a{}^{b\gC}\,T_{+\gC} - f_{ac}{}^b\,T_{+}{}^c + 2\,Z_a\,T_+{}^b\,,
	\nn \\
	T_{+a}\circ T_{-}{}^b &= f_a{}^{bc}\,T_{-c} + f_a{}^{b\gC}\,T_{-\gC} + f_{a-}{}^+\,T_{+}{}^b - f_{ac}{}^b\,T_{-}{}^c + (2\,Z_a-f_a)\,T_-{}^b\,,
	\nn \\
	T_{-a}\circ T_{+b} &= -f_{b-}{}^+\,T_{+a} + f_a\,T_{-b} \,,
	\nn \\
	T_{-a}\circ T_{-b} &= -f_{b-}{}^+\,T_{-a} + f_{a-}{}^+\,T_{-b} \,,
	\nn \\
	T_{-a}\circ T_{+\gB} &= f_a\,T_{-\gB} \,,
	\nn \\
	T_{-a}\circ T_{-\gB} &= f_{a-}{}^+\,T_{-\gB} \,,
	\nn \\
	T_{-a}\circ T_{+}{}^b &= \delta_a^b\,f_{c-}{}^+\,T_+{}^c + f_a\,T_-{}^b \,,
	\nn \\
	T_{-a}\circ T_{-}{}^b &= f_{a-}{}^+\,T_-{}^b + \delta_a^b\,f_{c-}{}^+\,T_-{}^c \,,
	\nn \\
	T_{+\gA} \circ T_{+b} &= f_b{}^{c}{}_{\gA}\,T_{+c} - f_b{}_I{}^{\gC}\,T_{+\gC} -Z_b\,T_{+\gA} + f_{bc\gA}\,T_{+}{}^{c}\,,
	\nn \\
	T_{+\gA} \circ T_{-b} &= f_b{}^{c}{}_{\gA}\,T_{-c} -f_c{}^{c}{}_{\gA}\,T_{-b} - f_b{}_{\gA}{}^{\gC}\,T_{-\gC} -Z_b\,T_{-\gA} + f_{bc\gA}\,T_{-}{}^{c}\,,
	\nn \\
	T_{+\gA}\circ T_{+\gB} &= f_{c\gA\gB}\,T_+{}^c + \delta_{\gA\gB}\,Z_c\,T_+{}^c\,,
	\nn \\
	T_{+\gA}\circ T_{-\gB} &= -f_c{}^c{}_{\gA}\,T_{-\gB} + f_{c\gA\gB}\,T_-{}^c + \delta_{\gA\gB}\,Z_c\,T_-{}^c\,,
	\nn \\
	T_{+\gA}\circ T_{+}{}^b &= -f_c{}^b{}_{\gA}\,T_+{}^c\,,
	\nn \\
	T_{+\gA}\circ T_{-}{}^b &= -f_c{}^b{}_{\gA}\,T_-{}^c - f_c{}^{c}{}_{\gA}\,T_-{}^b\,,
	\nn \\
	T_{-\gA}\circ T_{+b} &= f_c{}^c{}_{\gA}\,T_{-b} -f_{b-}{}^+\,T_{+\gA} \,,
	\nn \\
	T_{-\gA}\circ T_{-b} &= -f_{b-}{}^+\,T_{-\gA} \,,
	\nn \\
	T_{-\gA}\circ T_{+\gB} &= f_c{}^c{}_{\gA}\,T_{-\gB} + \delta_{\gA\gB}\,f_{c-}{}^+\,T_{+}{}^c\,,
	\nn \\
	T_{-\gA}\circ T_{-\gB} &= \delta_{\gA\gB}\,f_{c-}{}^+\,T_{-}{}^c \,,
	\nn \\
	T_{-\gA}\circ T_{+}{}^b &= f_c{}^c{}_{\gA}\,T_{-}{}^b\,,
	\nn \\
	T_{-\gA}\circ T_{-}{}^b &= 0\,,
	\nn \\
	T_{+}{}^a\circ T_{+b} &= -f_b{}^{ac}\,T_{+c} - f_b{}^{a\gC}\,T_{+\gC} + \bigl(f_{bc}{}^a+2\,\delta^a_b\,Z_c-2\,\delta^a_c\,Z_b\bigr)\,T_{+}{}^c \,,
	\nn \\
	T_{+}{}^a\circ T_{-b} &= -f_b{}^{ac}\,T_{-c} - f_c{}^{ca}\,T_{-b} - f_b{}^{a\gC}\,T_{-\gC} + \bigl(f_{bc}{}^a+2\,\delta^a_b\,Z_c-2\,\delta^a_c\,Z_b\bigr)\,T_{-}{}^c \,,
	\nn \\
	T_{+}{}^a\circ T_{+\gB} &= f_c{}^{a}{}_{\gB}\,T_{+}{}^c \,,
	\nn \\
	T_{+}{}^a\circ T_{-\gB} &= -f_c{}^{ca}\,T_{-\gB} + f_c{}^{a}{}_{\gB}\,T_{-}{}^c\,,
	\nn \\
	T_{+}{}^a\circ T_{+}{}^b &= f_c{}^{ab}\,T_{+}{}^c \,,
	\nn \\
	T_{+}{}^a\circ T_{-}{}^b &= f_c{}^{ab}\,T_{-}{}^c - f_c{}^{ca}\,T_{-}{}^b \,,
	\nn \\
	T_{-}{}^a\circ T_{+b} &= f_c{}^{ca}\,T_{-b} + \bigl(\delta^a_b\,f_{c-}{}^+ - \delta_c^a\,f_{b-}{}^+\bigr)\,T_{+}{}^c \,,
	\nn \\
	T_{-}{}^a\circ T_{-b} &= \bigl(\delta^a_b\,f_{c-}{}^+ - \delta_c^a\,f_{b-}{}^+\bigr)\,T_{-}{}^c \,,
	\nn \\
	T_{-}{}^a\circ T_{+\gB} &= f_c{}^{ca}\,T_{-\gB} \,,
	\nn \\
	T_{-}{}^a\circ T_{-\gB} &= 0 \,,
	\nn \\
	T_{-}{}^a\circ T_{+}{}^b &= f_c{}^{ca}\,T_{-}{}^b \,,
	\nn \\
	T_{-}{}^a\circ T_{-}{}^b &= 0 \,.
\end{align}

\bibliography{literature}

\bibliographystyle{JHEP}

\end{document}